\documentclass[aps,pra,showpacs,twoside,twocolumn,10pt]{revtex4-2}
\usepackage[colorlinks=true, citecolor=blue, urlcolor=blue]{hyperref}
\usepackage{epsfig,newlfont,amssymb,amsfonts,amsmath,bm,subfigure,palatino,mathtools,amsthm,braket,soul,enumitem,color,graphics,graphicx,times,physics,bbold,comment}
\usepackage{xcolor}
\usepackage{dsfont}
\usepackage{float}
\usepackage{tabularx}
\usepackage{array}
\usepackage[export]{adjustbox}
\usepackage{mathtools}
\usepackage{silence}
\newtheorem{proposition}{Proposition}

\newtheorem{observation}{Observation}

\newcolumntype{Y}{>{\centering\arraybackslash}X}
\begin{document}
\title{
Tradeoff between noise and banding in a quantum adder with qudits}

\author{Gaurang Agrawal$^{1,~2}$, Tanoy Kanti Konar$^2$, Leela Ganesh Chandra Lakkaraju$^2$, Aditi Sen(De)$^2$}
\affiliation{$^1$Indian Institute of Science Education and Research, Homi Bhabha Rd, Pashan, Pune 411 008, India}
\affiliation{$^2$Harish-Chandra Research Institute, A CI of Homi Bhabha National Institute,  Chhatnag Road, Jhunsi, Allahabad - 211019, India}

\begin{abstract}
    Quantum addition based on the quantum Fourier transform can be an integral part of a quantum circuit and proved to be more efficient than the existing classical ripple carry adder. Our study includes identifying the quantum resource required in a quantum adder in any arbitrary dimension and its relationship with the performance indicator in the presence of local noise acting on the circuit and when a limited number of controlled rotation operations is permitted, a procedure known as banding. We analytically prove an upper bound on the number of the controlled rotation gates required to accomplish the quantum addition up to an arbitrary defect in the fidelity between the desired and imperfect output. When the environment interacts with individual qudits, we establish a connection between quantum coherence and fidelity of the output.  Interestingly, we demonstrate that when banding is employed in the presence of noise, approximate circuits of constant depth outperform circuits with a higher number of controlled rotations, establishing a complementary relationship between the approximate quantum adder and the strength of the noise.   We exhibit that utilizing magnetic fields to prepare an initial state that evolves according to a one-dimensional spin chain for a specific amount of time can be a potential technique to implement quantum addition circuits in many-body systems.
\end{abstract}

\maketitle

\section{Introduction}

A quantum algorithm consists of a series of quantum gates that act on single or multiple sites, represented by unitary operations, resulting in more efficient execution of tasks as compared to existing classical algorithms \cite{nielsen_chuang_2010}. 
The imperative for device miniaturization and the groundbreaking discoveries of quantum algorithms, such as  Deutsch-Josza \cite{Deutsch_Jozsa}, Shor's factorization \cite{shor_1994, shor_pra_1995}  and Grover's search \cite{grover_arxiv_1996} algorithms which offer speedups over classical schemes \cite{mermin_book}, highlight the significance of building quantum computers. Following these theoretical proposals, they have been demonstrated in laboratories using physical substrates \cite{shor_experi_photonics_2007, silicon_experimental_deutsch_grover_2018}, despite the fact that quantum computers with a larger number of qubits \cite{Cirac2000Apr} are required to establish supremacy over their classical analogues \cite{arute_nature_2019, wu_prl_2021}.


On the other hand, determining the resources necessary for achieving quantum benefits in all quantum protocols, particularly in quantum algorithms, is a fascinating topic. 
The benefits of quantum states, such as entanglement \cite{horodecki2009}, coherence \cite{streltsov_rmp_2017} have been found to be the resource of these advantages; nevertheless, unlike quantum communication \cite{quantum_crypto_gisin_rmp_2002, aditi_quantum_comm_review_2011}, the quantum characteristics, i.e.,  quantum resources that are generated during the execution of a quantum algorithm responsible for quantum advantage over the classical algorithm still remains unsettled. 
To accomplish this, the role of entanglement in Shor's \cite{shors_entanglement_2006, multi_ent_grover_2012}, Grover's search \cite{multi_ent_grover_2012, global_ent_grover_2017}, Bernstein-Vazirani \cite{entanglement_bernstein_pra_2022}, and Harrow-Hassidim-Lloyd (HHL) \cite{aditi_hhl_2023} algorithms have recently been examined, yielding various counter-intuitive results.  For example,  it has been found that multipartite entanglement prohibits efficient execution of Bernstein-Vazirani \cite{entanglement_bernstein_pra_2022} and  HHL algorithms \cite{aditi_hhl_2023}, exhibiting that entanglement is necessary but not sufficient. However, in Shor's algorithm,  coherence is connected with the success probability \cite{plenio_shor_coh_prl_2022} while in the Bernstein-Vazirani algorithm,  coherence in the initial state is related to its performance \cite{entanglement_bernstein_pra_2022} (in this regard, see Ref. \cite{hillery_coh_deutsch_2016} for Deutsch-Josza  and Ref. \cite{aditi_hhl_2023} for HHL algorithms). 

Computing a high number of arithmetic operations in a short period of time is crucial for any computing device, whether it is classical or quantum. Numerous efficient classical algorithms exist for computing arithmetic operations on a classical computer, such as addition, multiplication, and division, with respect to both time complexity and depth \cite{mano2017digital}. In the quantum domain, analogous approaches include quantum addition based on quantum Fourier transform (QFT) \cite{zalka_arxiv_1998, draper_arxiv_2000} and periodicity estimation in a noisy environment \cite{barenco_pra_1996}. These kinds of quantum algorithms are useful not only for simple math operations but also for discrete logarithm and Shor's factorization problems \cite{shor_1994,beauregard_qic_2003,pavlidis_qic_2014}. In this work, we focus on the quantum addition algorithm which uses the QFT \cite{draper_arxiv_2000, paler_reversible_qft_2022} rather than  reversible quantum gates such as CNOT and Toffoli gates \cite{vedral_barenco_ekert_pra_1996, preskill_pra_1996, kaur_IEEE_2012, draper_arxiv_2004, Remaud2025Jan}. Further, it was also shown that even when some gates from a QFT-based quantum adder are removed, the performance in terms of the success probability does not change, displaying fault-tolerance \cite{blumel_pra_2015}.  
The choice to make the QFT-based adder was influenced by these earlier findings regarding robustness against noise \cite{barenco_pra_1996, pavlidis_qic_2014}. Taking this into account can be crucial if one wants to create a quantum adder in the noisy intermediate-scale quantum (NISQ) computing that is currently available. We demonstrate that this is indeed  the case.  


Another key element addressed in this study is the higher dimensional quantum adder instead of involving two-dimensional systems, which has been found to be advantageous when compared to the analogous circuits using qubits \cite{pavlidis_pra_2021}.  It has been shown that quantum features can become prominent with an increase in dimension, despite the common belief that higher dimensional systems become macroscopic and behave like classical systems. Examples of these features include the increased violation of Bell inequality, key rates, fidelity in quantum teleportation, randomness generation, entanglement purification and efficient decomposition of the Toffoli gate \cite{Gisin1992Jan, PhysRevLett.49.901, bell_qudit, qkd_qudit1, qkd_qudit2, aditi_advanves_physics_2007, barry_qudit, random_qudit1, teleport_qudit1, teleport_qudit2, enta_qudit1, toffoli_qudit, wang_fp_2020}. 
These dimensional advantages have also been verified in numerous physical systems including   superconducting qudits \cite{superconducting_qudit}, silicon-photonic circuits \cite{nitrogen_qudit, silicon_qudit}, trapped ions \cite{qudit_iontrap1, qudit_iontrap2, qudit_iontrap3}, photonic circuits \cite{photon_qudit}. 



This paper examines a quantum addition circuit with d-dimensional quantum states that consists of three parts: QFT for encoding at the beginning and its inverse version for decoding at the end of the circuit and the controlled rotation gates, also known as the SUM circuit, (see Fig. \ref{fig:schematics_qft_adder}). Our study addresses three main questions after we modify the QFT and SUM circuits in two ways: (1) the effect of reducing the number of controlled rotations in the SUM circuit, known as the banded circuit or approximate addition, on the fidelity between the desired output and the affected one obtained after the reduction; (2) the impact on the performance when noise acts on the individual qudit after every controlled rotation in the QFT and SUM circuits; and (3) how performance is disturbed in the presence of dual imperfections, i.e., both the banding and noise present in the circuit. 

In addition to providing answers to these questions, we further establish a relationship between the fidelity and the quantum coherence formed following the QFT and SUM circuits, proving that, in the case of a quantum adder, the coherence of the output states is what makes the scheme work. More specifically, we present a bound on the minimum number of controlled rotations  required to reach the maximum fidelity in an approximate quantum addition , which decreases with increasing system dimensions.  We prove analytically that, in the presence of an arbitrary noise model, the change in coherence in a particular step of the circuit is directly proportional to the change in fidelity that occurred in that same step. We additionally illustrate that the adverse effect of noise can be minimised by increasing dimension, as evidenced by an increase in coherence. We provide an in-depth analysis of the number of controlled rotations necessary to attain increased fidelity in a noisy environment. Surprisingly, it reveals  a nonmonotonic behavior of fidelity with the number of controlled rotation gates, exhibiting the benefit of the banded circuit over the ideal one. Moreover, the order of controlled rotation for which the best fidelity is acquired saturates with the number of qudits, implying \textcolor{black}{ execution of SUM circuit and} subsequent addition of numbers in constant \textcolor{black}{depth $\mathcal{O}(1)$}.
 Additionally, 
 we present a spin chain architecture designed for laboratory implementation of quantum addition. 
Specifically, we demonstrate that the entire addition operation can be implemented on a one-dimensional spin model by time evolution and that a controlled rotation operation is possible to mimic if the state is evolved by a long-range cluster state Hamiltonian \cite{maciej_prl_longrangecluseter_2015}.

The paper is organized as follows. We first introduce the quantum addition algorithm in terms of qudits in Sec. \ref{sec:quantum_addition}. We study the role of coherence in the circuit in Sec. \ref{sec:coherence_addition}  and compare the performance between ideal and banded circuits in Sec. \ref{sec:banding}. The action of noise in terms of several CPTP channels is studied in Sec. \ref{sec:noiseandbanding}. Sec. \ref{sec:noisebanding}  contains the analysis of the dual  effects of noise and banding. We go onto showing how the quantum addition circuit can be implemented on a spin chain in Sec. \ref{sec:implementation} while the results are summarized  in Sec. \ref{sec:conclusion}.

\section{Quantum Addition and Coherence as its resource}

\label{sec:quantum_addition}

\begin{figure*}
\centering
\includegraphics[scale=0.23]{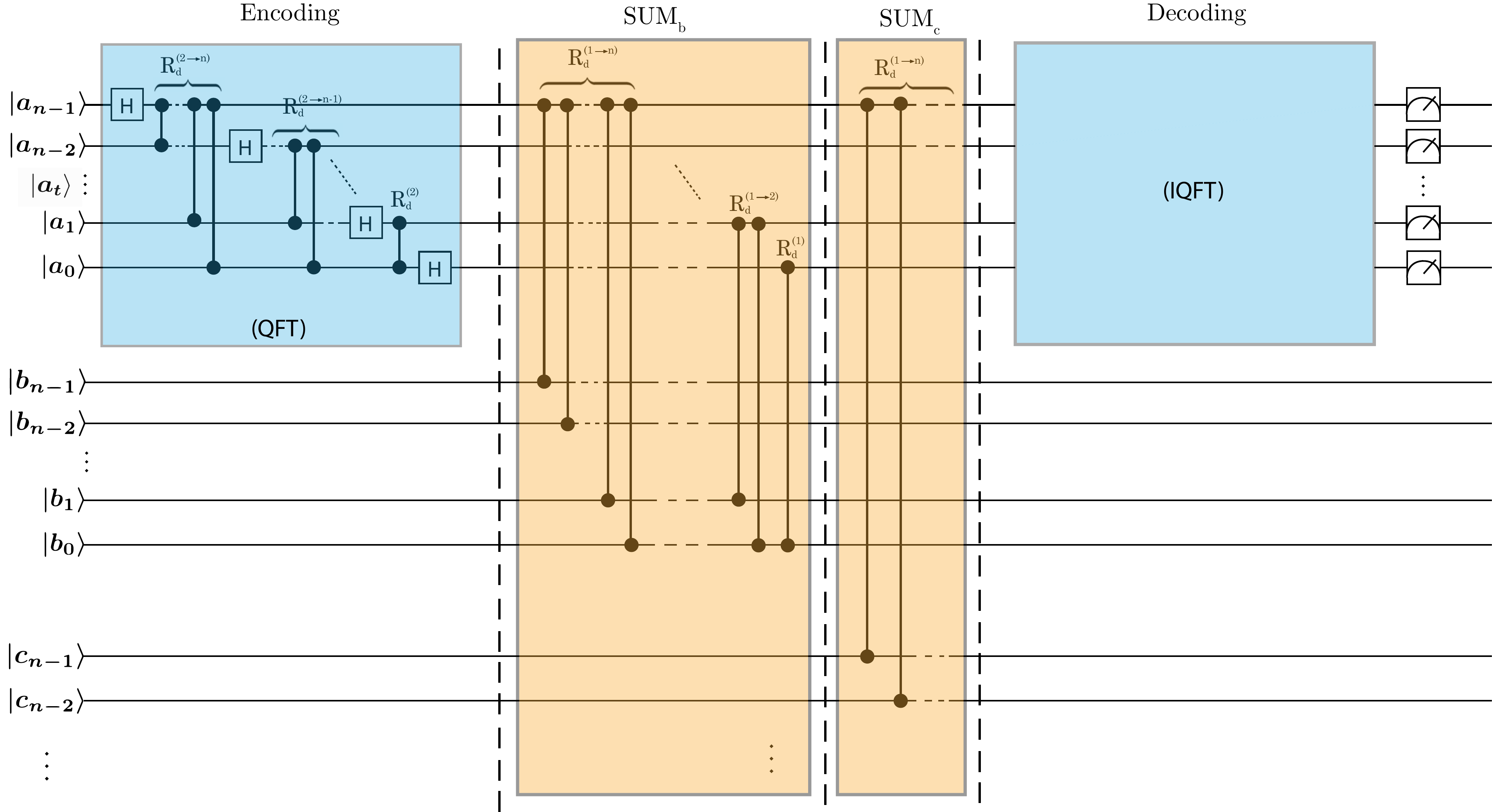}
\caption{\textbf{Layout of the  addition circuit with qudits based on QFT.} The first part of the circuit (highlighted in blue) shows the encoding where the input qudits are Fourier transformed as mentioned in Step 1. The second region, marked as SUM (highlighted in cream) represents the adder circuit where the second input is added to the Fourier transformed of the first. The last part (again in blue) is the decoding step which is carried out via  inverse Fourier transform (IQFT) to the states, and provides the sum of the input.}
\label{fig:schematics_qft_adder}
\end{figure*}

We briefly discuss here the quantum addition circuit in order to make the analysis self-sufficient. The initial proposal which is based on gate implementation using Toffoli and CNOT gates requires \(3n\)-qubits to add two \(n\)-bit numbers \cite{vedral_barenco_ekert_pra_1996, preskill_pra_1996, gossett_arxiv_1998}. Later on, the quantum adder gets modified in terms of the resources by using the quantum Fourier transform  which requires $2n+2$ number of qubits and  $O(n^2)$ depth for adding two \(n\)-bit numbers \cite{draper_arxiv_2000}. However,  subsequent addition of numbers can be done in linear time, since only a single input is required to be Fourier transformed. Moreover, note that the gate-based quantum adder uses more number of auxiliary qubits although it is better in terms of depth than the QFT-based ones  \cite{draper_arxiv_2004} (cf. Ref. \cite{Cuccaro2004Oct}). 
It is important that we perform our calculation using qudits, i.e., quantum systems where each subsystem has dimension $d$ in order to reduce the corresponding number of subsystems required to perform addition and to decrease the depth of the circuit \cite{pavlidis_pra_2021}. The quantum addition algorithm consists of three components \cite{draper_arxiv_2000} - (i) encoding involving QFT, (ii) quantum sum, and (iii) decoding carried out by inverse QFT (IQFT). 
In Fig.  \ref{fig:schematics_qft_adder}, we present a schematic diagram for the addition of two integers, \(a\) and \(b\) followed by an addition of integer $c$, represented in the base of \(d\) in a qudit circuit. Such a circuit can be generalized for an arbitrary number of addition of integers and for any real numbers although in order to represent these numbers in terms of number basis \(d\), we need to perform truncation. We choose addition between two integers for a simple illustration.
We elaborate the steps as follows.\\
{\it Step 1 -  Encoding.} The integers are encoded into the phase of the state by performing quantum Fourier transformation. The encoded integer \(a\) represented as \(a=a_0d^0+a_1d^1+a_2d^2+a_3d^3+...+a_{n-1}d^{n-1}\) in the number base \(d\),   \(a_i \in \{0,1,2,...,d-1\}\), each \(a_i\) is mapped to each element in the \(d\)-dimensional computational basis and the corresponding state after the QFT is given by
\begin{equation} 
    \mathcal{F}\ket{a_{n-1}...a_1a_0}=\frac{1}{\sqrt{d^n}}\bigotimes_{l = 0}^{n-1}\phi(a_l),
    \label{eq:fourier_transform}
\end{equation}
where \(\phi(a_l)=\sum_{k_l = 0}^{d-1}e^{i 2\pi \left( 0.a_{n-l-1}...a_0 \right)k_l}\ket{k_l}\). In addition, the second number, \(b\), is similarly encoded in the basis of another \(d\)-dimensional base. It can be realized in terms of quantum gates, \(d\)-dimensional Hadamard and controlled rotations respectively given by 
\begin{eqnarray}
    \nonumber H_d &=& \frac{1}{\sqrt{d}}\sum_m^{d-1}\sum_n^{d-1}e^{i \frac{ 2\pi a_m a_n }{d}}\ket{a_m}\bra{a_n}, \text{ and }\\
    R_d^{\tilde{q}} &=& \sum_m^{d-1}\sum_n^{d-1}e^{i \frac{ 2\pi a_m a_n }{d^{\tilde{q}}}}\ket{a_m}\bra{a_m} \otimes \ket{a_n}\bra{a_n},
    \label{eq:hadamard_d_cond_rot}
\end{eqnarray}
where order $\tilde{q} = a_m-a_n+1 $ is the difference of two digits (\(a_i \, \text{and}\, b_i\)) corresponding to either $a$ or $b$ for our case, although it can be arbitrary, in general.  Let us denote the maximum of \(\tilde{q}\) in the circuit by \(q\) which we use for all the calculations in  next sections. 
Summarizing,  after preparing the initial state as two integers in the computational basis, the QFT circuit can be implemented by applying the product of unitaries, $U_{step - 1} = \prod_{i = n-1 }^{0}\prod_{j = i-1}^{0}R_d^{a_i-a_j+1}H_d^{a_i}$, as given in Eq. (\ref{eq:hadamard_d_cond_rot}) (see Fig. \ref{fig:schematics_qft_adder})  where the controlled rotation depends on the values of the input $a_i, a_j$, and the Hadamard is applied on the $a_i$ qudit.
A possible implementation of these unitaries in laboratories is described in Sec. \ref{sec:implementation}.  

{\it Step 2 - Quantum Sum.} We employ the controlled rotation operation to add two numbers within the phases of their corresponding states in which  the Fourier-transformed qudits act as the target and  the other encoded integer, \(b\) serves as the control qudit. Suppose \(\ket{a_t}\) represents one of the Fourier-transformed qudits after passing through the QFT circuit. The state after the \((t+1)^{th}\) controlled rotation can be expressed as
\begin{eqnarray} 
&&\nonumber \frac{1}{\sqrt{d}} \sum_{k = 0}^{d-1}e^{i 2 \pi (0.a_ta_{t-1}...a_0)k}\ket{k} \xrightarrow{R_d^{(1)}\ldots R_d^{(t+1)}}\\&&
\frac{1}{\sqrt{d}}\sum_{k = 0}^{d-1} e^{i 2 \pi (0.a_ta_{t-1}...a_0 + 0.b_tb_{t-1}...b_0)k}\ket{k},
 \label
 {eq:summation}
 \end{eqnarray}
 where \(a_t a_{t-1}\ldots a_0\), and \(b_t b_{t-1}\ldots b_0\) are the bit strings of the integers to be added.
In the above equation, the information about the second integer \(b\) becomes encoded in the state of the first integer \(a\) by using the controlled rotation specified by $\tilde{q} = a_i - b_i +1, ~\forall i = \{n-1, \ldots, 1\}$. The output of the quantum sum circuit, as illustrated in Fig. \ref{fig:schematics_qft_adder}, labeled as ``SUM", essentially corresponds to the Fourier transform of the integer \(a+b\). Importantly, this operation does not alter the state of the integer \(b\). Thus, our focus remains on the state resulting from the controlled rotation of the integer \(a\), which is represented as \(\mathcal{F}\ket{a+b}\). 
The repeated application of controlled unitaries in this step can be represented  as $U_{step-2}= \prod_{i = n-1}^{0}\prod_{j = i}^{0}R_d^{a_i-b_j+1}$ such that the addition of  $b_i$ to the $a_i$ is performed in the phase of the qudits.

This circuit in Fig.  \ref{fig:schematics_qft_adder} can be generalized for addition of arbitrary numbers by subsequently adding another integer, say \(c\), which is useful in case of multiplication and in finding the mean of a given distribution \cite{beauregard_qic_2003, Ruiz-Perez2017Jun}.

\begin{figure}
    \centering
    \includegraphics[scale=0.21]{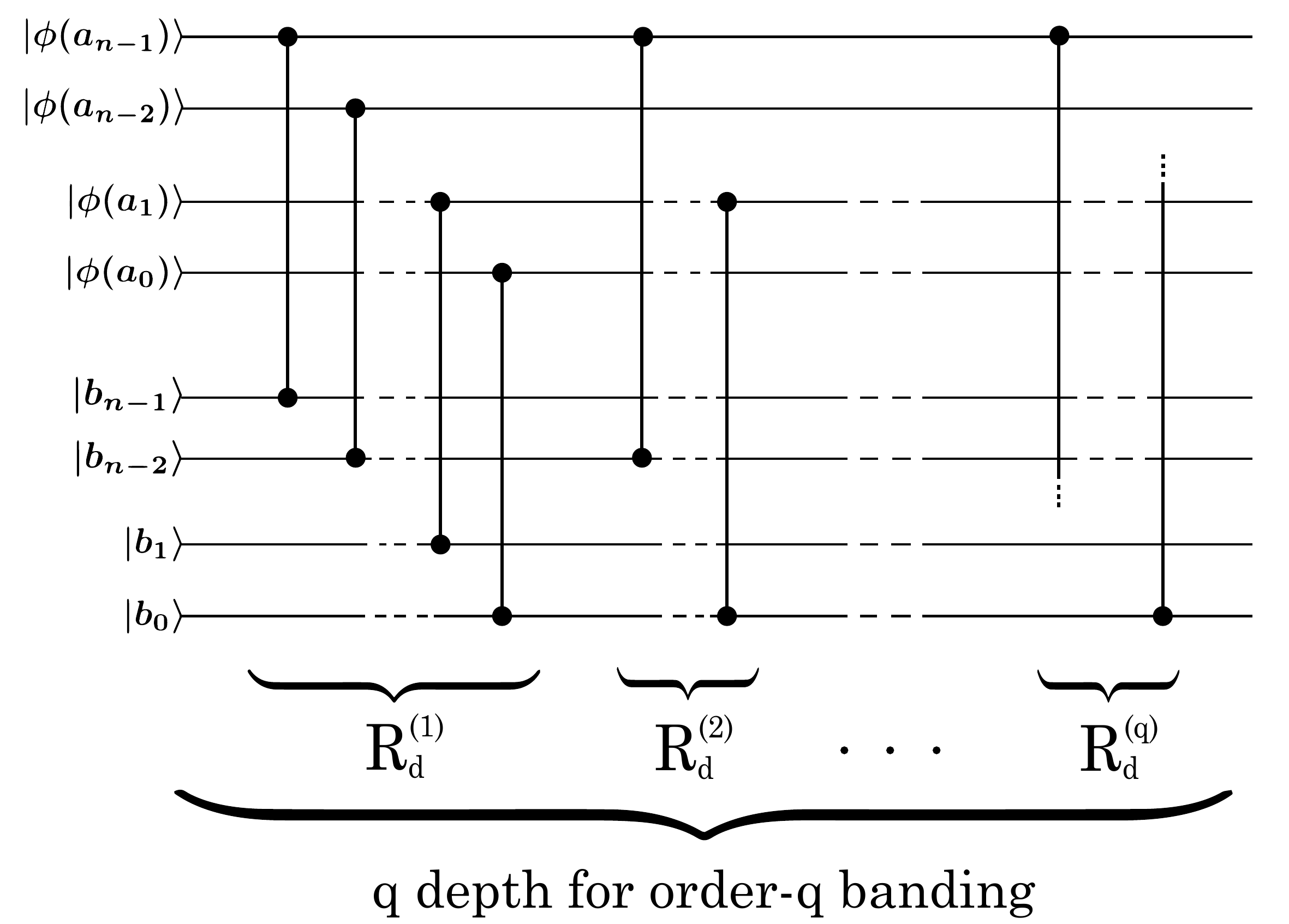}
    \caption{\textbf{Schematic diagram of the approximate SUM circuit.}  The circuit is a banding circuit when $q<n$, in which the controlled rotations  upto $q_{th}$ orders are applied. The total number of controlled rotations is $\frac{q(q+1)}{2}.$  
    As the controlled rotation gates can be implemented parallelly, the depth of the SUM circuit is equal to the highest order of controlled rotation, i.e., $q$.}
       \label{fig:qft_banding}
\end{figure}

{\it Step 3 - Decoding.} In order to retrieve the information of the summation, an inverse QFT (IQFT) is applied in \(\mathcal{F}\ket{a+b}\) which provides the state  $\ket{a+b} = \ket{(a+b)_{n-1}\ldots(a+b)_0}$, leaving the state $\ket{b_{n-1}\ldots b_{0}}$  unchanged. The decoding can be represented as $U_{step-3}  = U_{step-1}^\dagger$. 

Measurement on each qudit in their corresponding computational basis have to be performed to gain the information about each qudit. Note that $n$ qudits can encode an $n$ digit d-ary number which can lead to a modular summation (mod \((a+b)\)), while to perform the exact summation, we require $n+1$ qudits for an $n$ digit d-ary number. 
In this paper, we analyze the property of the state at different stages in the circuit.

\subsection{Coherence as a resource in quantum addition}
\label{sec:coherence_addition}

Quantum coherence \cite{streltsov_rmp_2017}, present in quantum states plays a vital role in several quantum information processing tasks, including quantum algorithms \cite{entanglement_bernstein_pra_2022,aditi_hhl_2023}, thermal devices like Otto engines \cite{camati_pra_2019}, quantum batteries \cite{Caravelli_quantum_2021}, quantum sensors \cite{matern_prb_2023} etc. In a quantum adder, all the controlled rotation operation commute with each other, and this parallelization immensely decreases the time required in the process, see Fig. \ref{fig:qft_banding}. Let us explore the behavior of quantum coherence in the step of quantum addition, both in the absence and presence of noise as well as with circuit deformation. Moreover, we connect the coherence created in quantum addition and the successful implementation of the scheme via fidelity. In contrast to algorithms such as Deutsch-Josza and Grover's search, where an initial superposition in states is essential, the quantum addition circuit does not require this condition. On the other hand, the QFT inherently generates the necessary coherence during its operation, enabling the desired quantum state manipulation without the need for an explicit initial superposition. This distinction underscores the unique role of the QFT in certain quantum algorithms, where coherence is achieved through circuit dynamics rather than state preparation.

The $l_1$-norm of coherence in the computational basis  \cite{Baumgratz_Prl_2014,streltsov_prl_2015} is defined as
\begin{equation}    
    C_{l_1}(\rho)= \sum_{i \neq j}\abs{\rho_{ij}},
    \label{eq:cohrence}
\end{equation}
where \(\rho_{ij}\) are the elements of the density matrix \(\rho\). The coherence of a maximally coherent state in the Hilbert space of dimension $D$ under this measure is $D-1$ where \(D=d^n\) with \(d\) being the Hilbert space dimension of individual subsystems  and \(n\) is the number of qudits. To ensure a fair comparison between all the dimensions, we scale the above measure as
\begin{equation}    
  C_{l_1}^\mathcal{N} \equiv C_{l_1}^\mathcal{N}(\rho) = \frac{\sum_{i \neq j}\abs{\rho_{ij}}}{D-1}.
\end{equation}

The results presented here also hold for the robustness of coherence defined in the \(d\)-dimensional Hilbert space, \(\mathbb{C}^d\) as
\begin{equation}
C_{R}(\rho)=\min _{\eta \in \mathcal{D}\left(\mathbb{C}^d\right)}\left\{s \geq 0 \mid \frac{\rho+s \eta}{1+s}=: \delta \in \mathcal{I}\right\},
\end{equation} 
where $\eta$ is an arbitrary state and $\mathcal{I}$ is the closed-convex set of all incoherent states. Robustness of coherence is the minimum weight of \(\eta\) that can be admixed with \(\rho\) such that the resulting state, \(\delta\), becomes incoherent \cite{napoli_prl_2016,piani_pra_2016}. 


\begin{observation} 
For the state produced at each step in the qudit quantum addition circuit,
there exists an incoherent unitary operation which can be applied 
to preserve the absolute values of all the elements in the state, thereby making \(l_1\)-norm coherence and robustness of coherence equal for noisy as well as noiseless scenario.
\end{observation}


\begin{proof}
Let us first illustrate it for a single qudit and then it can be easily generalized to all the qudits which are in a product state. Let us consider a single qudit state that appears in a circuit (with or without noise). It is typically of the form $ \rho_t = \sum_{k,l=0}^{d-1} C_{kl} {e^{i (\theta_k-\theta_l)}\ket{k}\bra{l}}$, where \(C_{kl}\) is some positive real number.  A single qudit unitary $U_t = \sum_{j=0}^{d-1}{e^{-i \theta_j}\ket{j}\bra{j}}$, which when applied on the state leads to  
\begin{align}
\tilde{\rho}_t=U_t\rho_t U_t^{\dagger} &= 
 \sum_{j=0}^{d-1}{e^{-i \theta_j}\ket{j}\bra{j}} \nonumber \\
&\times  \sum_{k,l=0}^{d-1}C_{kl}{e^{i (\theta_k-\theta_l)}\ket{k}\bra{l}} \nonumber  \\ 
&\times \sum_{m=0}^{d-1}{e^{i \theta_m}\ket{m}\bra{m}} \nonumber \\
&=\sum_{k,l = 0}^{d-1}C_{kl}{\ket{k}\bra{l}} = \sum_{k,l = 0}^{d-1}{\abs{\rho_{t_{kl}} }\ket{k}\bra{l}},
\end{align}
where $\rho_{t_{kl}}$ is the $(k,l)$ element of the state $\rho_t$. It is evident that coherence measures of the state before and after unitary operations coincide \cite{napoli_prl_2016,piani_pra_2016}, i.e., \(C_{l_1}(\tilde{\rho}_t)=C_{R}(\tilde{\rho}_t)=C_{l_1}({\rho}_t)=C_{R}({\rho}_t)\), although the corresponding states are not, in general, equal.
\end{proof}

\begin{figure}
    \centering
    \includegraphics[scale=0.35]{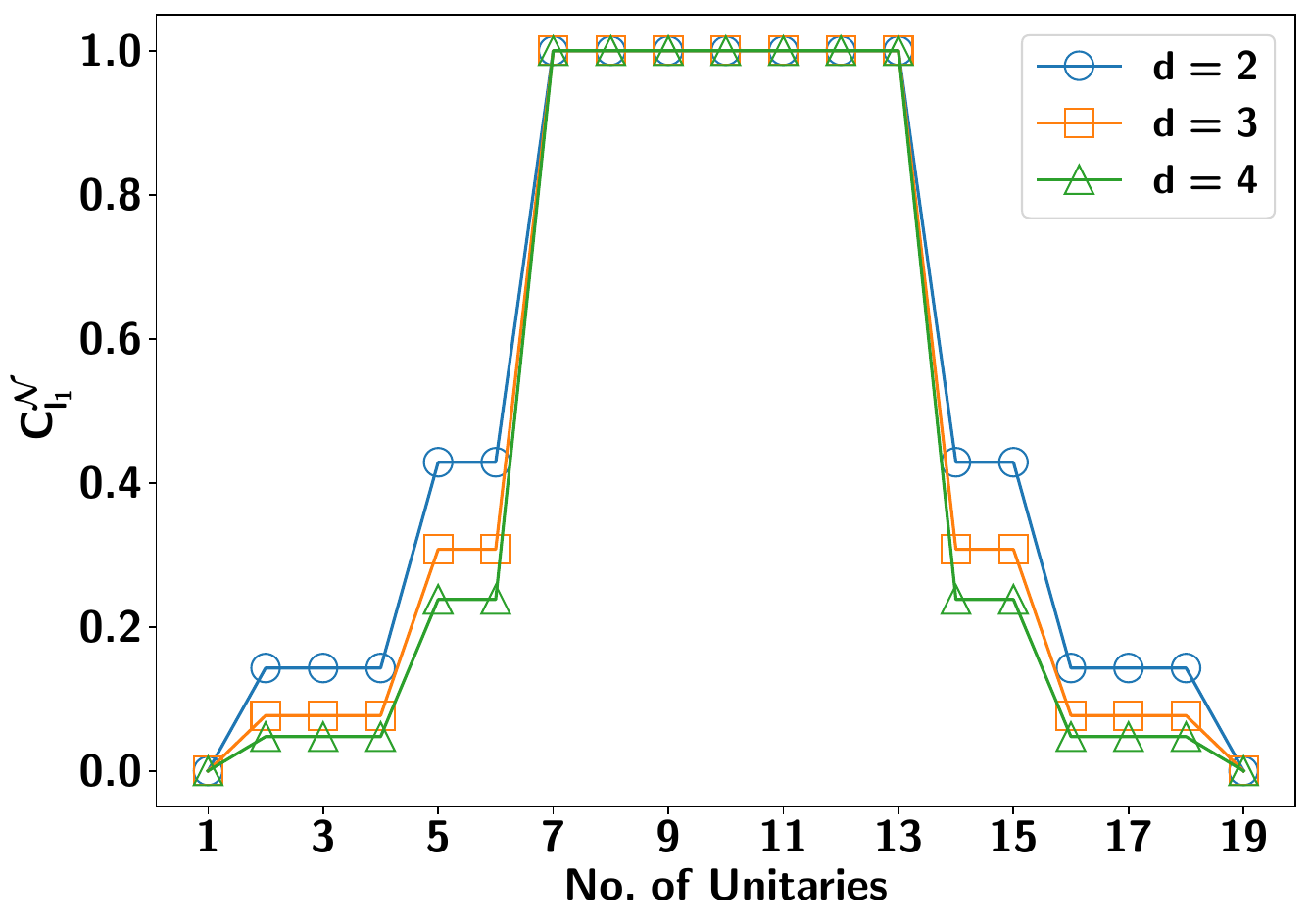}
    \caption{\textbf{Coherence (ordinate) vs the number of unitaries (abscissa).} \(C_{l_1}^{\mathcal{N}}\) is evaluated after the application of each unitary in the entire addition circuit for three qudits  (i.e., \(n=3\)) of dimensions \(2\), \(3\), and \(4\). In the absence of noise, the maximum occurs during the SUM circuit. All axes are dimensionless.}
    \label{fig:coherence_in_complete_circuit}
\end{figure}

In an ideal scenario, i.e., without noise and imperfection, we analyze the coherence of the state before and after the application of controlled rotation in different dimensions. Starting from the input state having vanishing quantum coherence, \(C_{l_1}\) increases with the number of unitaries applied, especially, after the application of Hadamard gates during QFT and IQFT (as shown in Fig. \ref{fig:coherence_in_complete_circuit}). For a perfect quantum adder, the coherence of an encoded state reaches its maximum value before entering the SUM circuit where it remains constant. Therefore, it shows that the production of a maximally coherent state can be used as an indicator to assess the perfect input for a quantum addition circuit. In the rest of the paper, the deviation from the maximally coherent state will be used to examine the noise or imperfection in the circuit.

\subsection{Effects on performance in approximate addition}
\label{sec:banding}

Due to the nature of the algorithm, as the number of qudits increases to encode larger numbers, the number as well as the order of the controlled rotation gates to be applied in the SUM circuit also increases and hence more decoherence can enter in the circuit. This leads to an undesirable outcome in terms of depth of the circuit as such precise control  with increasing circuit depth on the current NISQ hardware is hard to implement. As a remedy, the reduction in the controlled rotations corresponding to the highest order/finest angle, $\left(\frac{2 \pi}{d^{\tilde{q}}}\right)$, was proposed, known as {\it banding}, following the proposal of QFT banding \cite{barenco_pra_1996}. Such approximation results in the decrease of the success probability for obtaining the accurate sum of the two numbers \cite{draper_arxiv_2000}. We show that given the amount of allowed error \(\epsilon\), the order of the controlled rotations is upper bounded by a quantity which depends on the dimension. 


 We consider an approximate circuit in which the controlled rotation in the SUM circuit only goes up to the order \(q\), i.e.,  $\max \tilde{q} = q$ is applied as opposed to the order of n, which drastically reduces the number of the gates and, in turn, the depth of the circuit,  
 as shown in Fig. \ref{fig:qft_banding}. In our circuit, we require \(2\log_da+2\) number of qudits when \( a \geq b\) in order to perform non-modular quantum addition \footnote{The lowest bound for non-modular addition is $2n+1$ which can be achieved in the circuit we have analyzed by simply rejecting the $\ket{b_n} $ qudit} \cite{beauregard_qic_2003}. The approximate addition is more useful when the quantum circuit is affected by environmental noise as will be shown in the succeeding section.

To analyze the performance of the banded circuit, we first calculate the Uhlmann fidelity between the state after the banded SUM circuit and the output state of a non-banded SUM circuit, i.e., the circuit without approximation or banding. Mathematically, \(f=\expval{{\rho_{er}^\alpha}}{{\Psi^\alpha}}\), where $ \alpha \in \{in,out\}$,   the states before and after the non-banded SUM circuit are \(\ket{\Psi^{in}}\) and \(\ket{\Psi^{out}}\) respectively and in the presence of banding or any other circuit deformation, they are denoted as \({\rho_{er}^{in}}\) and \({\rho_{er}^{out}}\) respectively, with the subscript, ``er",  representing the errors. The fidelity after \(q\)-th controlled rotation for any arbitrary input is given as
\begin{align}
\label{eq:fidelity_banding}
    f^{q} &= \frac{1}{d} \prod_{t=0}^{n-1} \Bigg [ 1 + \frac{1}{d}\sum_{k = 0}^{d-1}2k \times \nonumber \\
   & \cos \Big (2\pi \times 0.\overbrace{00...00}^{m(q,t)}. b_{t-m(q,t)}...b_{0} \times (d-k)\Big )\Bigg ]  ,
\end{align}  
where \(m(q,t)=\min (q,t+1)\) and \textcolor{black}{\(b_{i}\left(i = 0,1,...\right)\)} are arbitrary inputs encoded in \(n\) qudits without performing Fourier transform.


 We address the question - ``what is the lower bound of \(q\) in order to obtain the final result up to an error \(\epsilon\) in arbitrary dimensions?" It was previously speculated  \cite{draper_arxiv_2000} that an approximate addition made of qubits can be efficiently computed with \(\log_2 n\) controlled rotations.

\begin{proposition}
    In an approximate quantum SUM circuit, the order of controlled rotations above which the fidelity \(f[w]\) can be obtained with an error \(\epsilon\) is given by
\begin{equation*}
    q\geq \frac{1}{2}\log_d \frac{(n-1)(d^2-1)\pi^2}{3\epsilon},
\end{equation*}
where \(d\) is the dimension of the subsystem and \(n\) is the number of qudits used in the implementation of the circuit.
\end{proposition}

\begin{proof}
We consider the input as the state for which \(b_i=d-1,~ \forall i\), denoted as $w$. It gives the worst fidelity from Eq. (\ref{eq:fidelity_banding}) and thus requires the highest order of controlled rotation gates as compared to any other input. The fidelity, $f[w]$, expression between the output of the banded and the non-banded circuits can be written as (see Appendix Eq. (\ref{eq:genfidelity}))
\begin{equation}
    f[w] = \prod_{i = q}^{n-1}\frac{1}{d}\left( 1+\frac{1}{d} \sum_{r = 1}^{d-1}2r\cos \left (\sum_{j=1}^{i-q+1}{\frac{x(d-1)(d-r)}{d^{j}}} \right) \right),
\end{equation}
where $q$ is the order of controlled rotation/ banding and \(x=\frac{2\pi}{d^q}\). We now check for the conditions \(f[w] \geq 1-\epsilon\) with \(\epsilon\) being small error. Note first that 
\begin{equation} 
\sum_{j = 1}^{i-q+1}{\frac{x(d-1)(d-r)}{d^{j}}} \leq \frac{x(d-1)(d-r)}{(d-1)} = (d-r)x,
\label{eq:angleineq}
\end{equation} 
and using the sum of the geometric series,  we get
\begin{equation} 
f[w] \geq \abs{\prod_{i = q}^{n-1}\frac{1}{d}\left (1+\frac{1}{d}\sum_{r = 1}^{d-1}2r\cos\left((d-r)x\right)  \right)}.
\label{eq:simp_inq}
\end{equation}
By taking Taylor's expansion around \(x=0\), we obtain 
\begin{align*} 
f[w] &\geq \abs{\prod_{i = q}^{n-1}\left({1-\frac{d^2-1}{12}x^2 + O(x^4)}\right)} \\ \nonumber &\geq  \left({1-\frac{d^2-1}{12}x^2 + O(x^4)}\right)^{(n-q)},
\end{align*}
where we ignore the fourth and other higher-order terms. A further simplified form of the inequality using the fact that \(q\geq 1\) is given as
\begin{multline} 
{f[w] \geq \left({1-(n-q)\frac{d^2-1}{12} x^2}\right)}\\ \geq
{\left({1-(n-1)\frac{d^2-1}{12}x^2}\right) \geq 1 - \epsilon}.
\label{eq:2nd_inq}
\end{multline}
Thus, the lower bound on fidelity is greater than or equal to $1-\epsilon$, when
\begin{eqnarray}
    \nonumber &&{(n-1)\frac{d^2-1}{12}\left( \frac{2\pi}{d^q} \right)^2} \leq \epsilon,\\
    && \Rightarrow q \geq \frac{1}{2} \log_d\left( \frac{(n-1)(d^2-1)\pi^2}{3 \epsilon} \right).
    \label{eq:boundtosimplify}
\end{eqnarray}
Note that the function is an increasing function of $n$ and scales logarithmically with \(n\) while it decreases with the increase of \(d\), thereby establishing an advantage of dimensions in terms of the circuit depth.
\end{proof}
\begin{figure} 
    \centering
    \includegraphics[scale=0.35]{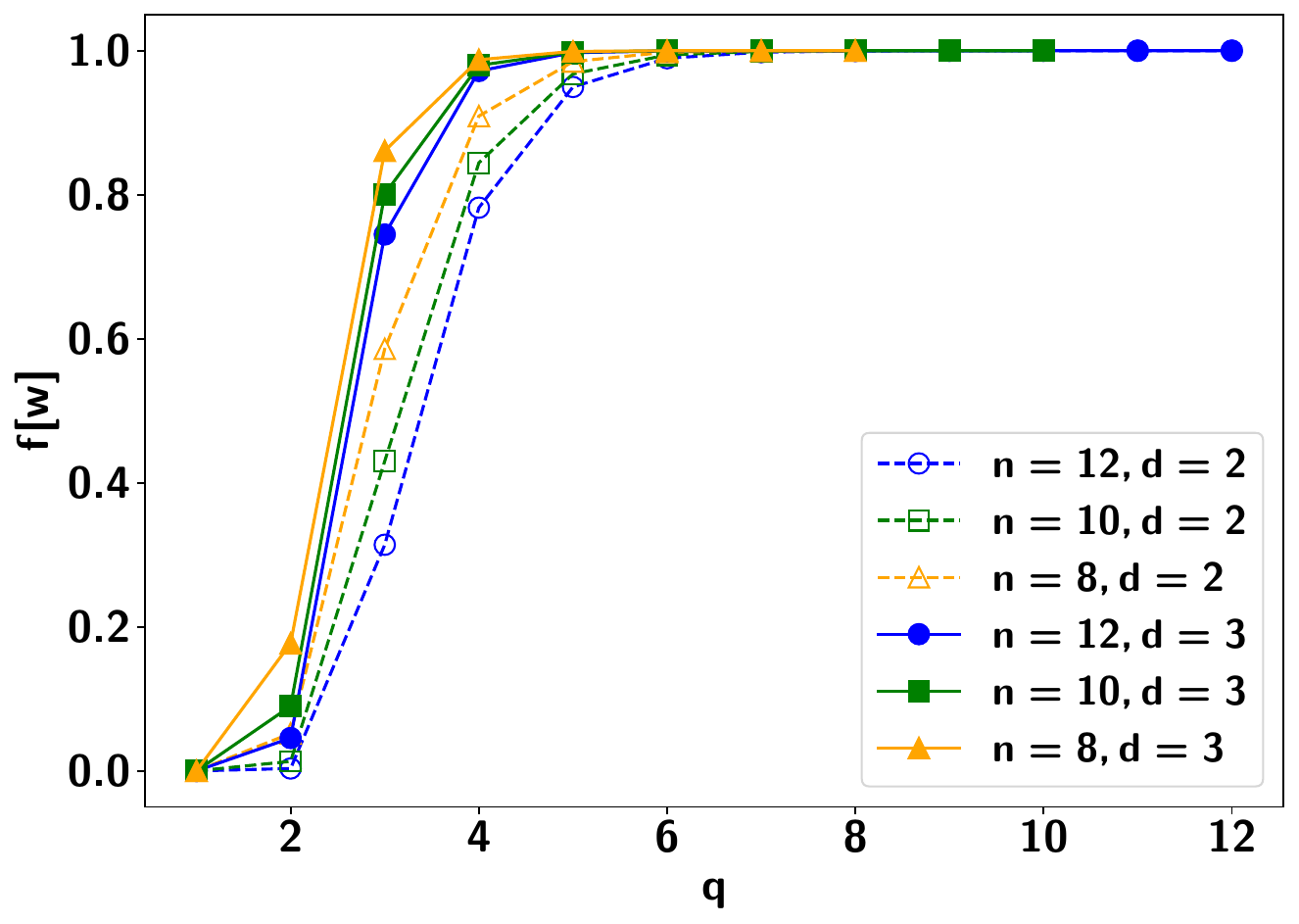}
    \caption{\textbf{Fidelity $f[w]$ against the order of controlled rotations/ banding, $q$ }.  Here no noise acts on the system and the fidelity is calculated for the worst possible input. We compute the fidelity where the summation is done on the qudit circuit, i.e., for two different $d$ values, $d=2 \text{ and } 3$ with different number of input qudits $n$.  
    All axes are dimensionless.}
    \label{fig:nonoiseonlybanding}
\end{figure}

This bound tells us the order of controlled rotations that, we require to implement in the approximate circuit to get the desired fidelity. Fig.  \ref{fig:nonoiseonlybanding} depicts \(f[w]\) with the increasing order of controlled rotation \(q\) by keeping  the dimension \(d\) and a number of qudits \(n\) fixed in the circuit. The fidelity gets saturated with an increment of the number of \(R_d^{q}\)s . Note also that the saturation is achieved faster with the increases of \(d\), which  establishes the usefulness of qudits in the quantum addition circuits.  Here we emphasize that we reduce the number of controlled gates popularly known as banding in the SUM circuit  as proposed in Ref. \cite{draper_arxiv_2000} while the QFT and IQFT steps are implemented exactly without any approximation. Note that in the previous work \cite{zilic2007scalingbetterapproximatingquantum},  the approximate QFT is studied in qudit circuits instead of the SUM circuit.  We now investigate if the dimensional advantage in terms of banding persists even in the presence of environmental noise.


\section{Quantum sum in  presence of noisy environment } 
\label{sec:noiseandbanding}

The order of controlled rotations in the quantum SUM circuit can be decreased in order to reduce the circuit depth  without compromising the efficiency in the output via the banding procedure. With the increase of the number of circuit operations, the effect of decoherence due to the presence of external environment also increases which can be suppressed by using banding. To understand the trade off between noise and the order of controlled rotations, we first study the action of environmental noise on the qudits, thereby the detrimental effect on a quantum adder.  In our analysis, we consider the case when the noise acts only after the controlled rotation on both controlled and target qudit, as implementation of the controlled gates takes much longer than the single qudit gate, like the Hadamard gate, involved in the circuit. Interestingly, the analysis of phase damping noise simplifies, since the control qudit in our circuit is always incoherent and hence such noise  has no effect on these states.

\subsection{Effects of noise without banding}
\label{subsec:noise_withoutbanding}

Let us first investigate how the fidelity between the noisy and ideal state, when noise is described by the completely positive trace preserving (CPTP) map that acts on the individual qudit. Since we are interested to connect these results with the banding step, we assume that the noise acts on both the quantum SUM and QFT circuit. Another aim is to establish a relation between the change in the fidelity and in  quantum coherence between the ideal and noisy scenarios.  We also demonstrate that quantum coherence is responsible for any change in the efficiency of the quantum addition circuit.



\begin{figure*}[ht] 
    \centering
    \includegraphics[scale=0.28]{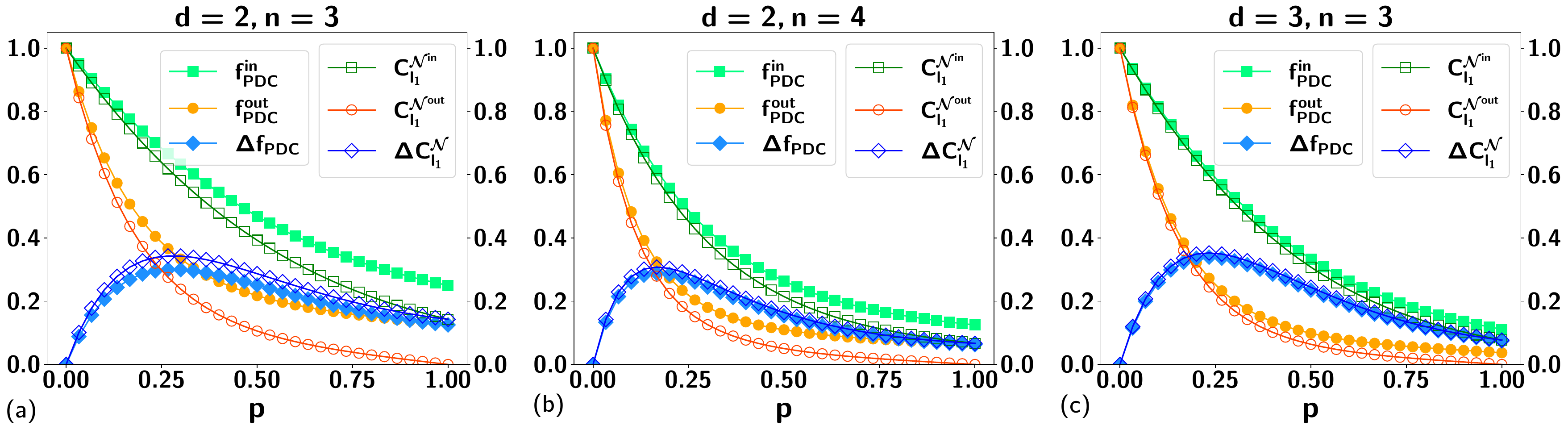}
    \caption{\textbf{Fidelity and coherence at the beginning and at the end of the SUM circuit along with their difference, defined in Eq. (\ref{eq:changeinfidelity}) (ordinate) vs the noise strength of the phase damping noise, $p$ (abscissa).} (a) 
    3 qubits, i.e., $d = 2, n = 3$, (b) for 4 qubits, i.e., $d= 2, n = 4$, and (c) for 3 qutrits, i.e., $d = 3, n = 3$. All axes are dimensionless. }
    \label{fig:dephasing_completeplot}
\end{figure*}

Without noise in the QFT circuit, the state having maximum coherence that enters the SUM circuit takes the form as
\begin{equation*}
    \rho^{in}=\ket{\Psi_s^i}\bra{\Psi_s^i} = \frac{1}{D}\sum_{j,k = 0}^{D-1}e^{i2 \pi (\phi_j-\phi_k)}\ket{j}\bra{k},
\end{equation*}
where \(\phi_j\) and \(\phi_k\) are arbitrary local phases and \(D=d^n\) is the total dimension of Hilbert space of the first integer. After the application of local noise on each qudit,  we consider that the state transforms as
\begin{equation} \label{eq:noise_model}
    \rho^{in} \xrightarrow{CPTP} \rho_{er}^{in}= \frac{1}{D}\sum_{j,k = 0}^{D-1}p_{jk}e^{i2 \pi (\phi_j-\phi_k)}\ket{j}\bra{k},
\end{equation}
where \(\sum_j p_{jj}=D\) and \(p_{jk}\geq 0 \). The above form of transformation occurs when phase damping, amplitude damping and depolarising noise among others act on states (see  Appendix \ref{app:matrix_element}). Evaluating the fidelity of the resulting state in Eq. (\ref{eq:noise_model}) with respect to the maximally coherent state obtained the ideal QFT circuit, we find 
\begin{equation}
\begin{split}
    f_{er}^{in}&=\expval{{\rho_{er}^{in}}}{{\Psi_s^i}}=\frac{1}{D^2}\sum_{j,k = 0}^{D-1}p_{jk} \\ & = \frac{1}{D^2} D \sum_{j,k= 0}^{D-1}{\abs{\rho^{in}_{er_{jk}}}}\\&=\frac{1}{D}\left(1+\sum_{j \neq k}^{D-1}{\abs{\rho^{in}_{er_{jk}}}}\right)\\&= \frac{(D-1)C_{l_1}^{\mathcal{N}}(\rho_{er}^{in})+1}{D}, 
\end{split}
\end{equation}
which is a linear relationship between coherence and fidelity. In a similar fashion, we again connect  coherence with the fidelity after the SUM circuit as
\begin{equation} \label{eq:fidcohrelation}
    f_{er}^{out}=\frac{(D-1)C_{l_1}^{\mathcal{N}}(\rho_{er}^{out})+1}{D}.
\end{equation}
From the above equations, we can safely conclude  that the coherence acts as a resource for obtaining a good performance in terms of  fidelity in a quantum adder, both in presence or absence of noise in the circuit. 

\subsection{Dephasing and amplitude damping channels on quantum adder}
\label{sec:dephasing}

To visualize the impact of noise on a quantum adder, we mainly focus on three types of paradigmatic noise models, namely, phase damping  (PDC), depolarising  (DPC), and amplitude damping channels (ADC) in qudit circuits. Apart from these local
noises, we investigate the effect of correlated Pauli noise channel (PC) in this circuit. These kinds of noise have been observed to occur in many experimental platforms \cite{Marques_natureSR_2015}. 

Before presenting the results, let us write down the Kraus operators. 
The Kraus operators of the PDC are given as  $M_0 = \sum_{i = 0}^{d-1}\sqrt{1-p}\ket{i}\bra{i}$ and $M_{i+1} = \sqrt{p} \ket{i}\bra{i} $ where $i \in \{ 0, \ldots, d-1 \}$ while in case of ADC \cite{dutta_pla_2016, Chessa_Giovannetti_Quantum_2023}, $M_0 = \sum_{k = 0}^{d-1}\sqrt{(1-kp)}\ket{k}\bra{k}$ and $M_i = \sum_{k = 0}^{d-i-1}\sqrt{p} \ket{k}\bra{k+i}$ where $i \in \{ 1, \ldots, d-1 \}$, and  $p$ is strength of system-environment interaction. However, unlike the PDC, analytical evaluation of the resulting state after the action of  local ADC  is cumbersome (see Appendix \ref{apdx:ampdamp_generalcriteria}).  

\begin{figure}
    \centering
    \includegraphics[scale=0.35]{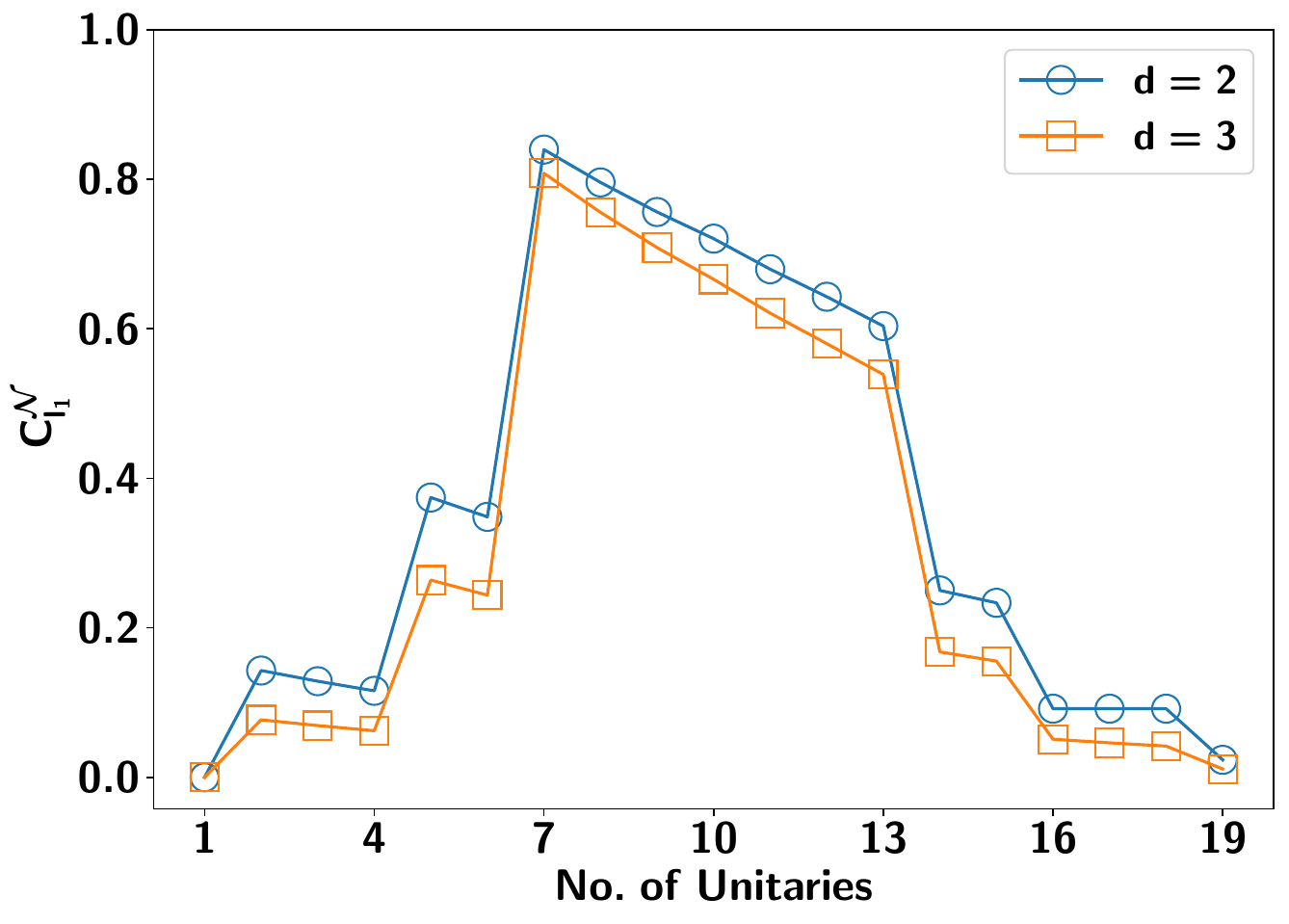}
    \caption{\textbf{Coherence (ordinate) with the number of unitaries  applied (abscissa) in the addition circuit of \(3\) qudit input (i.e., \(n=3\)).} Here phase damping noise of strength $p = 0.1$ acts on individual systems. 
    As shown in Fig. \ref{fig:coherence_in_complete_circuit} for the noiseless case, the maximum in presence of PDC also occurs   at the beginning of SUM circuit. Comparing \(C_{l_1}^{\mathcal{N}}\) in Fig. \ref{fig:coherence_in_complete_circuit} with this figure, it is evident that \(C_{l_1}^{\mathcal{N}}\) decreases with noise.    All axes are dimensionless.}
    \label{fig:coh_dep}
\end{figure}

When $nth$ order of controlled rotations are applied and the dephasing noise acts on both QFT and SUM circuits, the states before and after the SUM read respectfully as
\begin{align}
    \nonumber {\rho_{\text{PDC}}^{in}} &= \bigotimes_{t=0}^{n-1}{\frac{1}{{d}}\sum_{k,l = 0}^{d-1}(1-p)^{t(1-\delta_{k,l})}e^{i 2 \pi (0.a_ta_{t-1}...a_0)(k-l)}\ket{k}\bra{l}},\\
  \nonumber {\rho_{\text{PDC}}^{out}} &= \bigotimes_{t=0}^{n-1}\frac{1}{{d}}\sum_{k,l = 0}^{d-1}(1-p)^{(2t+1)(1-\delta_{k,l})}\\&\qquad \qquad \times e^{i 2 \pi (0.a_t...a_0+0.b_t...b_{0})(k-l)}\ket{k}\bra{l}.
\end{align}
The corresponding fidelities between the affected and the unaffected states are given by (see Appendix \ref{eq:gen_stateaftersum}) 
\begin{eqnarray}
    \nonumber f_{\text{PDC}}^{in} &=& \prod_{t = 0}^{n-1}{\frac{1}{d}\left( 1+ (d-1)(1-p)^{t} \right)}, ~ \text{and} \\
    \text{and} \, f_{\text{PDC}}^{out} &=& \prod_{t = 0}^{n-1}{\frac{1}{d}\left( 1+ (d-1)(1-p)^{2t+1} \right)}.
    \label{eq:outputfidelity}
\end{eqnarray}  
From Eq. (\ref{eq:fidcohrelation}), the \(l_1\)-norm of coherence before and after the SUM circuit 
are connected with respective fidelities as
\begin{equation} 
 C_{l_1}^{\mathcal{N}^\alpha} = \frac{d^n f_{PDC}^{\alpha} - 1}{d^{n}-1}, 
 \label{eq:cohfid_dephasing}
\end{equation}
where \(\alpha\in\{in, out\}\). As one expects, both the fidelity and the coherence decrease with the increase in the strength of the dephasing noise, $p$ after the application of each unitary (see Figs. \ref{fig:dephasing_completeplot} and \ref{fig:coh_dep}). A similar picture emerges also for other kinds of noise models. 


Although the trends of coherences and fidelities with noise are the same (comparing hollow symbols with solid ones), we define a quantity to exactly capture the change of fidelity or coherence that occurs due to the noise after the SUM circuit, i.e., 
\begin{equation}
    \Delta X = X^{in} - X^{out},
    \label{eq:changeinfidelity}
\end{equation} where $X$ is the fidelity $(f_{PDC})$ and coherence $(C_{l_1}^{\mathcal{N}^\alpha})$ of the state after the action of noise.  In this case, we obtain
\begin{equation}    \label{eq:changecohfid_dephasing}
\Delta f_{PDC} = \left(1-\frac{1}{d^n}\right)\Delta C_{l_1}^{\mathcal{N}} = \frac{\Delta C_{l_1}}{d^n},
\end{equation}
which implies that the proportionality constant gets closer to unity as the circuit size or the dimension of the system increases. This demonstrates that the loss of coherence in presence of noise is responsible for the loss of fidelity, and hence one can argue that  coherence is the resource for the performance of the SUM circuit even under noisy channels, satisfying Eq. (\ref{eq:noise_model}).

\begin{figure}
    \centering
    \includegraphics[scale=0.4]{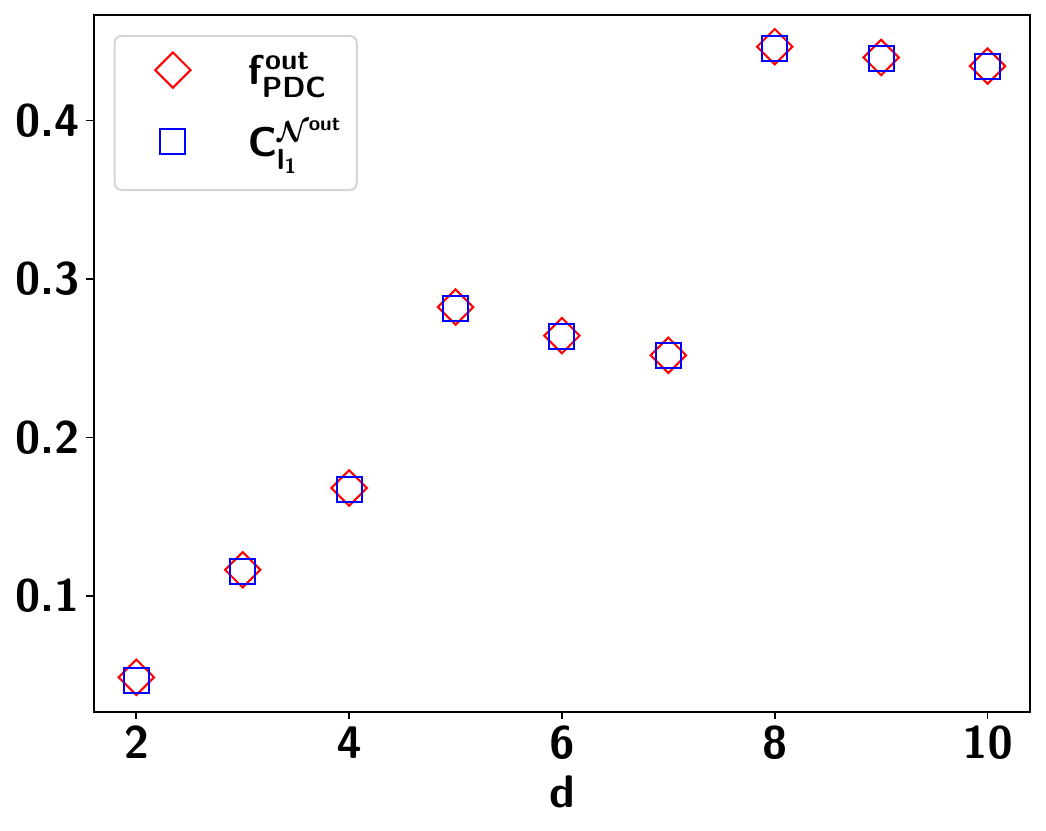}
    \caption{\textbf{Fidelity and coherence (vertical axis) evaluated after the SUM circuit as a function of dimensions, \(d\) (horizontal axis).} Here the circuits are made with minimum number of qudits to store and add $\max(a, b)=500$ under the phase damping noise of strength $0.1$.  Note that both the quantities coincide due to the large number of either $d$ or $n$. The increment in both fidelity and coherence demonstrate the dimensional advantage, since both fidelity and coherence increase with \(d\). All axes are dimensionless.}
    \label{dimcompare00}
\end{figure}

Instead of changing both the dimensions and the number of qudits for encoding, we fix the input and the strength of noise which actually fixes the number of qudits for encoding. E.g., for input  $v = \max(a,b) = 500$, with PDC having $p = 0.1$, one requires fewer qudits as the dimension increases to encode the number denoted as $ n = \lceil \log_d{v} \rceil$. In this respect, the output fidelity in Eq. (\ref{eq:outputfidelity}) modifies to be 
\begin{equation} 
f_{\text{PDC}}^{out} = \prod_{t = 0}^{\lceil \log_d{v} \rceil}{\frac{1}{d}\left( 1+ (d-1)(1-p)^{2t+1} \right)}.
\end{equation}

Both fidelity and coherence increase with the dimensions for circuits to add a constant value. As the dimensions increase, the circuit size reduces and the corresponding circuit of lower size can store the same value. This again signifies the dimensional advantage in the SUM circuit. \newline

\begin{figure} 
    \centering
    \includegraphics[width=\linewidth]{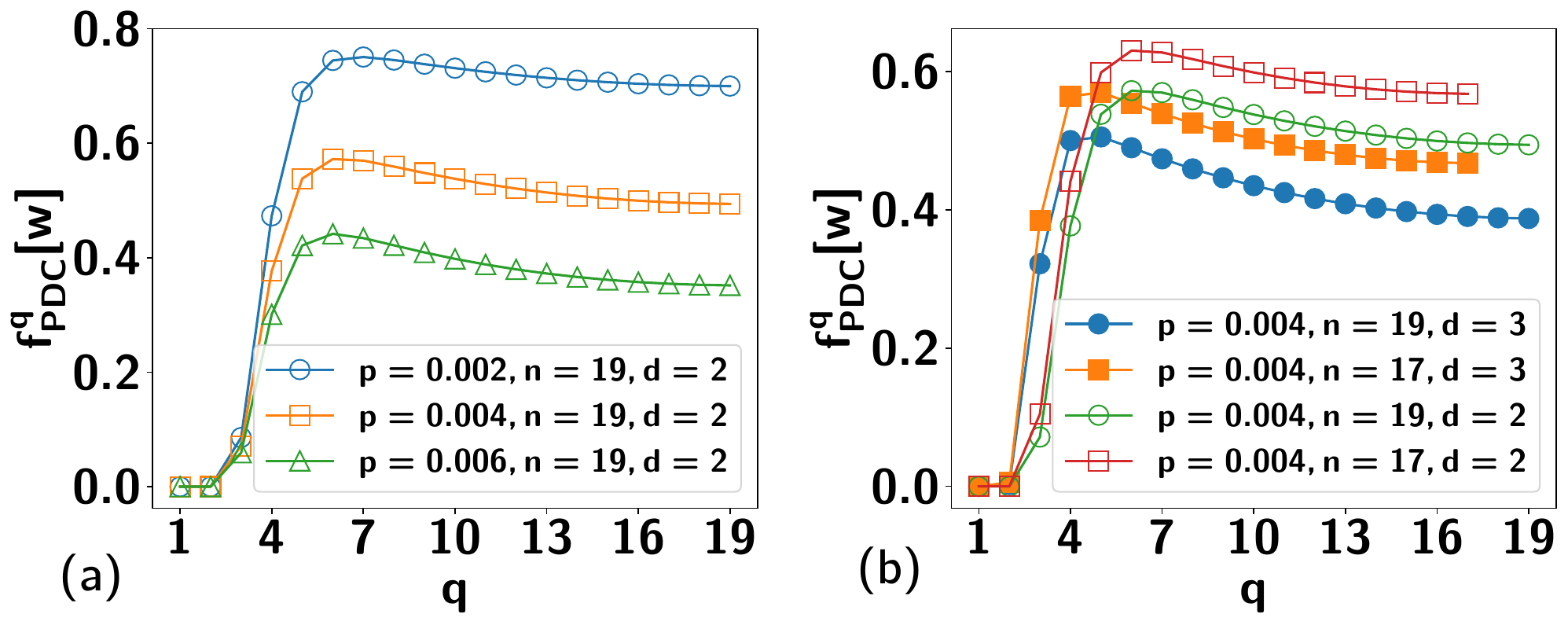}
    \caption{\textbf{Fidelity,  $f^q_{\text{{PDC}}}[w]$  (ordinate) of banded noisy addition circuits with respect to the banding order $q$ (abscissa) for the  worst input ($ \forall i ~ b_i = d-1 $).}  Here  phase damping noise $p$ acts locally  on the QFT and the SUM  circuits. The fidelity is computed between the desired output state and affected one. Note that the fidelity is independent of input $a$ as found in Eq. (\ref{eq:bandingfidelity}).   The plots show the advantage of banded circuit over the original circuit since the peak of the fidelity is achieved for banding order $q < n$.  (a) Fidelity of a  circuit with $19$ qubits in the presence of phase damping noises of different strengths $p$. (b) Comparison of  circuits between qubits and qutrits.  Fidelity of a circuit with $n$ qudits of dimension $d$ (\(d=2, 3\)) in the presence of phase damping noise of strength $0.004$. All axes are dimensionless.  }
    \label{fig:fidvscontrotqubit1}
\end{figure}

\begin{figure}[htb]
    \centering
    \includegraphics[width=\linewidth]{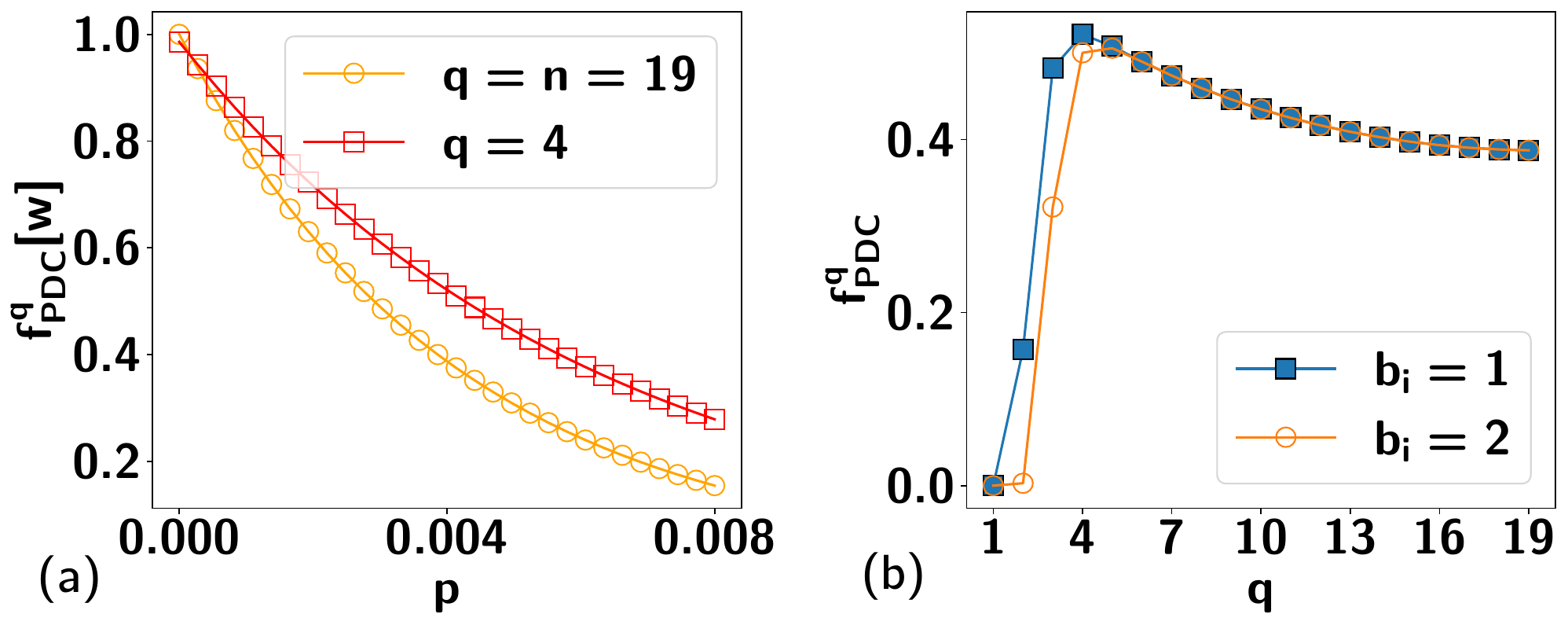}
    \caption{\textbf{Fidelity, $f^q_{\text{{PDC}}}[w]$  against the strength of noise $p$ (a) and $f^q_{\text{{PDC}}}$ with the   banding order $q$ (b) in a \(19\) qutrit circuit, showing the advantage of banding.}  The fidelity is evaluated for the worst possible input, i.e., $\forall i ~ b_i = 2$ and for \(19\) qutrit banding circuit. Note that the the banding circuit has consistently higher fidelity than the ideal circuit for all values of $p$ for a fixed \(q\) (in (a)). In (b), $p = 0.04$ with different inputs on 19 qutrits `worst' input -- all $ b_i = 2$, and `non worst' input -- all $b_i = 1$. The peak fidelity for `non worst' input is achieved for moderate value of \(q\) which is less than the value of $q$ for the worst input, showing that for all possible inputs, the maximum controlled rotations required are no more than the worst input case. All the axes are dimensionless.}
\label{fig:19qutritbanding}
\end{figure}

\section{Countering decoherence via banding in addition}
\label{sec:noisebanding}

We are now ready to study the juxtaposition of approximate addition in terms of banding in the SUM circuit and decoherence both in the QFT and the SUM circuit. Apparently, both approximate addition and noise (in Secs. \ref{sec:banding} and \ref{subsec:noise_withoutbanding}   respectively) individually can have a detrimental influence on the performance of quantum addition. However, we will now illustrate  that when they together act on the circuit, the effect may not always remain  destructive. We examine the worst possible input as it requires the highest order of controlled rotations to achieve the maximum fidelity as already done for banding alone. 

{\it Nonmonotonicity in fidelity.} We observe that for a fixed amount of local noise $p$ in the circuit, the fidelity does not increase monotonically with the increase in the order of the controlled rotation - nonmonotonicity with $q$ (as depicted in Fig. \ref{fig:fidvscontrotqubit1}). Specifically, we observe that for all the paradigmatic channels, $f^q_{\Lambda}[w]$ (\(\Lambda = \text{PDC, ADC, DPC, and PC}\)) reaches maximum with moderate value of $q$ before  almost saturating to a fixed value \textcolor{black}{or decreasing} with high $q$. Precisely, there exists an optimal order of controlled rotations $(q_{best})$ for which the maximum fidelity $(f_{\max})$ is achieved. Note that a similar behavior was found for the periodicity estimation algorithm which also uses QFT, where banding leads to a nonmonotonic behavior of fidelity with the order of controlled rotations \cite{barenco_pra_1996}. 



Let us try to analytically explain the above nonmonotonic nature of $f^q_{\Lambda}[w]$ with the variation of $q$, when the local dephasing channel of strength, $p$, acts on both QFT and sum circuits. Let us denote the states affected by noise and  controlled rotations upto  order \(q\)   as
\begin{widetext}
\begin{equation} 
{\rho_{PDC}^q} = \bigotimes_{t=0}^{n-1}\frac{1}{{d}}\sum_{k,l = 0}^{d-1}(1-p)^{(t+m(q,t))(1-\delta_{kl})}e^{i 2 \pi (0.a_t...a_0+0.b_t...b_{t-m(q,t)+1})(k-l)}\ket{k}\bra{l},
\end{equation}
and  in the ideal scenario (without noise and without banding) as 
\begin{equation} 
{\rho^0} = \bigotimes_{t=0}^{n-1}\frac{1}{{d}}\sum_{k,l = 0}^{d-1}e^{i 2 \pi (0.a_t...a_0+0.b_t...b_0(k-l)}\ket{k}\bra{l}.
\end{equation}  
The fidelity between the output state with phase damping noise and banding and the state obtained in the ideal case is given by
\begin{equation} 
\label{eq:bandingfidelity}
\begin{split}
    &f_{PDC}^q =\prod_{t=0}^{n-1}{\frac{1}{d}\Bigg(  1 + (1-p)^{(t+m(q+t))}\frac{1}{d}\sum_{k = 0}^{d-1}{2k\cos{(2\pi\times0.\overbrace{00...00}^{m(q,t)}b_{t-m(q,t)}...b_{0}(d-k))}} \Bigg )}.
\end{split}
\end{equation}
In case of the worst possible input as described before,  the fidelity expression reduces to
\begin{equation} \label{eq30}
\begin{split}    
    f_{PDC}^q[w] &= \prod_{i = 0}^{q-1}{\frac{1}{d}\left( 1+(d-1)(1-p)^{2i+1} \right)}\\
    &\times\prod_{i = q}^{n-1}\frac{1}{d}\Biggl( 1+(1-p)^{q+i}\frac{1}{d} \left(\sum_{r = 1}^{d-1}\left( 2r\cos\left( \sum_{c=1}^{i-q+1}{\frac{2\pi(d-1)(d-r)}{d^{q+c}}} \right) \right) \right) \Biggr).
\end{split}
\end{equation}
\end{widetext}
Analyzing the above expression, the nonmonotonicity observed in fidelity can be explained -- the cosine term increases as the banding order $q$ increases (see Fig. \ref{fig:nonoiseonlybanding}). It is responsible for the increase of fidelity, since the order of controlled rotations increases.  Along with that increment of \(q\),  the unfavorable influence of environment on the adder also increases leading to decoherence as evidenced from the $(1-p)^{q+i}$-term which leads to the decrease in fidelity. In case of moderate value of \(q\), the effects of noise on the fidelity is also moderate which leads to a high fidelity,  providing a high success probability in a quantum adder. This complementary effect is the reason for the nonmonotonic behavior of fidelity with the order $q$.   

\textcolor{black}{ Such a trade-off can be explained in terms of coherence and error accumulation. With a decrease in banding order, error accumulation increases as the circuit becomes more and more imperfect, while with an increase in banding order and thus with more controlled rotation gates, coherence decreases due to noise. This complementary behavior is the core reason behind the unimodality and nonmonotonicity. When banding order is very high, the higher ordered controlled rotations do not offer significant gains in fidelity but cause the same amount of noise, this is why they become redundant and removing them offers an advantage in fidelity. }


We perform the study for inputs that are worst from the perspective of banding or cosine parameter (i.e., they are chosen in such a way that the value of the cosine term is minimal) such that more controlled rotations are required for the best possible fidelity as compared to any other input (see Fig.  \ref{fig:19qutritbanding}). If we build a circuit with the order of controlled rotations greater than $q_{best}$, the fidelity will definitely be worse for all the inputs. However, if we consider the number of gates less than $q_{best}$, the fidelity for some of the inputs might be greater than  $f_{\max}$. Thus  $q_{best}$ is the optimal upper bound on the order of controlled gates in the noisy scenario. 

\begin{figure}[ht]
    \centering
    \includegraphics[width=\linewidth]{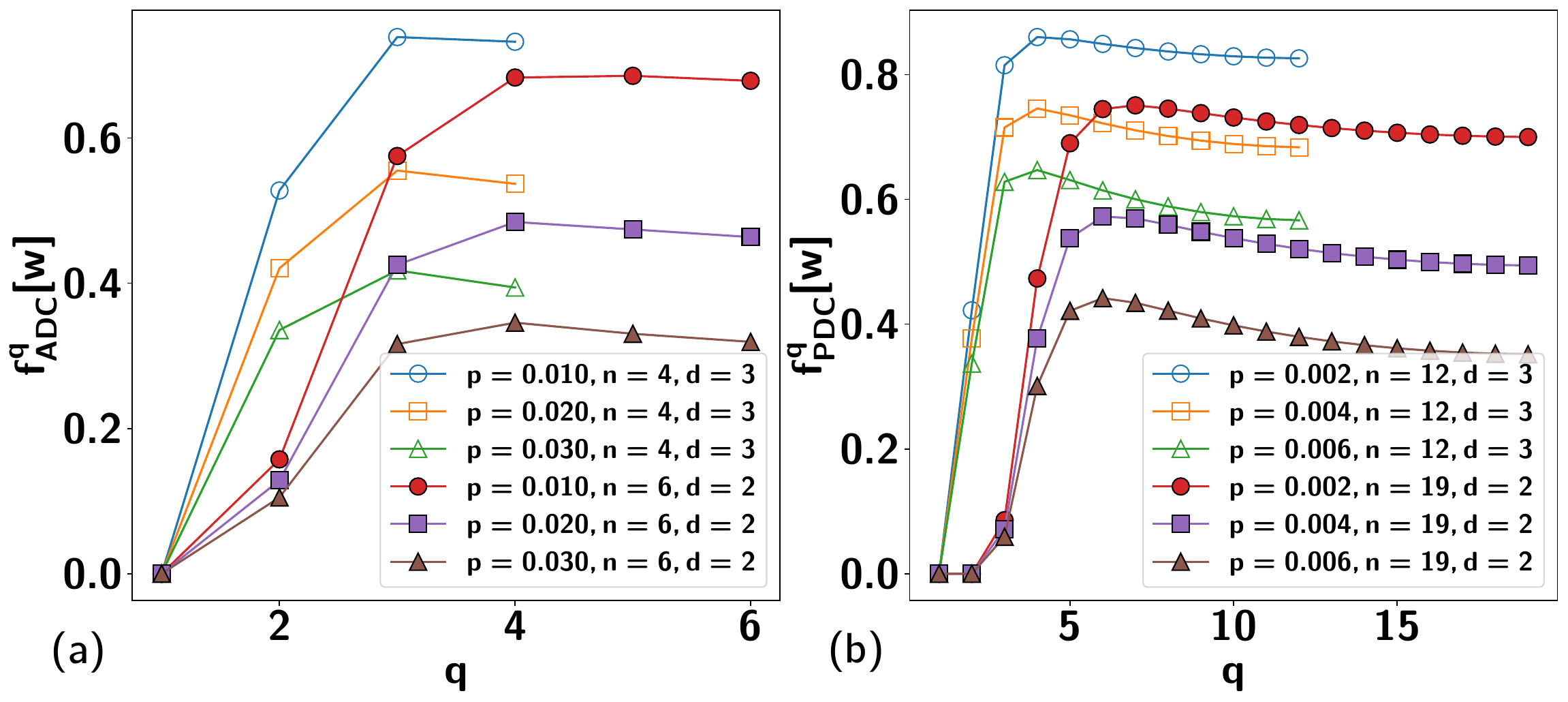}
    \caption{\textbf{ $f^q_{\Lambda}[w]$ (\(\Lambda\) representing ADC and PDC acting on circuits) (ordinate) 
   against $q$ (abscissa). } (a) Fidelity for the worst input on \(6\) qubits (solid symbols) and \(4\) qutrits (hollow symbols) under different noise strengths $p$ while in (b), it is for the worst input on \(19\) qubits and \(12\) qutrits with  different values of $p$. Note that while both the circuits can sum up to values of similar sizes, the performance of qutrit circuit is better for same value of noise strength.  Moreover, the maximum fidelity is achieved for moderate value of \(q\), reducing the circuit depth further. All axes are dimensionless. }
    \label{dimcompare}
\end{figure}

\begin{figure*}[ht]
\begin{subfigure}
    \centering
    \includegraphics[scale=0.3]{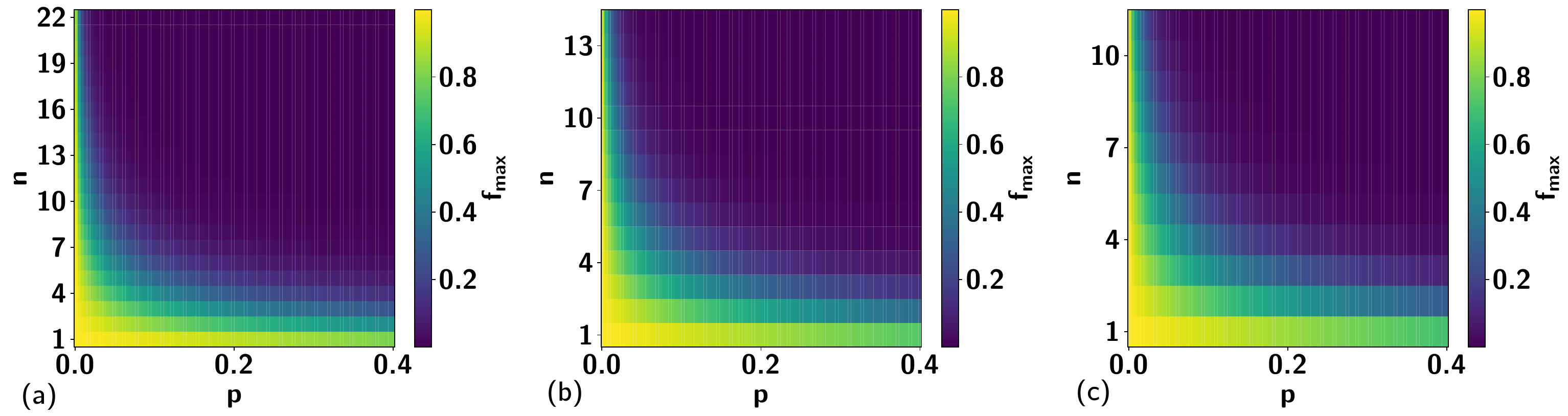}
    \label{finres1}
\end{subfigure}

\begin{subfigure}
    \centering
    \includegraphics[scale=0.3]{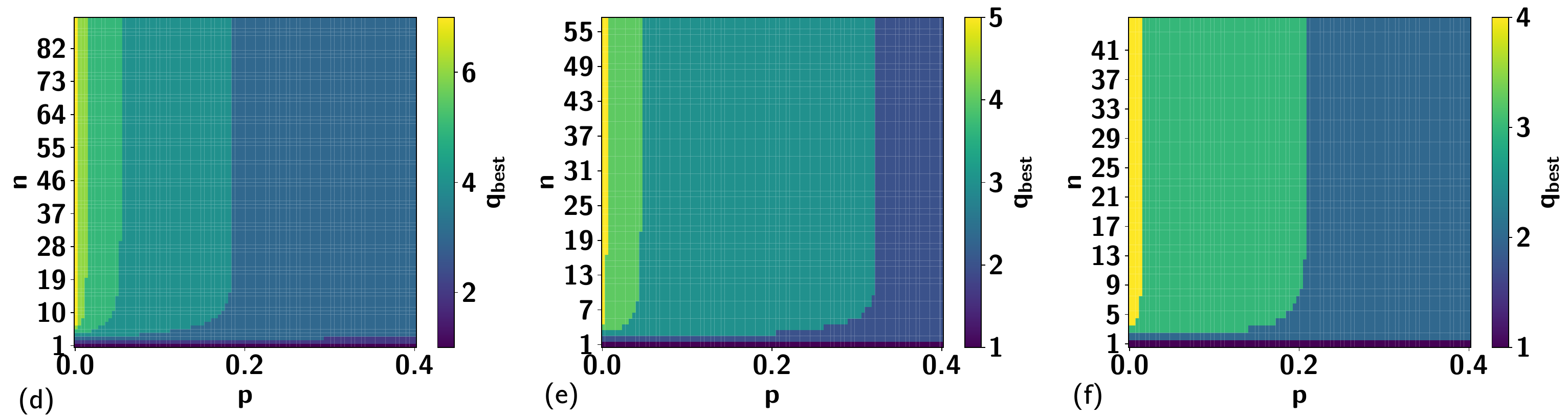}
\end{subfigure}
    \caption{\textbf{Map plot of maximum fidelity, $f_{max}$, ((a), (b) and (c)) and optimal banding order, $q_{best}$  ((d), (e) and (f)) for banding circuits in the presence of phase damping noise.} In the top row, the best achievable fidelity $(f_{\max})$ with respect to the ideal state at the end of the SUM circuit is plotted for the different input sizes, $n$ (ordinate), and noise values, $p$ (abscissa), for qubit circuits ($d = 2$). The corresponding $q_{best}$ that achieves $f_{max}$ is plotted in the bottom row. (a) and (d) are for qubits, (b) and (e) are for qutrits, and (c) and (f) are for ququarts. The different ranges for the $q_{best}$ are considered to establish distinguishability according to its color.
    All axes are dimensionless.}
    \label{fig:completebandinganalysis_dephasing}
\end{figure*}

{\it Dimensional gain.} It seems that $f_{\max}$  decreases as the dimension increases although $q_{best}$ improves (see Fig. \ref{fig:fidvscontrotqubit1}). To manifest the dimensional gain, we fix the total number of qudits depending on the input. For example, we notice that  $22$ qubits - $2^{22} = 4194304$, $14$ qutrits - $3^{14} = 4782969$ and $11$ ququarts - $4^{11} = 4194304$ require same dimensional complex Hilbert space. In this setting, there is a positive effect of decrement in $q_{best}$ and increment of $f_{\max}$ when the dimension increases [see Fig. \ref{dimcompare00}, \ref{dimcompare}(b), and Fig. \ref{fig:completebandinganalysis_dephasing}]. Although in a previous study \cite{zilic2007scalingbetterapproximatingquantum}, the number of gates are reduced in the AQFT circuit by increasing the dimension, we here show that the dimensional gain effects the optimal number of controlled rotations, $q_{best}$, and the fidelity in the SUM part of the circuit. 


{\it Variation of noise strength.} Let us analyze $q_{best}$ and $f_{\max}$ by varying noise strengths and total input qudits to establish their relationship. For a given range of noise strength, $q_{best}$ saturates with the number of input qudits, $n$. This can be a useful result in terms of building the hardware. Suppose, apriori, one knows the range of noise strength, i.e., $p \in \{p_{\min}, p_{\max}\}$. Since $q_{best}$ saturates with the variation of $n$ for a given $p$ value, one only needs to apply the said $q_{best}$ number of controlled rotations per qudit regardless of the number of qudits, leading to a constant depth and time complexity of the SUM circuit [see Fig. \ref{fig:completebandinganalysis_dephasing} (d)-(f)]. \textcolor{black}{Note that the depth is the order of controlled rotations in banding SUM circuit which is quite low (e.g. for noise strength $0.1-0.2$ for qubits, the order of controlled rotations, \(q\) goes only up to $5-6$ even for $90$ qubits and input number upto ~$10^{27}$). This finding represents a stronger and more novel claim than those known in literature.  To the best of our knowledge, this has not been explored in the existing literature on QFT-banding strategies. Specifically, we reduce the depth scaling from the existing $\mathcal{O}(n^2)$ and the speculative $\mathcal{O}(n \log{n})$ to $\mathcal{O}(1)$,} demonstrating that banding becomes significantly useful in the case of SUM circuit. \textcolor{black}{These circuits are particularly useful in situations where corrections to QFT-encoded inputs are required during the execution of a QFT-dependent algorithm. In such cases, when a QFT-encoded input is already provided, our method enables efficient corrections using only the SUM circuit. Rather than discarding the state after QFT and re-executing the entire QFT algorithm, our approach offers a correction mechanism with a circuit of constant depth.  Alternatively, our method can be described as a QFT-adder that approximately adds a QFT-encoded number $\mathcal{F} \ket{a}$ to another number $\ket{b}$, $(\mathcal{F}\ket{a}, \ket{b}) \rightarrow \mathcal{F}\ket{a+b}$, in constant depth and, therefore, in constant time. By repeated additions, the algorithm can also multiply a QFT-encoded number by another number, $(\mathcal{F}\ket{a}, \ket{b}) \rightarrow \mathcal{F}\ket{a \times b}$, in $\mathcal{O}(n)$ depth, achieving linear time complexity.} We again emphasize that  the advantage in terms of $q_{best}$ and $f_{max}$ is exhibited for the dephasing channel, all  the results presented here hold even for amplitude-damping and depolarizing noise. More specifically, in the case of a depolarizing channel, given by the transformation $\rho \longrightarrow \frac{p I_d}{d} + (1-p)\rho$, the corresponding maximally coherent state transforms as 
\begin{equation}
\begin{split}
&\sum_{i = 0}^{d-1}\frac{1}{d}\ket{i}\bra{i} + \sum_{i \neq j ~ i,j = 0}^{d-1}x_{ij}\ket{i}\bra{j}\\
&\longrightarrow \sum_{i = 0}^{d-1}\frac{p}{d}\ket{i}\bra{i} + (1-p)\left( \sum_{i = 0}^{d-1}\frac{1}{d}\ket{i}\bra{i} + \sum_{i \neq j ~ i,j = 0}^{d-1}x_{ij}\ket{i}\bra{j} \right)\\
& = \sum_{i = 0}^{d-1}\frac{1}{d}\ket{i}\bra{i} + (1-p)\left(\sum_{i \neq j ~ i,j = 0}^{d-1}x_{ij}\ket{i}\bra{j}\right),
\end{split}
\end{equation} 
which is the output of the dephasing channel. This directly shows why all the results also hold true for the depolarizing channel. In case of ADC, the similarity can be observed by comparing  Figs. \ref{dimcompare} (a) and (b) which we obtain by numerical simulation.

\begin{figure}[H]
\centering
\includegraphics[width=0.9\linewidth]{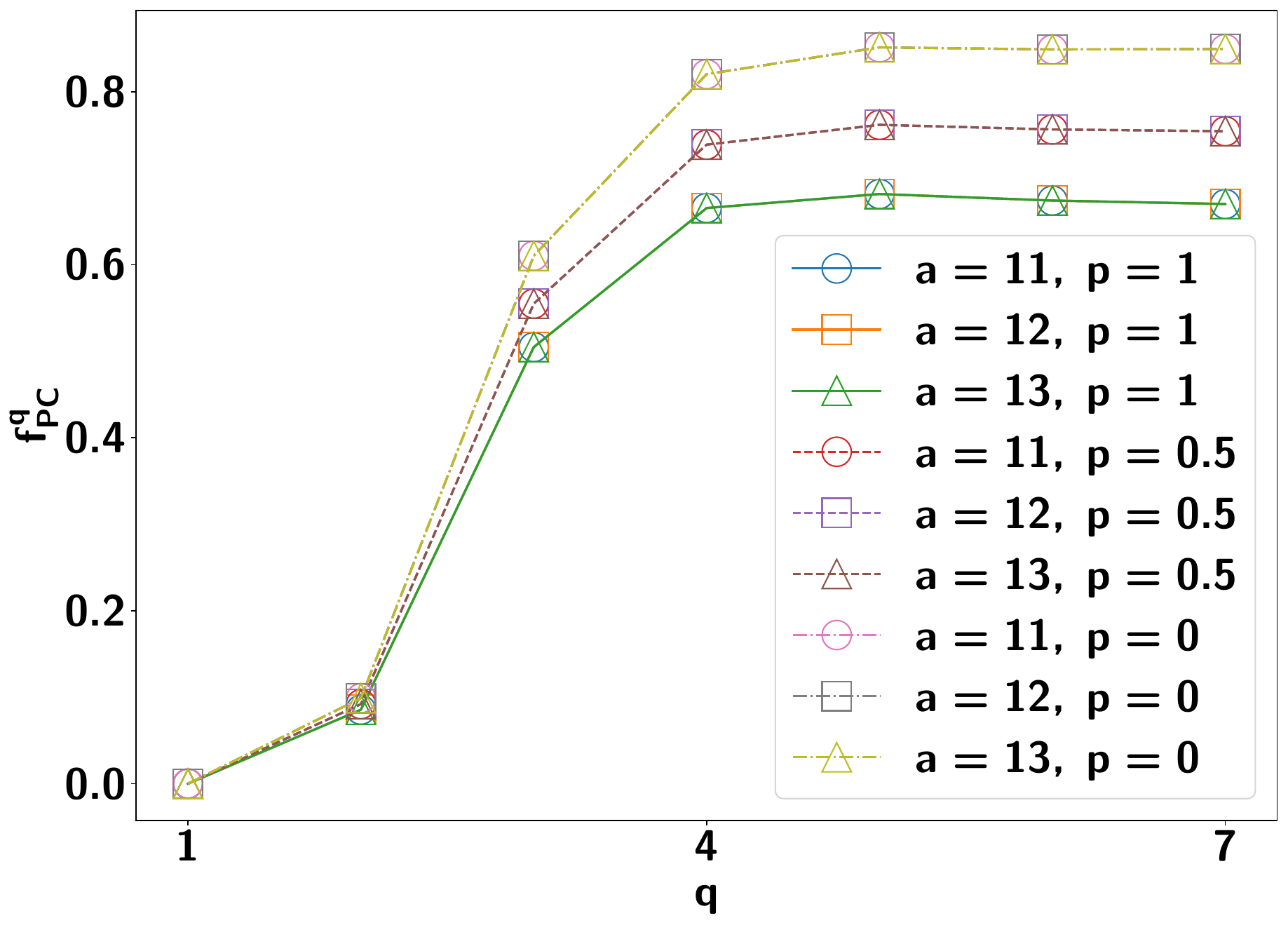}
\caption{\color{black} \textbf{Fidelity (ordinate) against different banding orders (abscissa) for different inputs values, $a$ and correlated noise strength, $p$ in a QFT.} Other parameters of figure are $p^{corr} = [0.99,0.01/3, 0.01/3, 0.01/3]$, $p^{uncr} = [\sqrt{0.99}, 1-\sqrt{0.99}, 0, 0]$ (corresponding to phase flip channel) is used for 7 qubits per input and second input (non QFT input), $b=63$. We employ noisy parameters such that there is equal probability of 0.99 for identity being applied and 0.01 probability of rest of the Kraus operators in both correlated and uncorrelated noise. All axes are dimensionless.}
\label{fig:aqft_vs_banding_corr_vs_uncorr}
\end{figure}

\textcolor{black}{\textit{Correlated Noise.} In practice, quantum circuits are also affected by other kinds of noises which may introduce correlations across qudits \cite{Macchiavello2001Jul}. Here we incorporate a probabilistic mixture of correlated and uncorrelated Pauli noise channel (PC) \cite{Macchiavello2001Jul} such that a two qubit density matrix transforms as $\rho_{AB} \xrightarrow{PC}  p\sum_{i = 0}^{4}p^{corr}_i\sigma_i\otimes\sigma_i \rho_{AB}\sigma_i\otimes\sigma_i + (1-p)\sum_{i,j=0}^4p_i^{uncr}p_j^{uncr}\sigma_i\otimes\sigma_j \rho_{AB}\sigma_i\otimes\sigma_j$ where $p$ is the correlation strength,  $p^{corr}$ and $p^{uncr}$ determine the nature of the correlated and uncorrelated parts respectively, and \(\sum_ip_i^{corr}=\sum_ip_i^{uncr}=1\). The noise is being applied after each of the controlled rotations in the QFT and SUM part. Note that our results hold well and generalize  for correlated noise with similar qualitative behavior to uncorrelated noise albeit with lower fidelity as shown in Fig. \ref{fig:aqft_vs_banding_corr_vs_uncorr}. We demonstrate this by employing noises of increasing correlation strength $p$. In all of the figures, we observe a unimodal behavior of fidelity that   increases sharply and starts to fall slowly with the banding order. We note that this correlated Pauli noise reduces the fidelity more strongly than its uncorrelated counterpart which is expected behavior since, correlated noises tend to affect two qubit gate performances to a higher degree. This is due to the fact that they introduce dependencies between errors on different qudits. We also note that for correlated noise, the fidelity looses its input agnostic behavior as it was found for phase damping, amplitude damping and other local noises. However, the dependence is very minimal and does not affect the qualitative analysis or the trade-off behavior between noise and banding.  }

\begin{figure}[H]
   \centering
    \includegraphics[width=\linewidth]{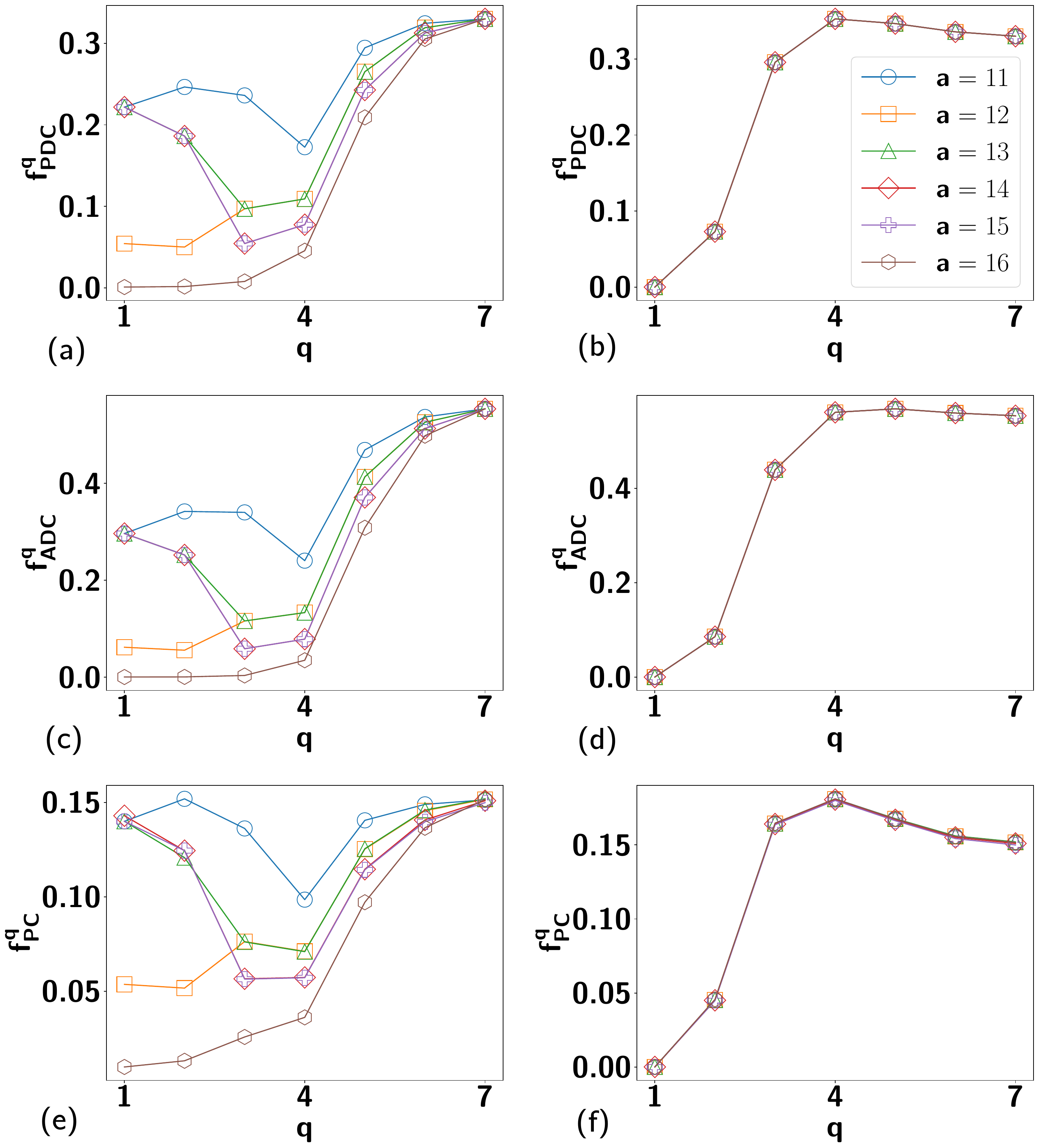}
\caption{\color{black} \textbf{Fidelity (ordinate) for different banding orders (abscissa) for different input numbers, \(a\)}. Other input is \(b=63\) and with 7 qubits per input. a) Phase damping noise with p = 0.05 c) Amplitude damping noise with p = 0.05 e) Completely correlated Pauli noise with parameters $p^{corr} = [0.95,0.05/3, 0.05/3, 0.05/3]$ are applied. Fidelity (oridante) against $q$ (abscissa) for a circuit in which AQFT has been performed by removing controlled rotation at QFT and IQFT steps, where $q$ refers to the maximum order of controlled rotations. b) Phase damping noise with p = 0.05 d) Amplitude damping noise with p = 0.05 f) Completely correlated Pauli noise with parameters $p^{corr} = [0.95,0.05/3, 0.05/3, 0.05/3]$ are applied. Left panel demonstrates banding on QFT (AQFT) part while right panel is for banding in SUM circuit.  Fidelity for a circuit in which the approximation is performed only at the SUM step by reducing the controlled rotations. All axes are dimensionless.}
\label{fig:aqft_vs_banding}
\end{figure}

\textcolor{black}{Furthermore, we incorporate completely correlated Pauli noise, i.e., at $p =1$ in our circuit for comparing SUM-banding and QFT-banding (AQFT) and demonstrate that the advantage obtained by the SUM-banding is not available for the QFT-banding  (see Fig. \ref{fig:aqft_vs_banding}). This result is along the line of other noise models that show similar behavior  with higher fidelity. While correlated noise does make the fidelity input-dependent, as can be observed in Fig. \ref{fig:aqft_vs_banding} (f), this dependence is much weaker and more uniform than what is observed in the QFT-banding case.}

\textit{\color{black}Difference between QFT-banding (AQFT) and SUM-banding.} The banding of the SUM circuit is different than the approximation of the QFT and IQFT circuit, which was performed in the case of periodicity estimation circuit \cite{barenco_pra_1996}. They are significantly different in terms of the implementations and efficiency. We compute the fidelity by fixing one of the input numbers and varying the others. The approximation is present in two scenarios under the presence of phase damping, amplitude damping, and completely correlated Pauli channels -- (a), (c), and (e) approximation in the  QFT and IQFT parts of the quantum adder  circuits and (b), (d), and (f) banding in the SUM part of the circuit (Eq. (\ref{eq:bandingfidelity})). It is evident that the approximation in the QFT circuit is not suitable due to the following reasons: (i) the behavior of fidelity in the case of QFT-banding (AQFT) is an input-dependent quantity, which negates the purpose of making a universal adder, whereas in case of SUM-banding, an input agnostic behavior is found. (ii) As already mentioned, in SUM-banding circuits, a non-monotonic behavior is observed with $q$, \textcolor{black}{ i.e., there is an optimal number of controlled rotations that gives the highest fidelity in presence of noise.} This proves that banding is also favorable from the experimental perspective as only a finite number of gates $q_{best} \ll N$ is required for obtaining the optimal fidelity under noise. \textcolor{black}{ On the other hand, there is a disadvantage in the case of AQFT since there is no value of banding for which the circuit maintains the highest fidelity as found for SUM-banding. Moreover, there is no unimodality and no value of banding, in general, at which local maxima or saturation in fidelity is reached universally for all inputs. Also, for certain inputs, there are values of banding order for which  the fidelity falls steeply, which is undesirable (see Fig. \ref{fig:aqft_vs_banding}) and the only way to prevent such a behavior is not to use the approximation or banding in the QFT and IQFT parts of the circuit in the first place. The QFT-banding case also highlights that a competition between the loss in fidelity due to decoherence in a longer circuit and approximation in a shorter circuit may not always be seen and strategies are required to manifest and exploit such a behavior.} 


 \textcolor{black}{Since, in the SUM part, the controlled rotations acting on different qudits commute and can be performed parallely, the circuit depth simplifies to be the highest order of controlled rotation $q$. In the presence of noise, the optimal banding order $q_{best}$ saturates and becomes constant with the number of qudits for a given value of noise. Thus, the circuit can be performed in constant time $\mathcal{O}(1)$. This reduction in depth is much better than a banded QFT circuit and other modern methods e.g. see Ref. \cite{Nam2020Mar}. Clearly, banding has a more prominent effect in reducing the circuit depth and time complexity in the SUM circuit (which reduces to \(\mathcal{O}(1)\)) as compared to the QFT circuit (which only reduces to $\mathcal{O}(n\log n$), leading to a more efficient noise tolerance.}

\textcolor{black}{We conclude that not only the adder circuit behaves differently than a  periodicity estimation circuit which uses QFT, we exhibit that if the same approximations are applied in the QFT and IQFT steps, the advantages found with banding on the SUM step are not replicated and it may be required to apply other strategies to mitigate the new issues. We suspect that this non-uniform behavior in fidelity and poor noise tolerance is simply because QFT and IQFT steps are part of encoding-decoding procedures. We can not expect to attain good fidelity in our addition circuit if encoding is performed for different or faulty numbers in the first place.}

\begin{figure}[!ht]
    \centering
    \includegraphics[width=\linewidth]{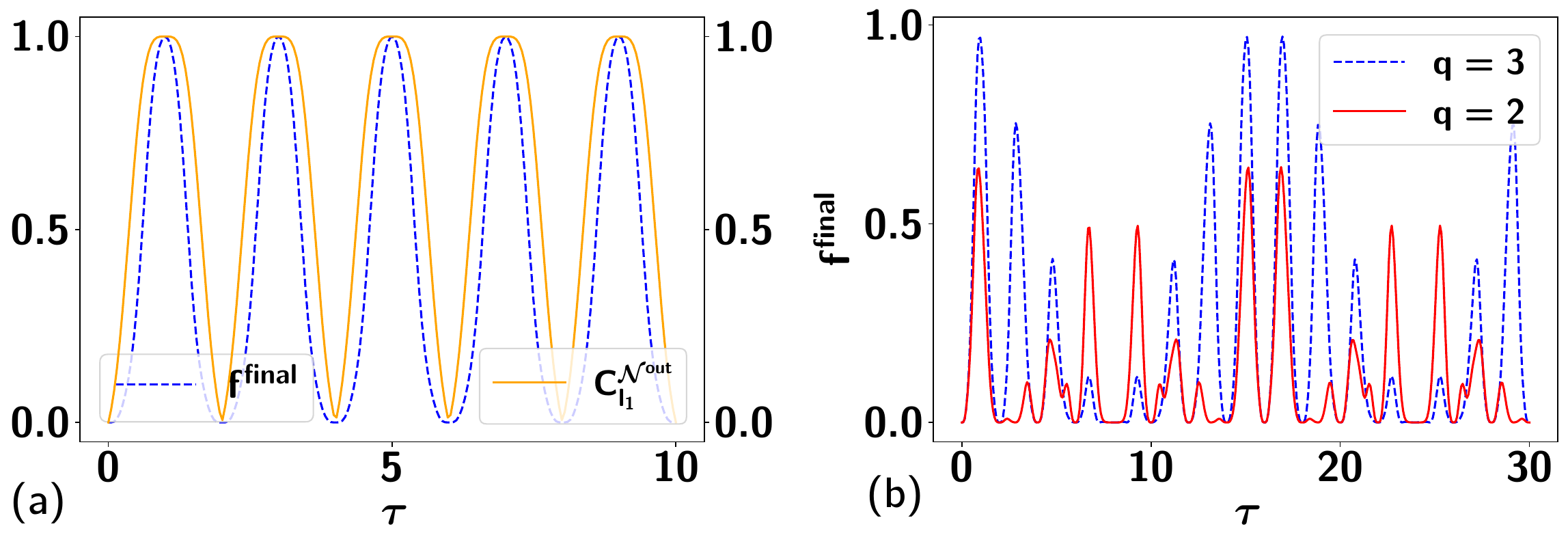}
    \caption{(a) \textbf{Fidelity $f^{final}$ and  $C_{l_1}^{\mathcal{N}^{out}}$ (vertical axis) vs time $\tau$ (horizontal axis).} 
    We add two integers, \(7\) and \(7\) via \(4\) qubits in the ideal addition circuit in which the gates are implemented by the Hamiltonian, presented in Eqs. (\ref{eq:Hadamard}) and (\ref{eq:ContRot}).  
    (b) Variation of \(f^{final}\) with \(\tau\) for banded circuit of banding order $q$. Note that while the maximum fidelity gets reduced, the time period for the periodic curves are same for different banding orders although the coherence remains  same as the unbanded case.
    All axes are dimensionless. }
    \label{fig:fidcohhardware}
\end{figure}

\section{Implementation of addition circuit in many-body systems}
\label{sec:implementation}

\textcolor{black}{We propose a possible implementation of the quantum addition algorithm. By leveraging long-range interacting spin chains and single-qubit operations, a quantum addition circuit can be simulated through time evolution starting from a prepared initial state. The implementation involves four key steps: (1) initial
state preparation, where the system is initialized in a state suitable for the addition operation; (2) QFT circuit
implementation; (3) SUM circuit implementation; and (4) Inverse QFT circuit implementation. Each step utilizes the capabilities of long-range interactions and single-qubit control to achieve efficient quantum operations. Below, we provide a detailed explanation of these steps.}

\begin{enumerate}
    \item \textcolor{black}{\textbf{\it Initial state preparation:} The algorithm’s first step involves preparing the initial state, comprising two numbers, \(\ket{v_1}\) and \(\ket{v_2}\), represented using the qubit element to enable further computational processes.}

    \item \textcolor{black}{\textbf{ \it QFT circuit implementation:}  In the subsequent step, the quantum Fourier transform (QFT) is applied to facilitate addition in the Fourier basis. The QFT implementation involves two key operations: a single-qubit operation and a two-qubit operation, with precise fine-tuning of local phases between the qubits. The single-qubit operation is achieved by carefully superposing longitudinal and transverse magnetic fields given by
    \begin{equation}    
        \mathcal{H}_{H_d}^{a_i} = -\frac{\pi}{\sqrt{8}}\left( \sigma_x^{a_i} +\sigma_z^{a_i} \right).
        \label{eq:Hadamard}
    \end{equation}
    where \(\sigma_x\) and \(\sigma_z\) are Pauli matrices and $a_i$ $(\forall 
 i \in \{0,\ldots, n-1\})$ is the qubit that is being subjected to a Hadamard gate. More precisely, the time evolution of the \(\mathcal{H}_{H_d}\) can mimic Hadamard gate, i.e., $U_{H_d}(\tau)=\exp(-\mathcal{H}_{H_d}\tau)$ at \(\tau=1,3,\ldots\).}

\textcolor{black}{On the other hand, any two-qubit controlled rotation by an angle $\phi = \frac{2 \pi}{2^k} $ 
can be realized by a two-body interacting Hamiltonian, written as
\begin{equation}    
\mathcal{H}_{R_d}^{a_i,a_j} = -J_{ij}\left(\mathrm{I} - \sigma_z^{a_i}  \right)\left(\mathrm{I} - \sigma_{z}^{a_j}  \right),
\label{eq:ContRot}
\end{equation}
where $a_i$ and $a_j$ represent control and target respectively and \(J_{ij}\) is chosen properly to implement the phase factor \(\phi\) accurately. By taking   \(J_{ij}=\pi/2^{q+1}\), where q is the order of the controlled rotation, $q = i - j + 1$,  the controlled rotation unitary is produced at \(\tau=2^qs+1\), \(s\in \{0,1,2,\ldots\}\) such that \(U_{R_d}(\tau)= diag \{1,1,1,e^{i\phi}\}\) \cite{chen_prxq_2023}. }

\textcolor{black}{\item \textbf{\it SUM circuit implementation:} The SUM circuit consists of two-qubit controlled rotation, which is similar to the QFT case. Hence, SUM circuit can be implemented via time evolution of Eq. (\ref{eq:ContRot}).}

\textcolor{black}{\item \textbf{\it Inverse quantum Fourier transform (IQFT):} The implementation of this part of the circuit is the inverse of QFT which can be similarly simulated as the QFT operation.}
    
\end{enumerate}

\textcolor{black}{Notice that one of the advantages of such unitary evolution is to obtain the result of the quantum addition at any arbitrary point of time, rather at any specific time. It may appear that while Hadamard gate is being performed in $\tau = odd$, the time period of controlled rotation grows exponentially with the order of controlled rotation $q$ as  \(\tau=2^qs+1\), \(s\in \{0,1,2,\ldots\}\). Here we show that the circuit can actually be realized for all $\tau = odd$. When $\tau = odd$, the controlled rotation gate is actually realized with the modified phase, $m \phi$, although the Hadamard gate is implemented perfectly. We can find out the modification factor \(m\) by equating \(\tau = 2^qr+z\), and then writing $z = \sum_{z_i \subset \mathbb{N}} 2^{q-z_i}$ to find $z_i$ which gives us $m = r+ \sum_{z_i \subset \mathbb{N}}(z_i+1)$ so that \(U_{R_d}(\tau)= diag \{1,1,1,e^{im\phi}\}\). The consequence of such scaling in controlled rotation is that instead of encoding of integer \(v_1\), a scaled number \(mv_1\) is encoded after the Fourier transform. Hence, in the encoding operation in the Fourier basis,  the number is scaled by a factor \(m\), i.e., if \(v_1\) and \(v_2\) are two integers to be added, \(v_1\) is scaled and encoded as \(\mathcal{F}\ket{mv_1}\) which are then summed to $ \mathcal{F}\ket{mv_1+mv_2}$.  Similarly, during decoding, the encoded input $\mathcal{F}\ket{mv_1+mv_2}$ is scaled down to give us the final result $\ket{v_1+ v_2}$.}

\textcolor{black}{This type of Hamiltonian can be implemented in a laboratory with different physical systems as it was already demonstrated in the case of measurement-based quantum computation (MBQC) \cite{raussendorf_prl_2001, pan_mbqc_exp_2011}. Note that instead of decomposing the circuit into possible single- and two-qubit universal gates, we propose an implementation strategy of this circuit on a physical substrate using naturally available interaction between the spins. Hence, this is inclined towards the analogue quantum computation. While spin-chain circuits have been implemented in several trapped ion setups, qudit circuits can be realized in various superconducting platforms where high-precision controlled rotations are achievable. However, such implementations tend to deteriorate due to environmental noise and imperfections in gate operations \cite{Goss2022Dec,liu_prx_2023,fischer_prx_2023}. }


We study the fidelity of the output state at the end of the circuit with the state obtained in case of  the ideal scenario, termed as $f^{final}$. As the unitary is an oscillatory function, the corresponding  fidelity, $f^{final}$, is a periodic function of  time for which the individual Hamiltonian is implemented, and it reaches the maximum value for a specified $\tau$. Thus, the Hamiltonian we prescribed can perform quantum addition and periodicity provides robustness to implement the quantum circuit (see Fig. \ref{fig:fidcohhardware}). The maximum coherence of the state in the SUM circuit is also periodic and attains maximum at the same time with the fidelity, thereby again establishing the coherence as a resource. We also simulate the Hamiltonian evolution for the banded circuits. Similar to the ideal case, the fidelity still remains periodic  with the modified time periods although the average fidelity becomes smaller in comparison to the perfect case.

\section{conclusion}
\label{sec:conclusion} 

Any quantum circuit that is used to execute certain tasks might need to perform addition, multiplication, and other basic arithmetic operations.  Therefore, a quantum adder based on the quantum Fourier transform can be useful in the implementation of quantum algorithms, particularly in Shor's factorization algorithm. The primary goals of this work are to first identify the quantum features that lead to an efficient quantum addition algorithm and then to ascertain the significance of dimension in achieving any improvement in performance when the circuit is disrupted by imperfections. By proving dimensional benefits and identifying quantum coherence as a resource for quantum addition, we accomplished both goals. 

A maximally coherent state is shown to be necessary as input for a quantum circuit to implement the quantum addition in an ideal situation when all operations are done flawlessly and noise does not damage the circuit. 
 Among the principal findings was the connection  between the quantum coherence and the fidelity between the targeted and impacted states.
  The performance gets affected either by the local environmental noise acting on each qudit, or  the reduced number of controlled rotation operations, or both. In particular, we found that there is a trade-off relation between the number of controlled rotation gates and noise strength. Precisely,  we  demonstrated that, for a given noise strength, fidelity reaches its maximum value for a moderate number of controlled rotations, and, surprisingly, that fidelity decreases as the number of such operations increases. 
  More crucially, in the presence of noise and in the case of an approximate addition, dimensional gain becomes more prominent. 
  Using quantum spin models, we proposed a potential scheme for implementing the algorithm: after suitably preparing the initial state, the system repeatedly evolves according to a quantum Ising Hamiltonian and magnetic fields towards realizing quantum gates. \textcolor{black}{Also, implementation of such long-range model is not easily achievable, although there are certain  development in this direction in recent years, e.g., {\it Superconducting circuit,} where recently long-range \(ZZ\) interaction is realized by Deng {\it et. al.} \cite{deng2024_zz}. On the other hand, in ion trap system, realization of long-range ising interaction are realized in different experrimantal set up \cite{Britton2012_nature,Keesling2019_nature}.} This suggests that the algorithm may be realized in physical systems such as superconducting circuits and trapped ions. 



\acknowledgements

We acknowledge the support from Interdisciplinary Cyber Physical Systems (ICPS) program of the Department of Science and Technology (DST), India, Grant No.: DST/ICPS/QuST/Theme- 1/2019/23. We acknowledge the use of \href{https://github.com/titaschanda/QIClib}{QIClib} -- a modern C++ library for general purpose quantum information processing and quantum computing (\url{https://titaschanda.github.io/QIClib}) and QuTiP library \cite{qutip}. This research was supported in part by the `INFOSYS' scholarship for senior
students’.

\bibliography{reference}

\begin{thebibliography}{84}%
\makeatletter
\providecommand \@ifxundefined [1]{%
 \@ifx{#1\undefined}
}%
\providecommand \@ifnum [1]{%
 \ifnum #1\expandafter \@firstoftwo
 \else \expandafter \@secondoftwo
 \fi
}%
\providecommand \@ifx [1]{%
 \ifx #1\expandafter \@firstoftwo
 \else \expandafter \@secondoftwo
 \fi
}%
\providecommand \natexlab [1]{#1}%
\providecommand \enquote  [1]{``#1''}%
\providecommand \bibnamefont  [1]{#1}%
\providecommand \bibfnamefont [1]{#1}%
\providecommand \citenamefont [1]{#1}%
\providecommand \href@noop [0]{\@secondoftwo}%
\providecommand \href [0]{\begingroup \@sanitize@url \@href}%
\providecommand \@href[1]{\@@startlink{#1}\@@href}%
\providecommand \@@href[1]{\endgroup#1\@@endlink}%
\providecommand \@sanitize@url [0]{\catcode `\\12\catcode `\$12\catcode
  `\&12\catcode `\#12\catcode `\^12\catcode `\_12\catcode `\%12\relax}%
\providecommand \@@startlink[1]{}%
\providecommand \@@endlink[0]{}%
\providecommand \url  [0]{\begingroup\@sanitize@url \@url }%
\providecommand \@url [1]{\endgroup\@href {#1}{\urlprefix }}%
\providecommand \urlprefix  [0]{URL }%
\providecommand \Eprint [0]{\href }%
\providecommand \doibase [0]{https://doi.org/}%
\providecommand \selectlanguage [0]{\@gobble}%
\providecommand \bibinfo  [0]{\@secondoftwo}%
\providecommand \bibfield  [0]{\@secondoftwo}%
\providecommand \translation [1]{[#1]}%
\providecommand \BibitemOpen [0]{}%
\providecommand \bibitemStop [0]{}%
\providecommand \bibitemNoStop [0]{.\EOS\space}%
\providecommand \EOS [0]{\spacefactor3000\relax}%
\providecommand \BibitemShut  [1]{\csname bibitem#1\endcsname}%
\let\auto@bib@innerbib\@empty
\bibitem [{\citenamefont {Nielsen}\ and\ \citenamefont
  {Chuang}(2010)}]{nielsen_chuang_2010}%
  \BibitemOpen
  \bibfield  {author} {\bibinfo {author} {\bibfnamefont {M.~A.}\ \bibnamefont
  {Nielsen}}\ and\ \bibinfo {author} {\bibfnamefont {I.~L.}\ \bibnamefont
  {Chuang}},\ }\href {https://doi.org/10.1017/CBO9780511976667} {\emph
  {\bibinfo {title} {Quantum Computation and Quantum Information: 10th
  Anniversary Edition}}}\ (\bibinfo  {publisher} {Cambridge University Press},\
  \bibinfo {year} {2010})\BibitemShut {NoStop}%
\bibitem [{\citenamefont {Deutsch}\ and\ \citenamefont
  {Jozsa}(1992)}]{Deutsch_Jozsa}%
  \BibitemOpen
  \bibfield  {author} {\bibinfo {author} {\bibfnamefont {D.}~\bibnamefont
  {Deutsch}}\ and\ \bibinfo {author} {\bibfnamefont {R.}~\bibnamefont
  {Jozsa}},\ }\bibfield  {title} {\bibinfo {title} {Rapid solution of problems
  by quantum computation},\ }\href {https://doi.org/10.1098/rspa.1992.0167}
  {\bibfield  {journal} {\bibinfo  {journal} {Proc. R. Soc. Lond. A}\ }\textbf
  {\bibinfo {volume} {439}},\ \bibinfo {pages} {553} (\bibinfo {year}
  {1992})}\BibitemShut {NoStop}%
\bibitem [{\citenamefont {Shor}(1994)}]{shor_1994}%
  \BibitemOpen
  \bibfield  {author} {\bibinfo {author} {\bibfnamefont {P.}~\bibnamefont
  {Shor}},\ }\bibfield  {title} {\bibinfo {title} {Algorithms for quantum
  computation: discrete logarithms and factoring},\ }in\ \href
  {https://doi.org/10.1109/SFCS.1994.365700} {\emph {\bibinfo {booktitle}
  {Proceedings 35th Annual Symposium on Foundations of Computer Science}}}\
  (\bibinfo {year} {1994})\ pp.\ \bibinfo {pages} {124--134}\BibitemShut
  {NoStop}%
\bibitem [{\citenamefont {Shor}(1995)}]{shor_pra_1995}%
  \BibitemOpen
  \bibfield  {author} {\bibinfo {author} {\bibfnamefont {P.~W.}\ \bibnamefont
  {Shor}},\ }\bibfield  {title} {\bibinfo {title} {Scheme for reducing
  decoherence in quantum computer memory},\ }\href
  {https://doi.org/10.1103/PhysRevA.52.R2493} {\bibfield  {journal} {\bibinfo
  {journal} {Phys. Rev. A}\ }\textbf {\bibinfo {volume} {52}},\ \bibinfo
  {pages} {R2493} (\bibinfo {year} {1995})}\BibitemShut {NoStop}%
\bibitem [{\citenamefont {Grover}(1996)}]{grover_arxiv_1996}%
  \BibitemOpen
  \bibfield  {author} {\bibinfo {author} {\bibfnamefont {L.~K.}\ \bibnamefont
  {Grover}},\ }\href@noop {} {\bibinfo {title} {A fast quantum mechanical
  algorithm for database search}} (\bibinfo {year} {1996}),\ \Eprint
  {https://arxiv.org/abs/quant-ph/9605043} {arXiv:quant-ph/9605043 [quant-ph]}
  \BibitemShut {NoStop}%
\bibitem [{\citenamefont {Mermin}(2007)}]{mermin_book}%
  \BibitemOpen
  \bibfield  {author} {\bibinfo {author} {\bibfnamefont {N.~D.}\ \bibnamefont
  {Mermin}},\ }\href {https://doi.org/10.1017/CBO9780511813870} {\emph
  {\bibinfo {title} {Quantum Computer Science: An Introduction}}}\ (\bibinfo
  {publisher} {Cambridge University Press},\ \bibinfo {year}
  {2007})\BibitemShut {NoStop}%
\bibitem [{\citenamefont {Lanyon}\ \emph {et~al.}(2007)\citenamefont {Lanyon},
  \citenamefont {Weinhold}, \citenamefont {Langford}, \citenamefont {Barbieri},
  \citenamefont {James}, \citenamefont {Gilchrist},\ and\ \citenamefont
  {White}}]{shor_experi_photonics_2007}%
  \BibitemOpen
  \bibfield  {author} {\bibinfo {author} {\bibfnamefont {B.~P.}\ \bibnamefont
  {Lanyon}}, \bibinfo {author} {\bibfnamefont {T.~J.}\ \bibnamefont
  {Weinhold}}, \bibinfo {author} {\bibfnamefont {N.~K.}\ \bibnamefont
  {Langford}}, \bibinfo {author} {\bibfnamefont {M.}~\bibnamefont {Barbieri}},
  \bibinfo {author} {\bibfnamefont {D.~F.~V.}\ \bibnamefont {James}}, \bibinfo
  {author} {\bibfnamefont {A.}~\bibnamefont {Gilchrist}},\ and\ \bibinfo
  {author} {\bibfnamefont {A.~G.}\ \bibnamefont {White}},\ }\bibfield  {title}
  {\bibinfo {title} {Experimental demonstration of a compiled version of shor's
  algorithm with quantum entanglement},\ }\href
  {https://doi.org/10.1103/PhysRevLett.99.250505} {\bibfield  {journal}
  {\bibinfo  {journal} {Phys. Rev. Lett.}\ }\textbf {\bibinfo {volume} {99}},\
  \bibinfo {pages} {250505} (\bibinfo {year} {2007})}\BibitemShut {NoStop}%
\bibitem [{\citenamefont {Watson}\ \emph {et~al.}(2018)\citenamefont {Watson},
  \citenamefont {Philips}, \citenamefont {Kawakami}, \citenamefont {Ward},
  \citenamefont {Scarlino}, \citenamefont {Veldhorst}, \citenamefont {Savage},
  \citenamefont {Lagally}, \citenamefont {Friesen}, \citenamefont
  {Coppersmith}, \citenamefont {Eriksson},\ and\ \citenamefont
  {Vandersypen}}]{silicon_experimental_deutsch_grover_2018}%
  \BibitemOpen
  \bibfield  {author} {\bibinfo {author} {\bibfnamefont {T.~F.}\ \bibnamefont
  {Watson}}, \bibinfo {author} {\bibfnamefont {S.~G.~J.}\ \bibnamefont
  {Philips}}, \bibinfo {author} {\bibfnamefont {E.}~\bibnamefont {Kawakami}},
  \bibinfo {author} {\bibfnamefont {D.~R.}\ \bibnamefont {Ward}}, \bibinfo
  {author} {\bibfnamefont {P.}~\bibnamefont {Scarlino}}, \bibinfo {author}
  {\bibfnamefont {M.}~\bibnamefont {Veldhorst}}, \bibinfo {author}
  {\bibfnamefont {D.~E.}\ \bibnamefont {Savage}}, \bibinfo {author}
  {\bibfnamefont {M.~G.}\ \bibnamefont {Lagally}}, \bibinfo {author}
  {\bibfnamefont {M.}~\bibnamefont {Friesen}}, \bibinfo {author} {\bibfnamefont
  {S.~N.}\ \bibnamefont {Coppersmith}}, \bibinfo {author} {\bibfnamefont
  {M.~A.}\ \bibnamefont {Eriksson}},\ and\ \bibinfo {author} {\bibfnamefont
  {L.~M.~K.}\ \bibnamefont {Vandersypen}},\ }\bibfield  {title} {\bibinfo
  {title} {{A programmable two-qubit quantum processor in silicon}},\ }\href
  {https://doi.org/10.1038/nature25766} {\bibfield  {journal} {\bibinfo
  {journal} {Nature}\ }\textbf {\bibinfo {volume} {555}},\ \bibinfo {pages}
  {633} (\bibinfo {year} {2018})}\BibitemShut {NoStop}%
\bibitem [{\citenamefont {Cirac}\ and\ \citenamefont
  {Zoller}(2000)}]{Cirac2000Apr}%
  \BibitemOpen
  \bibfield  {author} {\bibinfo {author} {\bibfnamefont {J.~I.}\ \bibnamefont
  {Cirac}}\ and\ \bibinfo {author} {\bibfnamefont {P.}~\bibnamefont {Zoller}},\
  }\bibfield  {title} {\bibinfo {title} {{A scalable quantum computer with ions
  in an array of microtraps}},\ }\href {https://doi.org/10.1038/35007021}
  {\bibfield  {journal} {\bibinfo  {journal} {Nature}\ }\textbf {\bibinfo
  {volume} {404}},\ \bibinfo {pages} {579} (\bibinfo {year}
  {2000})}\BibitemShut {NoStop}%
\bibitem [{\citenamefont {Arute}\ \emph {et~al.}(2019)\citenamefont {Arute},
  \citenamefont {Arya}, \citenamefont {Babbush}, \citenamefont {Bacon},
  \citenamefont {Bardin}, \citenamefont {Barends}, \citenamefont {Biswas},\
  and\ \citenamefont {{\it et al.}}}]{arute_nature_2019}%
  \BibitemOpen
  \bibfield  {author} {\bibinfo {author} {\bibfnamefont {F.}~\bibnamefont
  {Arute}}, \bibinfo {author} {\bibfnamefont {K.}~\bibnamefont {Arya}},
  \bibinfo {author} {\bibfnamefont {R.}~\bibnamefont {Babbush}}, \bibinfo
  {author} {\bibfnamefont {D.}~\bibnamefont {Bacon}}, \bibinfo {author}
  {\bibfnamefont {J.~C.}\ \bibnamefont {Bardin}}, \bibinfo {author}
  {\bibfnamefont {R.}~\bibnamefont {Barends}}, \bibinfo {author} {\bibfnamefont
  {R.}~\bibnamefont {Biswas}},\ and\ \bibinfo {author} {\bibnamefont {{\it et
  al.}}},\ }\bibfield  {title} {\bibinfo {title} {Quantum supremacy using a
  programmable superconducting processor},\ }\href
  {https://doi.org/10.1038/s41586-019-1666-5} {\bibfield  {journal} {\bibinfo
  {journal} {Nature}\ }\textbf {\bibinfo {volume} {574}},\ \bibinfo {pages}
  {505} (\bibinfo {year} {2019})}\BibitemShut {NoStop}%
\bibitem [{\citenamefont {Wu}\ \emph {et~al.}(2021)\citenamefont {Wu},
  \citenamefont {Bao}, \citenamefont {Cao},\ and\ \citenamefont {{\it et
  al}.}}]{wu_prl_2021}%
  \BibitemOpen
  \bibfield  {author} {\bibinfo {author} {\bibfnamefont {Y.}~\bibnamefont
  {Wu}}, \bibinfo {author} {\bibfnamefont {W.-S.}\ \bibnamefont {Bao}},
  \bibinfo {author} {\bibnamefont {Cao}},\ and\ \bibinfo {author} {\bibnamefont
  {{\it et al}.}},\ }\bibfield  {title} {\bibinfo {title} {Strong quantum
  computational advantage using a superconducting quantum processor},\ }\href
  {https://doi.org/10.1103/PhysRevLett.127.180501} {\bibfield  {journal}
  {\bibinfo  {journal} {Phys. Rev. Lett.}\ }\textbf {\bibinfo {volume} {127}},\
  \bibinfo {pages} {180501} (\bibinfo {year} {2021})}\BibitemShut {NoStop}%
\bibitem [{\citenamefont {Horodecki}\ \emph {et~al.}(2009)\citenamefont
  {Horodecki}, \citenamefont {Horodecki}, \citenamefont {Horodecki},\ and\
  \citenamefont {Horodecki}}]{horodecki2009}%
  \BibitemOpen
  \bibfield  {author} {\bibinfo {author} {\bibfnamefont {R.}~\bibnamefont
  {Horodecki}}, \bibinfo {author} {\bibfnamefont {P.}~\bibnamefont
  {Horodecki}}, \bibinfo {author} {\bibfnamefont {M.}~\bibnamefont
  {Horodecki}},\ and\ \bibinfo {author} {\bibfnamefont {K.}~\bibnamefont
  {Horodecki}},\ }\bibfield  {title} {\bibinfo {title} {Quantum entanglement},\
  }\href {https://doi.org/10.1103/RevModPhys.81.865} {\bibfield  {journal}
  {\bibinfo  {journal} {Rev. Mod. Phys.}\ }\textbf {\bibinfo {volume} {81}},\
  \bibinfo {pages} {865} (\bibinfo {year} {2009})}\BibitemShut {NoStop}%
\bibitem [{\citenamefont {Streltsov}\ \emph {et~al.}(2017)\citenamefont
  {Streltsov}, \citenamefont {Adesso},\ and\ \citenamefont
  {Plenio}}]{streltsov_rmp_2017}%
  \BibitemOpen
  \bibfield  {author} {\bibinfo {author} {\bibfnamefont {A.}~\bibnamefont
  {Streltsov}}, \bibinfo {author} {\bibfnamefont {G.}~\bibnamefont {Adesso}},\
  and\ \bibinfo {author} {\bibfnamefont {M.~B.}\ \bibnamefont {Plenio}},\
  }\bibfield  {title} {\bibinfo {title} {Colloquium: Quantum coherence as a
  resource},\ }\href {https://doi.org/10.1103/RevModPhys.89.041003} {\bibfield
  {journal} {\bibinfo  {journal} {Rev. Mod. Phys.}\ }\textbf {\bibinfo {volume}
  {89}},\ \bibinfo {pages} {041003} (\bibinfo {year} {2017})}\BibitemShut
  {NoStop}%
\bibitem [{\citenamefont {Gisin}\ \emph {et~al.}(2002)\citenamefont {Gisin},
  \citenamefont {Ribordy}, \citenamefont {Tittel},\ and\ \citenamefont
  {Zbinden}}]{quantum_crypto_gisin_rmp_2002}%
  \BibitemOpen
  \bibfield  {author} {\bibinfo {author} {\bibfnamefont {N.}~\bibnamefont
  {Gisin}}, \bibinfo {author} {\bibfnamefont {G.}~\bibnamefont {Ribordy}},
  \bibinfo {author} {\bibfnamefont {W.}~\bibnamefont {Tittel}},\ and\ \bibinfo
  {author} {\bibfnamefont {H.}~\bibnamefont {Zbinden}},\ }\bibfield  {title}
  {\bibinfo {title} {Quantum cryptography},\ }\href
  {https://doi.org/10.1103/RevModPhys.74.145} {\bibfield  {journal} {\bibinfo
  {journal} {Rev. Mod. Phys.}\ }\textbf {\bibinfo {volume} {74}},\ \bibinfo
  {pages} {145} (\bibinfo {year} {2002})}\BibitemShut {NoStop}%
\bibitem [{\citenamefont {De}\ and\ \citenamefont
  {Sen}(2011)}]{aditi_quantum_comm_review_2011}%
  \BibitemOpen
  \bibfield  {author} {\bibinfo {author} {\bibfnamefont {A.~S.}\ \bibnamefont
  {De}}\ and\ \bibinfo {author} {\bibfnamefont {U.}~\bibnamefont {Sen}},\
  }\href@noop {} {\bibinfo {title} {Quantum advantage in communication
  networks}} (\bibinfo {year} {2011}),\ \Eprint
  {https://arxiv.org/abs/1105.2412} {arXiv:1105.2412 [quant-ph]} \BibitemShut
  {NoStop}%
\bibitem [{\citenamefont {Kendon}\ and\ \citenamefont
  {Munro}(2006)}]{shors_entanglement_2006}%
  \BibitemOpen
  \bibfield  {author} {\bibinfo {author} {\bibfnamefont {V.~M.}\ \bibnamefont
  {Kendon}}\ and\ \bibinfo {author} {\bibfnamefont {W.~J.}\ \bibnamefont
  {Munro}},\ }\bibfield  {title} {\bibinfo {title} {{Entanglement and its role
  in Shor's algorithm}},\ }\href {https://doi.org/10.5555/2011698.2011704}
  {\bibfield  {journal} {\bibinfo  {journal} {Quantum Inf. Comput.}\ }\textbf
  {\bibinfo {volume} {6}},\ \bibinfo {pages} {630} (\bibinfo {year}
  {2006})}\BibitemShut {NoStop}%
\bibitem [{\citenamefont {Qu}\ \emph {et~al.}(2012)\citenamefont {Qu},
  \citenamefont {Wang}, \citenamefont {shang Li}, \citenamefont {li~Zhao},
  \citenamefont {ru~Bao},\ and\ \citenamefont {chun
  Cao}}]{multi_ent_grover_2012}%
  \BibitemOpen
  \bibfield  {author} {\bibinfo {author} {\bibfnamefont {R.}~\bibnamefont
  {Qu}}, \bibinfo {author} {\bibfnamefont {J.}~\bibnamefont {Wang}}, \bibinfo
  {author} {\bibfnamefont {Z.}~\bibnamefont {shang Li}}, \bibinfo {author}
  {\bibfnamefont {S.}~\bibnamefont {li~Zhao}}, \bibinfo {author} {\bibfnamefont
  {Y.}~\bibnamefont {ru~Bao}},\ and\ \bibinfo {author} {\bibfnamefont
  {X.}~\bibnamefont {chun Cao}},\ }\href@noop {} {\bibinfo {title}
  {Multipartite entanglement and grover's search algorithm}} (\bibinfo {year}
  {2012}),\ \Eprint {https://arxiv.org/abs/1210.3418} {arXiv:1210.3418
  [quant-ph]} \BibitemShut {NoStop}%
\bibitem [{\citenamefont {Pan}\ \emph {et~al.}(2017)\citenamefont {Pan},
  \citenamefont {Qiu},\ and\ \citenamefont {Zheng}}]{global_ent_grover_2017}%
  \BibitemOpen
  \bibfield  {author} {\bibinfo {author} {\bibfnamefont {M.}~\bibnamefont
  {Pan}}, \bibinfo {author} {\bibfnamefont {D.}~\bibnamefont {Qiu}},\ and\
  \bibinfo {author} {\bibfnamefont {S.}~\bibnamefont {Zheng}},\ }\bibfield
  {title} {\bibinfo {title} {{Global multipartite entanglement dynamics in
  Grover{'}s search algorithm}},\ }\href
  {https://doi.org/10.1007/s11128-017-1661-4} {\bibfield  {journal} {\bibinfo
  {journal} {Quantum Inf. Process.}\ }\textbf {\bibinfo {volume} {16}},\
  \bibinfo {pages} {1} (\bibinfo {year} {2017})}\BibitemShut {NoStop}%
\bibitem [{\citenamefont {Naseri}\ \emph {et~al.}(2022)\citenamefont {Naseri},
  \citenamefont {Kondra}, \citenamefont {Goswami}, \citenamefont
  {Fellous-Asiani},\ and\ \citenamefont
  {Streltsov}}]{entanglement_bernstein_pra_2022}%
  \BibitemOpen
  \bibfield  {author} {\bibinfo {author} {\bibfnamefont {M.}~\bibnamefont
  {Naseri}}, \bibinfo {author} {\bibfnamefont {T.~V.}\ \bibnamefont {Kondra}},
  \bibinfo {author} {\bibfnamefont {S.}~\bibnamefont {Goswami}}, \bibinfo
  {author} {\bibfnamefont {M.}~\bibnamefont {Fellous-Asiani}},\ and\ \bibinfo
  {author} {\bibfnamefont {A.}~\bibnamefont {Streltsov}},\ }\bibfield  {title}
  {\bibinfo {title} {Entanglement and coherence in the bernstein-vazirani
  algorithm},\ }\href {https://doi.org/10.1103/PhysRevA.106.062429} {\bibfield
  {journal} {\bibinfo  {journal} {Phys. Rev. A}\ }\textbf {\bibinfo {volume}
  {106}},\ \bibinfo {pages} {062429} (\bibinfo {year} {2022})}\BibitemShut
  {NoStop}%
\bibitem [{\citenamefont {Kumar}\ \emph {et~al.}(2023)\citenamefont {Kumar},
  \citenamefont {Konar}, \citenamefont {Lakkaraju},\ and\ \citenamefont
  {De}}]{aditi_hhl_2023}%
  \BibitemOpen
  \bibfield  {author} {\bibinfo {author} {\bibfnamefont {P.}~\bibnamefont
  {Kumar}}, \bibinfo {author} {\bibfnamefont {T.~K.}\ \bibnamefont {Konar}},
  \bibinfo {author} {\bibfnamefont {L.~G.~C.}\ \bibnamefont {Lakkaraju}},\ and\
  \bibinfo {author} {\bibfnamefont {A.~S.}\ \bibnamefont {De}},\ }\href@noop {}
  {\bibinfo {title} {Quantum resources in harrow-hassidim-lloyd algorithm}}
  (\bibinfo {year} {2023}),\ \Eprint {https://arxiv.org/abs/2308.04021}
  {arXiv:2308.04021 [quant-ph]} \BibitemShut {NoStop}%
\bibitem [{\citenamefont {Ahnefeld}\ \emph {et~al.}(2022)\citenamefont
  {Ahnefeld}, \citenamefont {Theurer}, \citenamefont {Egloff}, \citenamefont
  {Matera},\ and\ \citenamefont {Plenio}}]{plenio_shor_coh_prl_2022}%
  \BibitemOpen
  \bibfield  {author} {\bibinfo {author} {\bibfnamefont {F.}~\bibnamefont
  {Ahnefeld}}, \bibinfo {author} {\bibfnamefont {T.}~\bibnamefont {Theurer}},
  \bibinfo {author} {\bibfnamefont {D.}~\bibnamefont {Egloff}}, \bibinfo
  {author} {\bibfnamefont {J.~M.}\ \bibnamefont {Matera}},\ and\ \bibinfo
  {author} {\bibfnamefont {M.~B.}\ \bibnamefont {Plenio}},\ }\bibfield  {title}
  {\bibinfo {title} {Coherence as a resource for shor's algorithm},\ }\href
  {https://doi.org/10.1103/PhysRevLett.129.120501} {\bibfield  {journal}
  {\bibinfo  {journal} {Phys. Rev. Lett.}\ }\textbf {\bibinfo {volume} {129}},\
  \bibinfo {pages} {120501} (\bibinfo {year} {2022})}\BibitemShut {NoStop}%
\bibitem [{\citenamefont {Hillery}(2016)}]{hillery_coh_deutsch_2016}%
  \BibitemOpen
  \bibfield  {author} {\bibinfo {author} {\bibfnamefont {M.}~\bibnamefont
  {Hillery}},\ }\bibfield  {title} {\bibinfo {title} {Coherence as a resource
  in decision problems: The deutsch-jozsa algorithm and a variation},\ }\href
  {https://doi.org/10.1103/PhysRevA.93.012111} {\bibfield  {journal} {\bibinfo
  {journal} {Phys. Rev. A}\ }\textbf {\bibinfo {volume} {93}},\ \bibinfo
  {pages} {012111} (\bibinfo {year} {2016})}\BibitemShut {NoStop}%
\bibitem [{\citenamefont {Mano}(2017)}]{mano2017digital}%
  \BibitemOpen
  \bibfield  {author} {\bibinfo {author} {\bibfnamefont {M.~M.}\ \bibnamefont
  {Mano}},\ }\href@noop {} {\emph {\bibinfo {title} {Digital logic and computer
  design}}}\ (\bibinfo  {publisher} {Pearson Education India},\ \bibinfo {year}
  {2017})\BibitemShut {NoStop}%
\bibitem [{\citenamefont {Zalka}(1998)}]{zalka_arxiv_1998}%
  \BibitemOpen
  \bibfield  {author} {\bibinfo {author} {\bibfnamefont {C.}~\bibnamefont
  {Zalka}},\ }\bibfield  {title} {\bibinfo {title} {{Fast versions of Shor's
  quantum factoring algorithm}},\ }\bibfield  {journal} {\bibinfo  {journal}
  {arXiv}\ }\href {https://doi.org/10.48550/arXiv.quant-ph/9806084}
  {10.48550/arXiv.quant-ph/9806084} (\bibinfo {year} {1998}),\ \Eprint
  {https://arxiv.org/abs/quant-ph/9806084} {quant-ph/9806084} \BibitemShut
  {NoStop}%
\bibitem [{\citenamefont {Draper}(2000)}]{draper_arxiv_2000}%
  \BibitemOpen
  \bibfield  {author} {\bibinfo {author} {\bibfnamefont {T.~G.}\ \bibnamefont
  {Draper}},\ }\bibfield  {title} {\bibinfo {title} {{Addition on a Quantum
  Computer}},\ }\bibfield  {journal} {\bibinfo  {journal} {arXiv}\ }\href
  {https://doi.org/10.48550/arXiv.quant-ph/0008033}
  {10.48550/arXiv.quant-ph/0008033} (\bibinfo {year} {2000}),\ \Eprint
  {https://arxiv.org/abs/quant-ph/0008033} {quant-ph/0008033} \BibitemShut
  {NoStop}%
\bibitem [{\citenamefont {Barenco}\ \emph {et~al.}(1996)\citenamefont
  {Barenco}, \citenamefont {Ekert}, \citenamefont {Suominen},\ and\
  \citenamefont {T\"orm\"a}}]{barenco_pra_1996}%
  \BibitemOpen
  \bibfield  {author} {\bibinfo {author} {\bibfnamefont {A.}~\bibnamefont
  {Barenco}}, \bibinfo {author} {\bibfnamefont {A.}~\bibnamefont {Ekert}},
  \bibinfo {author} {\bibfnamefont {K.-A.}\ \bibnamefont {Suominen}},\ and\
  \bibinfo {author} {\bibfnamefont {P.}~\bibnamefont {T\"orm\"a}},\ }\bibfield
  {title} {\bibinfo {title} {Approximate quantum fourier transform and
  decoherence},\ }\href {https://doi.org/10.1103/PhysRevA.54.139} {\bibfield
  {journal} {\bibinfo  {journal} {Phys. Rev. A}\ }\textbf {\bibinfo {volume}
  {54}},\ \bibinfo {pages} {139} (\bibinfo {year} {1996})}\BibitemShut
  {NoStop}%
\bibitem [{\citenamefont {Beauregard}(2003)}]{beauregard_qic_2003}%
  \BibitemOpen
  \bibfield  {author} {\bibinfo {author} {\bibfnamefont {S.}~\bibnamefont
  {Beauregard}},\ }\bibfield  {title} {\bibinfo {title} {Circuit for shor's
  algorithm using 2n+3 qubits},\ }\href {https://doi.org/10.26421/QIC3.2-8}
  {\bibfield  {journal} {\bibinfo  {journal} {Quantum Inf. Comput.}\ }\textbf
  {\bibinfo {volume} {3}},\ \bibinfo {pages} {175} (\bibinfo {year}
  {2003})}\BibitemShut {NoStop}%
\bibitem [{\citenamefont {Pavlidis}\ and\ \citenamefont
  {Gizopoulos}(2014)}]{pavlidis_qic_2014}%
  \BibitemOpen
  \bibfield  {author} {\bibinfo {author} {\bibfnamefont {A.}~\bibnamefont
  {Pavlidis}}\ and\ \bibinfo {author} {\bibfnamefont {D.}~\bibnamefont
  {Gizopoulos}},\ }\bibfield  {title} {\bibinfo {title} {{Fast quantum modular
  exponentiation architecture for Shor's factoring algorithm}},\ }\href
  {https://doi.org/10.5555/2638682.2638690} {\bibfield  {journal} {\bibinfo
  {journal} {Quantum Inf. Comput.}\ }\textbf {\bibinfo {volume} {14}},\
  \bibinfo {pages} {649} (\bibinfo {year} {2014})}\BibitemShut {NoStop}%
\bibitem [{\citenamefont {Paler}(2022)}]{paler_reversible_qft_2022}%
  \BibitemOpen
  \bibfield  {author} {\bibinfo {author} {\bibfnamefont {A.}~\bibnamefont
  {Paler}},\ }\bibfield  {title} {\bibinfo {title} {Quantum fourier addition
  simplified to toffoli addition},\ }\href
  {https://doi.org/10.1103/PhysRevA.106.042444} {\bibfield  {journal} {\bibinfo
   {journal} {Phys. Rev. A}\ }\textbf {\bibinfo {volume} {106}},\ \bibinfo
  {pages} {042444} (\bibinfo {year} {2022})}\BibitemShut {NoStop}%
\bibitem [{\citenamefont {Vedral}\ \emph {et~al.}(1996)\citenamefont {Vedral},
  \citenamefont {Barenco},\ and\ \citenamefont
  {Ekert}}]{vedral_barenco_ekert_pra_1996}%
  \BibitemOpen
  \bibfield  {author} {\bibinfo {author} {\bibfnamefont {V.}~\bibnamefont
  {Vedral}}, \bibinfo {author} {\bibfnamefont {A.}~\bibnamefont {Barenco}},\
  and\ \bibinfo {author} {\bibfnamefont {A.}~\bibnamefont {Ekert}},\ }\bibfield
   {title} {\bibinfo {title} {Quantum networks for elementary arithmetic
  operations},\ }\href {https://doi.org/10.1103/PhysRevA.54.147} {\bibfield
  {journal} {\bibinfo  {journal} {Phys. Rev. A}\ }\textbf {\bibinfo {volume}
  {54}},\ \bibinfo {pages} {147} (\bibinfo {year} {1996})}\BibitemShut
  {NoStop}%
\bibitem [{\citenamefont {Beckman}\ \emph {et~al.}(1996)\citenamefont
  {Beckman}, \citenamefont {Chari}, \citenamefont {Devabhaktuni},\ and\
  \citenamefont {Preskill}}]{preskill_pra_1996}%
  \BibitemOpen
  \bibfield  {author} {\bibinfo {author} {\bibfnamefont {D.}~\bibnamefont
  {Beckman}}, \bibinfo {author} {\bibfnamefont {A.~N.}\ \bibnamefont {Chari}},
  \bibinfo {author} {\bibfnamefont {S.}~\bibnamefont {Devabhaktuni}},\ and\
  \bibinfo {author} {\bibfnamefont {J.}~\bibnamefont {Preskill}},\ }\bibfield
  {title} {\bibinfo {title} {Efficient networks for quantum factoring},\ }\href
  {https://doi.org/10.1103/PhysRevA.54.1034} {\bibfield  {journal} {\bibinfo
  {journal} {Phys. Rev. A}\ }\textbf {\bibinfo {volume} {54}},\ \bibinfo
  {pages} {1034} (\bibinfo {year} {1996})}\BibitemShut {NoStop}%
\bibitem [{\citenamefont {Kaur}\ and\ \citenamefont
  {Dhaliwal}(2012)}]{kaur_IEEE_2012}%
  \BibitemOpen
  \bibfield  {author} {\bibinfo {author} {\bibfnamefont {P.}~\bibnamefont
  {Kaur}}\ and\ \bibinfo {author} {\bibfnamefont {B.~S.}\ \bibnamefont
  {Dhaliwal}},\ }\bibfield  {title} {\bibinfo {title} {{Design of fault
  tolearnt full Adder/Subtarctor using reversible gates}},\ }in\ \href
  {https://doi.org/10.1109/ICCCI.2012.6158883} {\emph {\bibinfo {booktitle}
  {{2012 International Conference on Computer Communication and
  Informatics}}}}\ (\bibinfo  {publisher} {IEEE},\ \bibinfo {year} {2012})\
  pp.\ \bibinfo {pages} {10--12}\BibitemShut {NoStop}%
\bibitem [{\citenamefont {Draper}\ \emph {et~al.}(2004)\citenamefont {Draper},
  \citenamefont {Kutin}, \citenamefont {Rains},\ and\ \citenamefont
  {Svore}}]{draper_arxiv_2004}%
  \BibitemOpen
  \bibfield  {author} {\bibinfo {author} {\bibfnamefont {T.~G.}\ \bibnamefont
  {Draper}}, \bibinfo {author} {\bibfnamefont {S.~A.}\ \bibnamefont {Kutin}},
  \bibinfo {author} {\bibfnamefont {E.~M.}\ \bibnamefont {Rains}},\ and\
  \bibinfo {author} {\bibfnamefont {K.~M.}\ \bibnamefont {Svore}},\ }\bibfield
  {title} {\bibinfo {title} {{A logarithmic-depth quantum carry-lookahead
  adder}},\ }\bibfield  {journal} {\bibinfo  {journal} {arXiv}\ }\href
  {https://doi.org/10.48550/arXiv.quant-ph/0406142}
  {10.48550/arXiv.quant-ph/0406142} (\bibinfo {year} {2004}),\ \Eprint
  {https://arxiv.org/abs/quant-ph/0406142} {quant-ph/0406142} \BibitemShut
  {NoStop}%
\bibitem [{\citenamefont {Remaud}\ and\ \citenamefont
  {Vandaele}(2025)}]{Remaud2025Jan}%
  \BibitemOpen
  \bibfield  {author} {\bibinfo {author} {\bibfnamefont {M.}~\bibnamefont
  {Remaud}}\ and\ \bibinfo {author} {\bibfnamefont {V.}~\bibnamefont
  {Vandaele}},\ }\bibfield  {title} {\bibinfo {title} {{Ancilla-free Quantum
  Adder with Sublinear Depth}},\ }\bibfield  {journal} {\bibinfo  {journal}
  {arXiv}\ }\href {https://doi.org/10.48550/arXiv.2501.16802}
  {10.48550/arXiv.2501.16802} (\bibinfo {year} {2025}),\ \Eprint
  {https://arxiv.org/abs/2501.16802} {2501.16802} \BibitemShut {NoStop}%
\bibitem [{\citenamefont {Nam}\ and\ \citenamefont
  {Bl\"umel}(2015)}]{blumel_pra_2015}%
  \BibitemOpen
  \bibfield  {author} {\bibinfo {author} {\bibfnamefont {Y.~S.}\ \bibnamefont
  {Nam}}\ and\ \bibinfo {author} {\bibfnamefont {R.}~\bibnamefont {Bl\"umel}},\
  }\bibfield  {title} {\bibinfo {title} {Analytical formulas for the
  performance scaling of quantum processors with a large number of defective
  gates},\ }\href {https://doi.org/10.1103/PhysRevA.92.042301} {\bibfield
  {journal} {\bibinfo  {journal} {Phys. Rev. A}\ }\textbf {\bibinfo {volume}
  {92}},\ \bibinfo {pages} {042301} (\bibinfo {year} {2015})}\BibitemShut
  {NoStop}%
\bibitem [{\citenamefont {Pavlidis}\ and\ \citenamefont
  {Floratos}(2021)}]{pavlidis_pra_2021}%
  \BibitemOpen
  \bibfield  {author} {\bibinfo {author} {\bibfnamefont {A.}~\bibnamefont
  {Pavlidis}}\ and\ \bibinfo {author} {\bibfnamefont {E.}~\bibnamefont
  {Floratos}},\ }\bibfield  {title} {\bibinfo {title}
  {Quantum-fourier-transform-based quantum arithmetic with qudits},\ }\href
  {https://doi.org/10.1103/PhysRevA.103.032417} {\bibfield  {journal} {\bibinfo
   {journal} {Phys. Rev. A}\ }\textbf {\bibinfo {volume} {103}},\ \bibinfo
  {pages} {032417} (\bibinfo {year} {2021})}\BibitemShut {NoStop}%
\bibitem [{\citenamefont {Gisin}\ and\ \citenamefont
  {Peres}(1992)}]{Gisin1992Jan}%
  \BibitemOpen
  \bibfield  {author} {\bibinfo {author} {\bibfnamefont {N.}~\bibnamefont
  {Gisin}}\ and\ \bibinfo {author} {\bibfnamefont {A.}~\bibnamefont {Peres}},\
  }\bibfield  {title} {\bibinfo {title} {{Maximal violation of Bell's
  inequality for arbitrarily large spin}},\ }\href
  {https://doi.org/10.1016/0375-9601(92)90949-M} {\bibfield  {journal}
  {\bibinfo  {journal} {Phys. Lett. A}\ }\textbf {\bibinfo {volume} {162}},\
  \bibinfo {pages} {15} (\bibinfo {year} {1992})}\BibitemShut {NoStop}%
\bibitem [{\citenamefont {Garg}\ and\ \citenamefont
  {Mermin}(1982)}]{PhysRevLett.49.901}%
  \BibitemOpen
  \bibfield  {author} {\bibinfo {author} {\bibfnamefont {A.}~\bibnamefont
  {Garg}}\ and\ \bibinfo {author} {\bibfnamefont {N.~D.}\ \bibnamefont
  {Mermin}},\ }\bibfield  {title} {\bibinfo {title} {Bell inequalities with a
  range of violation that does not diminish as the spin becomes arbitrarily
  large},\ }\href {https://doi.org/10.1103/PhysRevLett.49.901} {\bibfield
  {journal} {\bibinfo  {journal} {Phys. Rev. Lett.}\ }\textbf {\bibinfo
  {volume} {49}},\ \bibinfo {pages} {901} (\bibinfo {year} {1982})}\BibitemShut
  {NoStop}%
\bibitem [{\citenamefont {V\'ertesi}\ \emph {et~al.}(2010)\citenamefont
  {V\'ertesi}, \citenamefont {Pironio},\ and\ \citenamefont
  {Brunner}}]{bell_qudit}%
  \BibitemOpen
  \bibfield  {author} {\bibinfo {author} {\bibfnamefont {T.}~\bibnamefont
  {V\'ertesi}}, \bibinfo {author} {\bibfnamefont {S.}~\bibnamefont {Pironio}},\
  and\ \bibinfo {author} {\bibfnamefont {N.}~\bibnamefont {Brunner}},\
  }\bibfield  {title} {\bibinfo {title} {Closing the detection loophole in bell
  experiments using qudits},\ }\href
  {https://doi.org/10.1103/PhysRevLett.104.060401} {\bibfield  {journal}
  {\bibinfo  {journal} {Phys. Rev. Lett.}\ }\textbf {\bibinfo {volume} {104}},\
  \bibinfo {pages} {060401} (\bibinfo {year} {2010})}\BibitemShut {NoStop}%
\bibitem [{\citenamefont {Durt}\ \emph {et~al.}(2004)\citenamefont {Durt},
  \citenamefont {Kaszlikowski}, \citenamefont {Chen},\ and\ \citenamefont
  {Kwek}}]{qkd_qudit1}%
  \BibitemOpen
  \bibfield  {author} {\bibinfo {author} {\bibfnamefont {T.}~\bibnamefont
  {Durt}}, \bibinfo {author} {\bibfnamefont {D.}~\bibnamefont {Kaszlikowski}},
  \bibinfo {author} {\bibfnamefont {J.-L.}\ \bibnamefont {Chen}},\ and\
  \bibinfo {author} {\bibfnamefont {L.~C.}\ \bibnamefont {Kwek}},\ }\bibfield
  {title} {\bibinfo {title} {Security of quantum key distributions with
  entangled qudits},\ }\href {https://doi.org/10.1103/PhysRevA.69.032313}
  {\bibfield  {journal} {\bibinfo  {journal} {Phys. Rev. A}\ }\textbf {\bibinfo
  {volume} {69}},\ \bibinfo {pages} {032313} (\bibinfo {year}
  {2004})}\BibitemShut {NoStop}%
\bibitem [{\citenamefont {Durt}\ \emph {et~al.}(2003)\citenamefont {Durt},
  \citenamefont {Cerf}, \citenamefont {Gisin},\ and\ \citenamefont
  {\ifmmode~\dot{Z}\else \.{Z}\fi{}ukowski}}]{qkd_qudit2}%
  \BibitemOpen
  \bibfield  {author} {\bibinfo {author} {\bibfnamefont {T.}~\bibnamefont
  {Durt}}, \bibinfo {author} {\bibfnamefont {N.~J.}\ \bibnamefont {Cerf}},
  \bibinfo {author} {\bibfnamefont {N.}~\bibnamefont {Gisin}},\ and\ \bibinfo
  {author} {\bibfnamefont {M.}~\bibnamefont {\ifmmode~\dot{Z}\else
  \.{Z}\fi{}ukowski}},\ }\bibfield  {title} {\bibinfo {title} {Security of
  quantum key distribution with entangled qutrits},\ }\href
  {https://doi.org/10.1103/PhysRevA.67.012311} {\bibfield  {journal} {\bibinfo
  {journal} {Phys. Rev. A}\ }\textbf {\bibinfo {volume} {67}},\ \bibinfo
  {pages} {012311} (\bibinfo {year} {2003})}\BibitemShut {NoStop}%
\bibitem [{\citenamefont {Lewenstein}\ \emph {et~al.}(2007)\citenamefont
  {Lewenstein}, \citenamefont {Sanpera}, \citenamefont {Ahufinger},
  \citenamefont {Damski}, \citenamefont {Sen(De)},\ and\ \citenamefont
  {Sen}}]{aditi_advanves_physics_2007}%
  \BibitemOpen
  \bibfield  {author} {\bibinfo {author} {\bibfnamefont {M.}~\bibnamefont
  {Lewenstein}}, \bibinfo {author} {\bibfnamefont {A.}~\bibnamefont {Sanpera}},
  \bibinfo {author} {\bibfnamefont {V.}~\bibnamefont {Ahufinger}}, \bibinfo
  {author} {\bibfnamefont {B.}~\bibnamefont {Damski}}, \bibinfo {author}
  {\bibfnamefont {A.}~\bibnamefont {Sen(De)}},\ and\ \bibinfo {author}
  {\bibfnamefont {U.}~\bibnamefont {Sen}},\ }\bibfield  {title} {\bibinfo
  {title} {{Ultracold atomic gases in optical lattices: mimicking condensed
  matter physics and beyond}},\ }\href
  {https://doi.org/10.1080/00018730701223200} {\bibfield  {journal} {\bibinfo
  {journal} {Adv. Phys.}\ }\textbf {\bibinfo {volume} {56}},\ \bibinfo {pages}
  {243} (\bibinfo {year} {2007})}\BibitemShut {NoStop}%
\bibitem [{\citenamefont {Bartlett}\ \emph {et~al.}(2002)\citenamefont
  {Bartlett}, \citenamefont {de~Guise},\ and\ \citenamefont
  {Sanders}}]{barry_qudit}%
  \BibitemOpen
  \bibfield  {author} {\bibinfo {author} {\bibfnamefont {S.~D.}\ \bibnamefont
  {Bartlett}}, \bibinfo {author} {\bibfnamefont {H.}~\bibnamefont {de~Guise}},\
  and\ \bibinfo {author} {\bibfnamefont {B.~C.}\ \bibnamefont {Sanders}},\
  }\bibfield  {title} {\bibinfo {title} {Quantum encodings in spin systems and
  harmonic oscillators},\ }\href {https://doi.org/10.1103/PhysRevA.65.052316}
  {\bibfield  {journal} {\bibinfo  {journal} {Phys. Rev. A}\ }\textbf {\bibinfo
  {volume} {65}},\ \bibinfo {pages} {052316} (\bibinfo {year}
  {2002})}\BibitemShut {NoStop}%
\bibitem [{\citenamefont {Skrzypczyk}\ and\ \citenamefont
  {Cavalcanti}(2018)}]{random_qudit1}%
  \BibitemOpen
  \bibfield  {author} {\bibinfo {author} {\bibfnamefont {P.}~\bibnamefont
  {Skrzypczyk}}\ and\ \bibinfo {author} {\bibfnamefont {D.}~\bibnamefont
  {Cavalcanti}},\ }\bibfield  {title} {\bibinfo {title} {Maximal randomness
  generation from steering inequality violations using qudits},\ }\href
  {https://doi.org/10.1103/PhysRevLett.120.260401} {\bibfield  {journal}
  {\bibinfo  {journal} {Phys. Rev. Lett.}\ }\textbf {\bibinfo {volume} {120}},\
  \bibinfo {pages} {260401} (\bibinfo {year} {2018})}\BibitemShut {NoStop}%
\bibitem [{\citenamefont {Roy}\ \emph {et~al.}(2022)\citenamefont {Roy},
  \citenamefont {Mal},\ and\ \citenamefont {Sen(De)}}]{teleport_qudit1}%
  \BibitemOpen
  \bibfield  {author} {\bibinfo {author} {\bibfnamefont {S.}~\bibnamefont
  {Roy}}, \bibinfo {author} {\bibfnamefont {S.}~\bibnamefont {Mal}},\ and\
  \bibinfo {author} {\bibfnamefont {A.}~\bibnamefont {Sen(De)}},\ }\bibfield
  {title} {\bibinfo {title} {Gain in performance of teleportation with
  uniformity-breaking distributions},\ }\href
  {https://doi.org/10.1103/PhysRevA.105.022610} {\bibfield  {journal} {\bibinfo
   {journal} {Phys. Rev. A}\ }\textbf {\bibinfo {volume} {105}},\ \bibinfo
  {pages} {022610} (\bibinfo {year} {2022})}\BibitemShut {NoStop}%
\bibitem [{\citenamefont {Fonseca}(2019)}]{teleport_qudit2}%
  \BibitemOpen
  \bibfield  {author} {\bibinfo {author} {\bibfnamefont {A.}~\bibnamefont
  {Fonseca}},\ }\bibfield  {title} {\bibinfo {title} {High-dimensional quantum
  teleportation under noisy environments},\ }\href
  {https://doi.org/10.1103/PhysRevA.100.062311} {\bibfield  {journal} {\bibinfo
   {journal} {Phys. Rev. A}\ }\textbf {\bibinfo {volume} {100}},\ \bibinfo
  {pages} {062311} (\bibinfo {year} {2019})}\BibitemShut {NoStop}%
\bibitem [{\citenamefont {Miguel-Ramiro}\ and\ \citenamefont
  {D\"ur}(2018)}]{enta_qudit1}%
  \BibitemOpen
  \bibfield  {author} {\bibinfo {author} {\bibfnamefont {J.}~\bibnamefont
  {Miguel-Ramiro}}\ and\ \bibinfo {author} {\bibfnamefont {W.}~\bibnamefont
  {D\"ur}},\ }\bibfield  {title} {\bibinfo {title} {Efficient entanglement
  purification protocols for $d$-level systems},\ }\href
  {https://doi.org/10.1103/PhysRevA.98.042309} {\bibfield  {journal} {\bibinfo
  {journal} {Phys. Rev. A}\ }\textbf {\bibinfo {volume} {98}},\ \bibinfo
  {pages} {042309} (\bibinfo {year} {2018})}\BibitemShut {NoStop}%
\bibitem [{\citenamefont {Ralph}\ \emph {et~al.}(2007)\citenamefont {Ralph},
  \citenamefont {Resch},\ and\ \citenamefont {Gilchrist}}]{toffoli_qudit}%
  \BibitemOpen
  \bibfield  {author} {\bibinfo {author} {\bibfnamefont {T.~C.}\ \bibnamefont
  {Ralph}}, \bibinfo {author} {\bibfnamefont {K.~J.}\ \bibnamefont {Resch}},\
  and\ \bibinfo {author} {\bibfnamefont {A.}~\bibnamefont {Gilchrist}},\
  }\bibfield  {title} {\bibinfo {title} {Efficient toffoli gates using
  qudits},\ }\href {https://doi.org/10.1103/PhysRevA.75.022313} {\bibfield
  {journal} {\bibinfo  {journal} {Phys. Rev. A}\ }\textbf {\bibinfo {volume}
  {75}},\ \bibinfo {pages} {022313} (\bibinfo {year} {2007})}\BibitemShut
  {NoStop}%
\bibitem [{\citenamefont {Wang}\ \emph {et~al.}(2020)\citenamefont {Wang},
  \citenamefont {Hu}, \citenamefont {Sanders},\ and\ \citenamefont
  {Kais}}]{wang_fp_2020}%
  \BibitemOpen
  \bibfield  {author} {\bibinfo {author} {\bibfnamefont {Y.}~\bibnamefont
  {Wang}}, \bibinfo {author} {\bibfnamefont {Z.}~\bibnamefont {Hu}}, \bibinfo
  {author} {\bibfnamefont {B.~C.}\ \bibnamefont {Sanders}},\ and\ \bibinfo
  {author} {\bibfnamefont {S.}~\bibnamefont {Kais}},\ }\bibfield  {title}
  {\bibinfo {title} {{Qudits and High-Dimensional Quantum Computing}},\ }\href
  {https://doi.org/10.3389/fphy.2020.589504} {\bibfield  {journal} {\bibinfo
  {journal} {Front. Phys.}\ }\textbf {\bibinfo {volume} {8}},\ \bibinfo {pages}
  {589504} (\bibinfo {year} {2020})}\BibitemShut {NoStop}%
\bibitem [{\citenamefont {Neeley}\ \emph {et~al.}(2009)\citenamefont {Neeley},
  \citenamefont {Ansmann}, \citenamefont {Bialczak}, \citenamefont {Hofheinz},
  \citenamefont {Lucero}, \citenamefont {O'Connell}, \citenamefont {Sank},
  \citenamefont {Wang}, \citenamefont {Wenner}, \citenamefont {Cleland},
  \citenamefont {Geller},\ and\ \citenamefont
  {Martinis}}]{superconducting_qudit}%
  \BibitemOpen
  \bibfield  {author} {\bibinfo {author} {\bibfnamefont {M.}~\bibnamefont
  {Neeley}}, \bibinfo {author} {\bibfnamefont {M.}~\bibnamefont {Ansmann}},
  \bibinfo {author} {\bibfnamefont {R.~C.}\ \bibnamefont {Bialczak}}, \bibinfo
  {author} {\bibfnamefont {M.}~\bibnamefont {Hofheinz}}, \bibinfo {author}
  {\bibfnamefont {E.}~\bibnamefont {Lucero}}, \bibinfo {author} {\bibfnamefont
  {A.~D.}\ \bibnamefont {O'Connell}}, \bibinfo {author} {\bibfnamefont
  {D.}~\bibnamefont {Sank}}, \bibinfo {author} {\bibfnamefont {H.}~\bibnamefont
  {Wang}}, \bibinfo {author} {\bibfnamefont {J.}~\bibnamefont {Wenner}},
  \bibinfo {author} {\bibfnamefont {A.~N.}\ \bibnamefont {Cleland}}, \bibinfo
  {author} {\bibfnamefont {M.~R.}\ \bibnamefont {Geller}},\ and\ \bibinfo
  {author} {\bibfnamefont {J.~M.}\ \bibnamefont {Martinis}},\ }\bibfield
  {title} {\bibinfo {title} {{Emulation of a Quantum Spin with a
  Superconducting Phase Qudit}},\ }\href
  {https://doi.org/10.1126/science.1173440} {\bibfield  {journal} {\bibinfo
  {journal} {Science}\ }\textbf {\bibinfo {volume} {325}},\ \bibinfo {pages}
  {722} (\bibinfo {year} {2009})}\BibitemShut {NoStop}%
\bibitem [{\citenamefont {Soltamov}\ \emph {et~al.}(2019)\citenamefont
  {Soltamov}, \citenamefont {Kasper}, \citenamefont {Poshakinskiy},
  \citenamefont {Anisimov}, \citenamefont {Mokhov}, \citenamefont {Sperlich},
  \citenamefont {Tarasenko}, \citenamefont {Baranov}, \citenamefont
  {Astakhov},\ and\ \citenamefont {Dyakonov}}]{nitrogen_qudit}%
  \BibitemOpen
  \bibfield  {author} {\bibinfo {author} {\bibfnamefont {V.~A.}\ \bibnamefont
  {Soltamov}}, \bibinfo {author} {\bibfnamefont {C.}~\bibnamefont {Kasper}},
  \bibinfo {author} {\bibfnamefont {A.~V.}\ \bibnamefont {Poshakinskiy}},
  \bibinfo {author} {\bibfnamefont {A.~N.}\ \bibnamefont {Anisimov}}, \bibinfo
  {author} {\bibfnamefont {E.~N.}\ \bibnamefont {Mokhov}}, \bibinfo {author}
  {\bibfnamefont {A.}~\bibnamefont {Sperlich}}, \bibinfo {author}
  {\bibfnamefont {S.~A.}\ \bibnamefont {Tarasenko}}, \bibinfo {author}
  {\bibfnamefont {P.~G.}\ \bibnamefont {Baranov}}, \bibinfo {author}
  {\bibfnamefont {G.~V.}\ \bibnamefont {Astakhov}},\ and\ \bibinfo {author}
  {\bibfnamefont {V.}~\bibnamefont {Dyakonov}},\ }\bibfield  {title} {\bibinfo
  {title} {{Excitation and coherent control of spin qudit modes in silicon
  carbide at room temperature}},\ }\href
  {https://doi.org/10.1038/s41467-019-09429-x} {\bibfield  {journal} {\bibinfo
  {journal} {Nat. Commun.}\ }\textbf {\bibinfo {volume} {10}},\ \bibinfo
  {pages} {1} (\bibinfo {year} {2019})}\BibitemShut {NoStop}%
\bibitem [{\citenamefont {Chi}\ \emph {et~al.}(2022)\citenamefont {Chi},
  \citenamefont {Huang}, \citenamefont {Zhang}, \citenamefont {Mao},
  \citenamefont {Zhou}, \citenamefont {Chen}, \citenamefont {Zhai},
  \citenamefont {Bao}, \citenamefont {Dai}, \citenamefont {Yuan}, \citenamefont
  {Zhang}, \citenamefont {Dai}, \citenamefont {Tang}, \citenamefont {Yang},
  \citenamefont {Li}, \citenamefont {Ding}, \citenamefont {Oxenl{\o}we},
  \citenamefont {Thompson}, \citenamefont {O{'}Brien}, \citenamefont {Li},
  \citenamefont {Gong},\ and\ \citenamefont {Wang}}]{silicon_qudit}%
  \BibitemOpen
  \bibfield  {author} {\bibinfo {author} {\bibfnamefont {Y.}~\bibnamefont
  {Chi}}, \bibinfo {author} {\bibfnamefont {J.}~\bibnamefont {Huang}}, \bibinfo
  {author} {\bibfnamefont {Z.}~\bibnamefont {Zhang}}, \bibinfo {author}
  {\bibfnamefont {J.}~\bibnamefont {Mao}}, \bibinfo {author} {\bibfnamefont
  {Z.}~\bibnamefont {Zhou}}, \bibinfo {author} {\bibfnamefont {X.}~\bibnamefont
  {Chen}}, \bibinfo {author} {\bibfnamefont {C.}~\bibnamefont {Zhai}}, \bibinfo
  {author} {\bibfnamefont {J.}~\bibnamefont {Bao}}, \bibinfo {author}
  {\bibfnamefont {T.}~\bibnamefont {Dai}}, \bibinfo {author} {\bibfnamefont
  {H.}~\bibnamefont {Yuan}}, \bibinfo {author} {\bibfnamefont {M.}~\bibnamefont
  {Zhang}}, \bibinfo {author} {\bibfnamefont {D.}~\bibnamefont {Dai}}, \bibinfo
  {author} {\bibfnamefont {B.}~\bibnamefont {Tang}}, \bibinfo {author}
  {\bibfnamefont {Y.}~\bibnamefont {Yang}}, \bibinfo {author} {\bibfnamefont
  {Z.}~\bibnamefont {Li}}, \bibinfo {author} {\bibfnamefont {Y.}~\bibnamefont
  {Ding}}, \bibinfo {author} {\bibfnamefont {L.~K.}\ \bibnamefont
  {Oxenl{\o}we}}, \bibinfo {author} {\bibfnamefont {M.~G.}\ \bibnamefont
  {Thompson}}, \bibinfo {author} {\bibfnamefont {J.~L.}\ \bibnamefont
  {O{'}Brien}}, \bibinfo {author} {\bibfnamefont {Y.}~\bibnamefont {Li}},
  \bibinfo {author} {\bibfnamefont {Q.}~\bibnamefont {Gong}},\ and\ \bibinfo
  {author} {\bibfnamefont {J.}~\bibnamefont {Wang}},\ }\bibfield  {title}
  {\bibinfo {title} {{A programmable qudit-based quantum processor}},\ }\href
  {https://doi.org/10.1038/s41467-022-28767-x} {\bibfield  {journal} {\bibinfo
  {journal} {Nat. Commun.}\ }\textbf {\bibinfo {volume} {13}},\ \bibinfo
  {pages} {1} (\bibinfo {year} {2022})}\BibitemShut {NoStop}%
\bibitem [{\citenamefont {Low}\ \emph {et~al.}(2020)\citenamefont {Low},
  \citenamefont {White}, \citenamefont {Cox}, \citenamefont {Day},\ and\
  \citenamefont {Senko}}]{qudit_iontrap1}%
  \BibitemOpen
  \bibfield  {author} {\bibinfo {author} {\bibfnamefont {P.~J.}\ \bibnamefont
  {Low}}, \bibinfo {author} {\bibfnamefont {B.~M.}\ \bibnamefont {White}},
  \bibinfo {author} {\bibfnamefont {A.~A.}\ \bibnamefont {Cox}}, \bibinfo
  {author} {\bibfnamefont {M.~L.}\ \bibnamefont {Day}},\ and\ \bibinfo {author}
  {\bibfnamefont {C.}~\bibnamefont {Senko}},\ }\bibfield  {title} {\bibinfo
  {title} {Practical trapped-ion protocols for universal qudit-based quantum
  computing},\ }\href {https://doi.org/10.1103/PhysRevResearch.2.033128}
  {\bibfield  {journal} {\bibinfo  {journal} {Phys. Rev. Res.}\ }\textbf
  {\bibinfo {volume} {2}},\ \bibinfo {pages} {033128} (\bibinfo {year}
  {2020})}\BibitemShut {NoStop}%
\bibitem [{\citenamefont {Hrmo}\ \emph {et~al.}(2023)\citenamefont {Hrmo},
  \citenamefont {Wilhelm}, \citenamefont {Gerster}, \citenamefont {van Mourik},
  \citenamefont {Huber}, \citenamefont {Blatt}, \citenamefont {Schindler},
  \citenamefont {Monz},\ and\ \citenamefont {Ringbauer}}]{qudit_iontrap2}%
  \BibitemOpen
  \bibfield  {author} {\bibinfo {author} {\bibfnamefont {P.}~\bibnamefont
  {Hrmo}}, \bibinfo {author} {\bibfnamefont {B.}~\bibnamefont {Wilhelm}},
  \bibinfo {author} {\bibfnamefont {L.}~\bibnamefont {Gerster}}, \bibinfo
  {author} {\bibfnamefont {M.~W.}\ \bibnamefont {van Mourik}}, \bibinfo
  {author} {\bibfnamefont {M.}~\bibnamefont {Huber}}, \bibinfo {author}
  {\bibfnamefont {R.}~\bibnamefont {Blatt}}, \bibinfo {author} {\bibfnamefont
  {P.}~\bibnamefont {Schindler}}, \bibinfo {author} {\bibfnamefont
  {T.}~\bibnamefont {Monz}},\ and\ \bibinfo {author} {\bibfnamefont
  {M.}~\bibnamefont {Ringbauer}},\ }\bibfield  {title} {\bibinfo {title}
  {{Native qudit entanglement in a trapped ion quantum processor}},\ }\href
  {https://doi.org/10.1038/s41467-023-37375-2} {\bibfield  {journal} {\bibinfo
  {journal} {Nat. Commun.}\ }\textbf {\bibinfo {volume} {14}},\ \bibinfo
  {pages} {1} (\bibinfo {year} {2023})}\BibitemShut {NoStop}%
\bibitem [{\citenamefont {Ringbauer}\ \emph {et~al.}(2022)\citenamefont
  {Ringbauer}, \citenamefont {Meth}, \citenamefont {Postler}, \citenamefont
  {Stricker}, \citenamefont {Blatt}, \citenamefont {Schindler},\ and\
  \citenamefont {Monz}}]{qudit_iontrap3}%
  \BibitemOpen
  \bibfield  {author} {\bibinfo {author} {\bibfnamefont {M.}~\bibnamefont
  {Ringbauer}}, \bibinfo {author} {\bibfnamefont {M.}~\bibnamefont {Meth}},
  \bibinfo {author} {\bibfnamefont {L.}~\bibnamefont {Postler}}, \bibinfo
  {author} {\bibfnamefont {R.}~\bibnamefont {Stricker}}, \bibinfo {author}
  {\bibfnamefont {R.}~\bibnamefont {Blatt}}, \bibinfo {author} {\bibfnamefont
  {P.}~\bibnamefont {Schindler}},\ and\ \bibinfo {author} {\bibfnamefont
  {T.}~\bibnamefont {Monz}},\ }\bibfield  {title} {\bibinfo {title} {{A
  universal qudit quantum processor with trapped ions}},\ }\href
  {https://doi.org/10.1038/s41567-022-01658-0} {\bibfield  {journal} {\bibinfo
  {journal} {Nat. Phys.}\ }\textbf {\bibinfo {volume} {18}},\ \bibinfo {pages}
  {1053} (\bibinfo {year} {2022})}\BibitemShut {NoStop}%
\bibitem [{\citenamefont {Erhard}\ \emph {et~al.}(2018)\citenamefont {Erhard},
  \citenamefont {Fickler}, \citenamefont {Krenn},\ and\ \citenamefont
  {Zeilinger}}]{photon_qudit}%
  \BibitemOpen
  \bibfield  {author} {\bibinfo {author} {\bibfnamefont {M.}~\bibnamefont
  {Erhard}}, \bibinfo {author} {\bibfnamefont {R.}~\bibnamefont {Fickler}},
  \bibinfo {author} {\bibfnamefont {M.}~\bibnamefont {Krenn}},\ and\ \bibinfo
  {author} {\bibfnamefont {A.}~\bibnamefont {Zeilinger}},\ }\bibfield  {title}
  {\bibinfo {title} {{Twisted photons: new quantum perspectives in high
  dimensions}},\ }\href {https://doi.org/10.1038/lsa.2017.146} {\bibfield
  {journal} {\bibinfo  {journal} {Light Sci. Appl.}\ }\textbf {\bibinfo
  {volume} {7}},\ \bibinfo {pages} {17146} (\bibinfo {year}
  {2018})}\BibitemShut {NoStop}%
\bibitem [{\citenamefont {D\"ur}\ \emph {et~al.}(2005)\citenamefont {D\"ur},
  \citenamefont {Hartmann}, \citenamefont {Hein}, \citenamefont {Lewenstein},\
  and\ \citenamefont {Briegel}}]{maciej_prl_longrangecluseter_2015}%
  \BibitemOpen
  \bibfield  {author} {\bibinfo {author} {\bibfnamefont {W.}~\bibnamefont
  {D\"ur}}, \bibinfo {author} {\bibfnamefont {L.}~\bibnamefont {Hartmann}},
  \bibinfo {author} {\bibfnamefont {M.}~\bibnamefont {Hein}}, \bibinfo {author}
  {\bibfnamefont {M.}~\bibnamefont {Lewenstein}},\ and\ \bibinfo {author}
  {\bibfnamefont {H.-J.}\ \bibnamefont {Briegel}},\ }\bibfield  {title}
  {\bibinfo {title} {Entanglement in spin chains and lattices with long-range
  ising-type interactions},\ }\href
  {https://doi.org/10.1103/PhysRevLett.94.097203} {\bibfield  {journal}
  {\bibinfo  {journal} {Phys. Rev. Lett.}\ }\textbf {\bibinfo {volume} {94}},\
  \bibinfo {pages} {097203} (\bibinfo {year} {2005})}\BibitemShut {NoStop}%
\bibitem [{\citenamefont {Gossett}(1998)}]{gossett_arxiv_1998}%
  \BibitemOpen
  \bibfield  {author} {\bibinfo {author} {\bibfnamefont {P.}~\bibnamefont
  {Gossett}},\ }\href@noop {} {\bibinfo {title} {Quantum carry-save
  arithmetic}} (\bibinfo {year} {1998}),\ \Eprint
  {https://arxiv.org/abs/quant-ph/9808061} {arXiv:quant-ph/9808061 [quant-ph]}
  \BibitemShut {NoStop}%
\bibitem [{\citenamefont {Cuccaro}\ \emph {et~al.}(2004)\citenamefont
  {Cuccaro}, \citenamefont {Draper}, \citenamefont {Kutin},\ and\ \citenamefont
  {Moulton}}]{Cuccaro2004Oct}%
  \BibitemOpen
  \bibfield  {author} {\bibinfo {author} {\bibfnamefont {S.~A.}\ \bibnamefont
  {Cuccaro}}, \bibinfo {author} {\bibfnamefont {T.~G.}\ \bibnamefont {Draper}},
  \bibinfo {author} {\bibfnamefont {S.~A.}\ \bibnamefont {Kutin}},\ and\
  \bibinfo {author} {\bibfnamefont {D.~P.}\ \bibnamefont {Moulton}},\
  }\bibfield  {title} {\bibinfo {title} {{A new quantum ripple-carry addition
  circuit}},\ }\bibfield  {journal} {\bibinfo  {journal} {arXiv}\ }\href
  {https://doi.org/10.48550/arXiv.quant-ph/0410184}
  {10.48550/arXiv.quant-ph/0410184} (\bibinfo {year} {2004}),\ \Eprint
  {https://arxiv.org/abs/quant-ph/0410184} {quant-ph/0410184} \BibitemShut
  {NoStop}%
\bibitem [{\citenamefont {Ruiz-Perez}\ and\ \citenamefont
  {Garcia-Escartin}(2017)}]{Ruiz-Perez2017Jun}%
  \BibitemOpen
  \bibfield  {author} {\bibinfo {author} {\bibfnamefont {L.}~\bibnamefont
  {Ruiz-Perez}}\ and\ \bibinfo {author} {\bibfnamefont {J.~C.}\ \bibnamefont
  {Garcia-Escartin}},\ }\bibfield  {title} {\bibinfo {title} {{Quantum
  arithmetic with the quantum Fourier transform}},\ }\href
  {https://doi.org/10.1007/s11128-017-1603-1} {\bibfield  {journal} {\bibinfo
  {journal} {Quantum Inf. Process.}\ }\textbf {\bibinfo {volume} {16}},\
  \bibinfo {pages} {1} (\bibinfo {year} {2017})}\BibitemShut {NoStop}%
\bibitem [{\citenamefont {Camati}\ \emph {et~al.}(2019)\citenamefont {Camati},
  \citenamefont {Santos},\ and\ \citenamefont {Serra}}]{camati_pra_2019}%
  \BibitemOpen
  \bibfield  {author} {\bibinfo {author} {\bibfnamefont {P.~A.}\ \bibnamefont
  {Camati}}, \bibinfo {author} {\bibfnamefont {J.~F.~G.}\ \bibnamefont
  {Santos}},\ and\ \bibinfo {author} {\bibfnamefont {R.~M.}\ \bibnamefont
  {Serra}},\ }\bibfield  {title} {\bibinfo {title} {Coherence effects in the
  performance of the quantum otto heat engine},\ }\href
  {https://doi.org/10.1103/PhysRevA.99.062103} {\bibfield  {journal} {\bibinfo
  {journal} {Phys. Rev. A}\ }\textbf {\bibinfo {volume} {99}},\ \bibinfo
  {pages} {062103} (\bibinfo {year} {2019})}\BibitemShut {NoStop}%
\bibitem [{\citenamefont {Caravelli}\ \emph {et~al.}(2021)\citenamefont
  {Caravelli}, \citenamefont {Yan}, \citenamefont {Garc{\'{i}}a-Pintos},\ and\
  \citenamefont {Hamma}}]{Caravelli_quantum_2021}%
  \BibitemOpen
  \bibfield  {author} {\bibinfo {author} {\bibfnamefont {F.}~\bibnamefont
  {Caravelli}}, \bibinfo {author} {\bibfnamefont {B.}~\bibnamefont {Yan}},
  \bibinfo {author} {\bibfnamefont {L.~P.}\ \bibnamefont
  {Garc{\'{i}}a-Pintos}},\ and\ \bibinfo {author} {\bibfnamefont
  {A.}~\bibnamefont {Hamma}},\ }\bibfield  {title} {\bibinfo {title} {Energy
  storage and coherence in closed and open quantum batteries},\ }\href
  {https://doi.org/10.22331/q-2021-07-15-505} {\bibfield  {journal} {\bibinfo
  {journal} {{Quantum}}\ }\textbf {\bibinfo {volume} {5}},\ \bibinfo {pages}
  {505} (\bibinfo {year} {2021})}\BibitemShut {NoStop}%
\bibitem [{\citenamefont {Matern}\ \emph {et~al.}(2023)\citenamefont {Matern},
  \citenamefont {Macieszczak}, \citenamefont {Wozny},\ and\ \citenamefont
  {Leijnse}}]{matern_prb_2023}%
  \BibitemOpen
  \bibfield  {author} {\bibinfo {author} {\bibfnamefont {S.}~\bibnamefont
  {Matern}}, \bibinfo {author} {\bibfnamefont {K.}~\bibnamefont {Macieszczak}},
  \bibinfo {author} {\bibfnamefont {S.}~\bibnamefont {Wozny}},\ and\ \bibinfo
  {author} {\bibfnamefont {M.}~\bibnamefont {Leijnse}},\ }\bibfield  {title}
  {\bibinfo {title} {Metastability and quantum coherence assisted sensing in
  interacting parallel quantum dots},\ }\href
  {https://doi.org/10.1103/PhysRevB.107.125424} {\bibfield  {journal} {\bibinfo
   {journal} {Phys. Rev. B}\ }\textbf {\bibinfo {volume} {107}},\ \bibinfo
  {pages} {125424} (\bibinfo {year} {2023})}\BibitemShut {NoStop}%
\bibitem [{\citenamefont {Baumgratz}\ \emph {et~al.}(2014)\citenamefont
  {Baumgratz}, \citenamefont {Cramer},\ and\ \citenamefont
  {Plenio}}]{Baumgratz_Prl_2014}%
  \BibitemOpen
  \bibfield  {author} {\bibinfo {author} {\bibfnamefont {T.}~\bibnamefont
  {Baumgratz}}, \bibinfo {author} {\bibfnamefont {M.}~\bibnamefont {Cramer}},\
  and\ \bibinfo {author} {\bibfnamefont {M.~B.}\ \bibnamefont {Plenio}},\
  }\bibfield  {title} {\bibinfo {title} {Quantifying coherence},\ }\href
  {https://doi.org/10.1103/PhysRevLett.113.140401} {\bibfield  {journal}
  {\bibinfo  {journal} {Phys. Rev. Lett.}\ }\textbf {\bibinfo {volume} {113}},\
  \bibinfo {pages} {140401} (\bibinfo {year} {2014})}\BibitemShut {NoStop}%
\bibitem [{\citenamefont {Streltsov}\ \emph {et~al.}(2015)\citenamefont
  {Streltsov}, \citenamefont {Singh}, \citenamefont {Dhar}, \citenamefont
  {Bera},\ and\ \citenamefont {Adesso}}]{streltsov_prl_2015}%
  \BibitemOpen
  \bibfield  {author} {\bibinfo {author} {\bibfnamefont {A.}~\bibnamefont
  {Streltsov}}, \bibinfo {author} {\bibfnamefont {U.}~\bibnamefont {Singh}},
  \bibinfo {author} {\bibfnamefont {H.~S.}\ \bibnamefont {Dhar}}, \bibinfo
  {author} {\bibfnamefont {M.~N.}\ \bibnamefont {Bera}},\ and\ \bibinfo
  {author} {\bibfnamefont {G.}~\bibnamefont {Adesso}},\ }\bibfield  {title}
  {\bibinfo {title} {Measuring quantum coherence with entanglement},\ }\href
  {https://doi.org/10.1103/PhysRevLett.115.020403} {\bibfield  {journal}
  {\bibinfo  {journal} {Phys. Rev. Lett.}\ }\textbf {\bibinfo {volume} {115}},\
  \bibinfo {pages} {020403} (\bibinfo {year} {2015})}\BibitemShut {NoStop}%
\bibitem [{\citenamefont {Napoli}\ \emph {et~al.}(2016)\citenamefont {Napoli},
  \citenamefont {Bromley}, \citenamefont {Cianciaruso}, \citenamefont {Piani},
  \citenamefont {Johnston},\ and\ \citenamefont {Adesso}}]{napoli_prl_2016}%
  \BibitemOpen
  \bibfield  {author} {\bibinfo {author} {\bibfnamefont {C.}~\bibnamefont
  {Napoli}}, \bibinfo {author} {\bibfnamefont {T.~R.}\ \bibnamefont {Bromley}},
  \bibinfo {author} {\bibfnamefont {M.}~\bibnamefont {Cianciaruso}}, \bibinfo
  {author} {\bibfnamefont {M.}~\bibnamefont {Piani}}, \bibinfo {author}
  {\bibfnamefont {N.}~\bibnamefont {Johnston}},\ and\ \bibinfo {author}
  {\bibfnamefont {G.}~\bibnamefont {Adesso}},\ }\bibfield  {title} {\bibinfo
  {title} {Robustness of coherence: An operational and observable measure of
  quantum coherence},\ }\href {https://doi.org/10.1103/PhysRevLett.116.150502}
  {\bibfield  {journal} {\bibinfo  {journal} {Phys. Rev. Lett.}\ }\textbf
  {\bibinfo {volume} {116}},\ \bibinfo {pages} {150502} (\bibinfo {year}
  {2016})}\BibitemShut {NoStop}%
\bibitem [{\citenamefont {Piani}\ \emph {et~al.}(2016)\citenamefont {Piani},
  \citenamefont {Cianciaruso}, \citenamefont {Bromley}, \citenamefont {Napoli},
  \citenamefont {Johnston},\ and\ \citenamefont {Adesso}}]{piani_pra_2016}%
  \BibitemOpen
  \bibfield  {author} {\bibinfo {author} {\bibfnamefont {M.}~\bibnamefont
  {Piani}}, \bibinfo {author} {\bibfnamefont {M.}~\bibnamefont {Cianciaruso}},
  \bibinfo {author} {\bibfnamefont {T.~R.}\ \bibnamefont {Bromley}}, \bibinfo
  {author} {\bibfnamefont {C.}~\bibnamefont {Napoli}}, \bibinfo {author}
  {\bibfnamefont {N.}~\bibnamefont {Johnston}},\ and\ \bibinfo {author}
  {\bibfnamefont {G.}~\bibnamefont {Adesso}},\ }\bibfield  {title} {\bibinfo
  {title} {Robustness of asymmetry and coherence of quantum states},\ }\href
  {https://doi.org/10.1103/PhysRevA.93.042107} {\bibfield  {journal} {\bibinfo
  {journal} {Phys. Rev. A}\ }\textbf {\bibinfo {volume} {93}},\ \bibinfo
  {pages} {042107} (\bibinfo {year} {2016})}\BibitemShut {NoStop}%
\bibitem [{Note1()}]{Note1}%
  \BibitemOpen
  \bibinfo {note} {The lowest bound for non-modular addition is $2n+1$ which
  can be achieved in the circuit we have analyzed by simply rejecting the $\ket
  {b_n} $ qudit}\BibitemShut {NoStop}%
\bibitem [{\citenamefont {Zilic}\ and\ \citenamefont
  {Radecka}(2007)}]{zilic2007scalingbetterapproximatingquantum}%
  \BibitemOpen
  \bibfield  {author} {\bibinfo {author} {\bibfnamefont {Z.}~\bibnamefont
  {Zilic}}\ and\ \bibinfo {author} {\bibfnamefont {K.}~\bibnamefont
  {Radecka}},\ }\href {https://arxiv.org/abs/quant-ph/0702195} {\bibinfo
  {title} {Scaling and better approximating quantum fourier transform by higher
  radices}} (\bibinfo {year} {2007}),\ \Eprint
  {https://arxiv.org/abs/quant-ph/0702195} {arXiv:quant-ph/0702195 [quant-ph]}
  \BibitemShut {NoStop}%
\bibitem [{\citenamefont {Marques}\ \emph {et~al.}(2015)\citenamefont
  {Marques}, \citenamefont {Matoso}, \citenamefont {Pimenta}, \citenamefont
  {Guti{\ifmmode\acute{e}\else\'{e}\fi}rrez-Esparza}, \citenamefont {Santos},\
  and\ \citenamefont
  {P{\ifmmode\acute{a}\else\'{a}\fi}dua}}]{Marques_natureSR_2015}%
  \BibitemOpen
  \bibfield  {author} {\bibinfo {author} {\bibfnamefont {B.}~\bibnamefont
  {Marques}}, \bibinfo {author} {\bibfnamefont {A.~A.}\ \bibnamefont {Matoso}},
  \bibinfo {author} {\bibfnamefont {W.~M.}\ \bibnamefont {Pimenta}}, \bibinfo
  {author} {\bibfnamefont {A.~J.}\ \bibnamefont
  {Guti{\ifmmode\acute{e}\else\'{e}\fi}rrez-Esparza}}, \bibinfo {author}
  {\bibfnamefont {M.~F.}\ \bibnamefont {Santos}},\ and\ \bibinfo {author}
  {\bibfnamefont {S.}~\bibnamefont {P{\ifmmode\acute{a}\else\'{a}\fi}dua}},\
  }\bibfield  {title} {\bibinfo {title} {{Experimental simulation of
  decoherence in photonics qudits}},\ }\href
  {https://doi.org/10.1038/srep16049} {\bibfield  {journal} {\bibinfo
  {journal} {Sci. Rep.}\ }\textbf {\bibinfo {volume} {5}},\ \bibinfo {pages}
  {1} (\bibinfo {year} {2015})}\BibitemShut {NoStop}%
\bibitem [{\citenamefont {Dutta}\ \emph {et~al.}(2016)\citenamefont {Dutta},
  \citenamefont {Ryu}, \citenamefont {Laskowski},\ and\ \citenamefont
  {Żukowski}}]{dutta_pla_2016}%
  \BibitemOpen
  \bibfield  {author} {\bibinfo {author} {\bibfnamefont {A.}~\bibnamefont
  {Dutta}}, \bibinfo {author} {\bibfnamefont {J.}~\bibnamefont {Ryu}}, \bibinfo
  {author} {\bibfnamefont {W.}~\bibnamefont {Laskowski}},\ and\ \bibinfo
  {author} {\bibfnamefont {M.}~\bibnamefont {Żukowski}},\ }\bibfield  {title}
  {\bibinfo {title} {Entanglement criteria for noise resistance of two-qudit
  states},\ }\href
  {https://doi.org/https://doi.org/10.1016/j.physleta.2016.04.043} {\bibfield
  {journal} {\bibinfo  {journal} {Physics Letters A}\ }\textbf {\bibinfo
  {volume} {380}},\ \bibinfo {pages} {2191} (\bibinfo {year}
  {2016})}\BibitemShut {NoStop}%
\bibitem [{\citenamefont {Chessa}\ and\ \citenamefont
  {Giovannetti}(2023)}]{Chessa_Giovannetti_Quantum_2023}%
  \BibitemOpen
  \bibfield  {author} {\bibinfo {author} {\bibfnamefont {S.}~\bibnamefont
  {Chessa}}\ and\ \bibinfo {author} {\bibfnamefont {V.}~\bibnamefont
  {Giovannetti}},\ }\bibfield  {title} {\bibinfo {title} {Resonant {M}ultilevel
  {A}mplitude {D}amping {C}hannels},\ }\href
  {https://doi.org/10.22331/q-2023-01-19-902} {\bibfield  {journal} {\bibinfo
  {journal} {{Quantum}}\ }\textbf {\bibinfo {volume} {7}},\ \bibinfo {pages}
  {902} (\bibinfo {year} {2023})}\BibitemShut {NoStop}%
\bibitem [{\citenamefont {Macchiavello}\ and\ \citenamefont
  {Palma}(2001)}]{Macchiavello2001Jul}%
  \BibitemOpen
  \bibfield  {author} {\bibinfo {author} {\bibfnamefont {C.}~\bibnamefont
  {Macchiavello}}\ and\ \bibinfo {author} {\bibfnamefont {G.~M.}\ \bibnamefont
  {Palma}},\ }\bibfield  {title} {\bibinfo {title} {{Entanglement enhanced
  information transmission over a quantum channel with correlated noise}},\
  }\bibfield  {journal} {\bibinfo  {journal} {arXiv}\ }\href
  {https://doi.org/10.1103/PhysRevA.65.050301} {10.1103/PhysRevA.65.050301}
  (\bibinfo {year} {2001}),\ \Eprint {https://arxiv.org/abs/quant-ph/0107052}
  {quant-ph/0107052} \BibitemShut {NoStop}%
\bibitem [{\citenamefont {Nam}\ \emph {et~al.}(2020)\citenamefont {Nam},
  \citenamefont {Su},\ and\ \citenamefont {Maslov}}]{Nam2020Mar}%
  \BibitemOpen
  \bibfield  {author} {\bibinfo {author} {\bibfnamefont {Y.}~\bibnamefont
  {Nam}}, \bibinfo {author} {\bibfnamefont {Y.}~\bibnamefont {Su}},\ and\
  \bibinfo {author} {\bibfnamefont {D.}~\bibnamefont {Maslov}},\ }\bibfield
  {title} {\bibinfo {title} {{Approximate quantum Fourier transform with O(n
  log(n)) T gates}},\ }\href {https://doi.org/10.1038/s41534-020-0257-5}
  {\bibfield  {journal} {\bibinfo  {journal} {npj Quantum Inf.}\ }\textbf
  {\bibinfo {volume} {6}},\ \bibinfo {pages} {1} (\bibinfo {year}
  {2020})}\BibitemShut {NoStop}%
\bibitem [{\citenamefont {Chen}\ \emph {et~al.}(2023)\citenamefont {Chen},
  \citenamefont {Stoudenmire},\ and\ \citenamefont {White}}]{chen_prxq_2023}%
  \BibitemOpen
  \bibfield  {author} {\bibinfo {author} {\bibfnamefont {J.}~\bibnamefont
  {Chen}}, \bibinfo {author} {\bibfnamefont {E.}~\bibnamefont {Stoudenmire}},\
  and\ \bibinfo {author} {\bibfnamefont {S.~R.}\ \bibnamefont {White}},\
  }\bibfield  {title} {\bibinfo {title} {Quantum fourier transform has small
  entanglement},\ }\href {https://doi.org/10.1103/PRXQuantum.4.040318}
  {\bibfield  {journal} {\bibinfo  {journal} {PRX Quantum}\ }\textbf {\bibinfo
  {volume} {4}},\ \bibinfo {pages} {040318} (\bibinfo {year}
  {2023})}\BibitemShut {NoStop}%
\bibitem [{\citenamefont {Raussendorf}\ and\ \citenamefont
  {Briegel}(2001)}]{raussendorf_prl_2001}%
  \BibitemOpen
  \bibfield  {author} {\bibinfo {author} {\bibfnamefont {R.}~\bibnamefont
  {Raussendorf}}\ and\ \bibinfo {author} {\bibfnamefont {H.~J.}\ \bibnamefont
  {Briegel}},\ }\bibfield  {title} {\bibinfo {title} {A one-way quantum
  computer},\ }\href {https://doi.org/10.1103/PhysRevLett.86.5188} {\bibfield
  {journal} {\bibinfo  {journal} {Phys. Rev. Lett.}\ }\textbf {\bibinfo
  {volume} {86}},\ \bibinfo {pages} {5188} (\bibinfo {year}
  {2001})}\BibitemShut {NoStop}%
\bibitem [{\citenamefont {Gao}\ \emph {et~al.}(2011)\citenamefont {Gao},
  \citenamefont {Yao}, \citenamefont {Cai}, \citenamefont {Lu}, \citenamefont
  {Xu}, \citenamefont {Yang}, \citenamefont {Lu}, \citenamefont {Chen},
  \citenamefont {Chen},\ and\ \citenamefont {Pan}}]{pan_mbqc_exp_2011}%
  \BibitemOpen
  \bibfield  {author} {\bibinfo {author} {\bibfnamefont {W.-B.}\ \bibnamefont
  {Gao}}, \bibinfo {author} {\bibfnamefont {X.-C.}\ \bibnamefont {Yao}},
  \bibinfo {author} {\bibfnamefont {J.-M.}\ \bibnamefont {Cai}}, \bibinfo
  {author} {\bibfnamefont {H.}~\bibnamefont {Lu}}, \bibinfo {author}
  {\bibfnamefont {P.}~\bibnamefont {Xu}}, \bibinfo {author} {\bibfnamefont
  {T.}~\bibnamefont {Yang}}, \bibinfo {author} {\bibfnamefont {C.-Y.}\
  \bibnamefont {Lu}}, \bibinfo {author} {\bibfnamefont {Y.-A.}\ \bibnamefont
  {Chen}}, \bibinfo {author} {\bibfnamefont {Z.-B.}\ \bibnamefont {Chen}},\
  and\ \bibinfo {author} {\bibfnamefont {J.-W.}\ \bibnamefont {Pan}},\
  }\bibfield  {title} {\bibinfo {title} {{Experimental measurement-based
  quantum computing beyond the cluster-state model}},\ }\href
  {https://doi.org/10.1038/nphoton.2010.283} {\bibfield  {journal} {\bibinfo
  {journal} {Nat. Photonics}\ }\textbf {\bibinfo {volume} {5}},\ \bibinfo
  {pages} {117} (\bibinfo {year} {2011})}\BibitemShut {NoStop}%
\bibitem [{\citenamefont {Goss}\ \emph {et~al.}(2022)\citenamefont {Goss},
  \citenamefont {Morvan}, \citenamefont {Marinelli} \emph
  {et~al.}}]{Goss2022Dec}%
  \BibitemOpen
  \bibfield  {author} {\bibinfo {author} {\bibfnamefont {N.}~\bibnamefont
  {Goss}}, \bibinfo {author} {\bibfnamefont {A.}~\bibnamefont {Morvan}},
  \bibinfo {author} {\bibfnamefont {B.}~\bibnamefont {Marinelli}}, \emph
  {et~al.},\ }\bibfield  {title} {\bibinfo {title} {{High-fidelity qutrit
  entangling gates for superconducting circuits}},\ }\href
  {https://doi.org/10.1038/s41467-022-34851-z} {\bibfield  {journal} {\bibinfo
  {journal} {Nat. Commun.}\ }\textbf {\bibinfo {volume} {13}},\ \bibinfo
  {pages} {1} (\bibinfo {year} {2022})}\BibitemShut {NoStop}%
\bibitem [{\citenamefont {Liu}\ \emph {et~al.}(2023)\citenamefont {Liu},
  \citenamefont {Wang}, \citenamefont {Zhang}, \citenamefont {Zhang},
  \citenamefont {Cai}, \citenamefont {Xu}, \citenamefont {Li}, \citenamefont
  {Han}, \citenamefont {Li}, \citenamefont {Xue}, \citenamefont {Liu},
  \citenamefont {You}, \citenamefont {Jin},\ and\ \citenamefont
  {Yu}}]{liu_prx_2023}%
  \BibitemOpen
  \bibfield  {author} {\bibinfo {author} {\bibfnamefont {P.}~\bibnamefont
  {Liu}}, \bibinfo {author} {\bibfnamefont {R.}~\bibnamefont {Wang}}, \bibinfo
  {author} {\bibfnamefont {J.-N.}\ \bibnamefont {Zhang}}, \bibinfo {author}
  {\bibfnamefont {Y.}~\bibnamefont {Zhang}}, \bibinfo {author} {\bibfnamefont
  {X.}~\bibnamefont {Cai}}, \bibinfo {author} {\bibfnamefont {H.}~\bibnamefont
  {Xu}}, \bibinfo {author} {\bibfnamefont {Z.}~\bibnamefont {Li}}, \bibinfo
  {author} {\bibfnamefont {J.}~\bibnamefont {Han}}, \bibinfo {author}
  {\bibfnamefont {X.}~\bibnamefont {Li}}, \bibinfo {author} {\bibfnamefont
  {G.}~\bibnamefont {Xue}}, \bibinfo {author} {\bibfnamefont {W.}~\bibnamefont
  {Liu}}, \bibinfo {author} {\bibfnamefont {L.}~\bibnamefont {You}}, \bibinfo
  {author} {\bibfnamefont {Y.}~\bibnamefont {Jin}},\ and\ \bibinfo {author}
  {\bibfnamefont {H.}~\bibnamefont {Yu}},\ }\bibfield  {title} {\bibinfo
  {title} {Performing $\mathrm{SU}(d)$ operations and rudimentary algorithms in
  a superconducting transmon qudit for $d=3$ and $d=4$},\ }\href
  {https://doi.org/10.1103/PhysRevX.13.021028} {\bibfield  {journal} {\bibinfo
  {journal} {Phys. Rev. X}\ }\textbf {\bibinfo {volume} {13}},\ \bibinfo
  {pages} {021028} (\bibinfo {year} {2023})}\BibitemShut {NoStop}%
\bibitem [{\citenamefont {Fischer}\ \emph {et~al.}(2023)\citenamefont
  {Fischer}, \citenamefont {Chiesa}, \citenamefont {Tacchino}, \citenamefont
  {Egger}, \citenamefont {Carretta},\ and\ \citenamefont
  {Tavernelli}}]{fischer_prx_2023}%
  \BibitemOpen
  \bibfield  {author} {\bibinfo {author} {\bibfnamefont {L.~E.}\ \bibnamefont
  {Fischer}}, \bibinfo {author} {\bibfnamefont {A.}~\bibnamefont {Chiesa}},
  \bibinfo {author} {\bibfnamefont {F.}~\bibnamefont {Tacchino}}, \bibinfo
  {author} {\bibfnamefont {D.~J.}\ \bibnamefont {Egger}}, \bibinfo {author}
  {\bibfnamefont {S.}~\bibnamefont {Carretta}},\ and\ \bibinfo {author}
  {\bibfnamefont {I.}~\bibnamefont {Tavernelli}},\ }\bibfield  {title}
  {\bibinfo {title} {Universal qudit gate synthesis for transmons},\ }\href
  {https://doi.org/10.1103/PRXQuantum.4.030327} {\bibfield  {journal} {\bibinfo
   {journal} {PRX Quantum}\ }\textbf {\bibinfo {volume} {4}},\ \bibinfo {pages}
  {030327} (\bibinfo {year} {2023})}\BibitemShut {NoStop}%
\bibitem [{\citenamefont {Deng}\ \emph {et~al.}(2024)\citenamefont {Deng},
  \citenamefont {Zheng}, \citenamefont {Liao}, \citenamefont {Zhou},
  \citenamefont {Ge}, \citenamefont {Zhao}, \citenamefont {Lan}, \citenamefont
  {Tan}, \citenamefont {Zhang}, \citenamefont {Li},\ and\ \citenamefont
  {Yu}}]{deng2024_zz}%
  \BibitemOpen
  \bibfield  {author} {\bibinfo {author} {\bibfnamefont {X.}~\bibnamefont
  {Deng}}, \bibinfo {author} {\bibfnamefont {W.}~\bibnamefont {Zheng}},
  \bibinfo {author} {\bibfnamefont {X.}~\bibnamefont {Liao}}, \bibinfo {author}
  {\bibfnamefont {H.}~\bibnamefont {Zhou}}, \bibinfo {author} {\bibfnamefont
  {Y.}~\bibnamefont {Ge}}, \bibinfo {author} {\bibfnamefont {J.}~\bibnamefont
  {Zhao}}, \bibinfo {author} {\bibfnamefont {D.}~\bibnamefont {Lan}}, \bibinfo
  {author} {\bibfnamefont {X.}~\bibnamefont {Tan}}, \bibinfo {author}
  {\bibfnamefont {Y.}~\bibnamefont {Zhang}}, \bibinfo {author} {\bibfnamefont
  {S.}~\bibnamefont {Li}},\ and\ \bibinfo {author} {\bibfnamefont
  {Y.}~\bibnamefont {Yu}},\ }\href {https://arxiv.org/abs/2408.16617} {\bibinfo
  {title} {Long-range $zz$ interaction via resonator-induced phase in
  superconducting qubits}} (\bibinfo {year} {2024}),\ \Eprint
  {https://arxiv.org/abs/2408.16617} {arXiv:2408.16617 [quant-ph]} \BibitemShut
  {NoStop}%
\bibitem [{\citenamefont {Britton}\ \emph {et~al.}(2012)\citenamefont
  {Britton}, \citenamefont {Sawyer}, \citenamefont {Keith}, \citenamefont
  {Wang}, \citenamefont {Freericks}, \citenamefont {Uys}, \citenamefont
  {Biercuk},\ and\ \citenamefont {Bollinger}}]{Britton2012_nature}%
  \BibitemOpen
  \bibfield  {author} {\bibinfo {author} {\bibfnamefont {J.~W.}\ \bibnamefont
  {Britton}}, \bibinfo {author} {\bibfnamefont {B.~C.}\ \bibnamefont {Sawyer}},
  \bibinfo {author} {\bibfnamefont {A.~C.}\ \bibnamefont {Keith}}, \bibinfo
  {author} {\bibfnamefont {C.-C.~J.}\ \bibnamefont {Wang}}, \bibinfo {author}
  {\bibfnamefont {J.~K.}\ \bibnamefont {Freericks}}, \bibinfo {author}
  {\bibfnamefont {H.}~\bibnamefont {Uys}}, \bibinfo {author} {\bibfnamefont
  {M.~J.}\ \bibnamefont {Biercuk}},\ and\ \bibinfo {author} {\bibfnamefont
  {J.~J.}\ \bibnamefont {Bollinger}},\ }\bibfield  {title} {\bibinfo {title}
  {{Engineered two-dimensional Ising interactions in a trapped-ion quantum
  simulator with hundreds of spins}},\ }\href
  {https://doi.org/10.1038/nature10981} {\bibfield  {journal} {\bibinfo
  {journal} {Nature}\ }\textbf {\bibinfo {volume} {484}},\ \bibinfo {pages}
  {489} (\bibinfo {year} {2012})}\BibitemShut {NoStop}%
\bibitem [{\citenamefont {Keesling}\ \emph {et~al.}(2019)\citenamefont
  {Keesling}, \citenamefont {Omran}, \citenamefont {Levine}, \citenamefont
  {Bernien}, \citenamefont {Pichler}, \citenamefont {Choi}, \citenamefont
  {Samajdar}, \citenamefont {Schwartz}, \citenamefont {Silvi}, \citenamefont
  {Sachdev}, \citenamefont {Zoller}, \citenamefont {Endres}, \citenamefont
  {Greiner}, \citenamefont {Vuleti{\ifmmode\acute{c}\else\'{c}\fi}},\ and\
  \citenamefont {Lukin}}]{Keesling2019_nature}%
  \BibitemOpen
  \bibfield  {author} {\bibinfo {author} {\bibfnamefont {A.}~\bibnamefont
  {Keesling}}, \bibinfo {author} {\bibfnamefont {A.}~\bibnamefont {Omran}},
  \bibinfo {author} {\bibfnamefont {H.}~\bibnamefont {Levine}}, \bibinfo
  {author} {\bibfnamefont {H.}~\bibnamefont {Bernien}}, \bibinfo {author}
  {\bibfnamefont {H.}~\bibnamefont {Pichler}}, \bibinfo {author} {\bibfnamefont
  {S.}~\bibnamefont {Choi}}, \bibinfo {author} {\bibfnamefont {R.}~\bibnamefont
  {Samajdar}}, \bibinfo {author} {\bibfnamefont {S.}~\bibnamefont {Schwartz}},
  \bibinfo {author} {\bibfnamefont {P.}~\bibnamefont {Silvi}}, \bibinfo
  {author} {\bibfnamefont {S.}~\bibnamefont {Sachdev}}, \bibinfo {author}
  {\bibfnamefont {P.}~\bibnamefont {Zoller}}, \bibinfo {author} {\bibfnamefont
  {M.}~\bibnamefont {Endres}}, \bibinfo {author} {\bibfnamefont
  {M.}~\bibnamefont {Greiner}}, \bibinfo {author} {\bibfnamefont
  {V.}~\bibnamefont {Vuleti{\ifmmode\acute{c}\else\'{c}\fi}}},\ and\ \bibinfo
  {author} {\bibfnamefont {M.~D.}\ \bibnamefont {Lukin}},\ }\bibfield  {title}
  {\bibinfo {title} {{Quantum Kibble{\textendash}Zurek mechanism and critical
  dynamics on a programmable Rydberg simulator}},\ }\href
  {https://doi.org/10.1038/s41586-019-1070-1} {\bibfield  {journal} {\bibinfo
  {journal} {Nature}\ }\textbf {\bibinfo {volume} {568}},\ \bibinfo {pages}
  {207} (\bibinfo {year} {2019})}\BibitemShut {NoStop}%
\bibitem [{\citenamefont {Johansson}\ \emph {et~al.}(2012)\citenamefont
  {Johansson}, \citenamefont {Nation},\ and\ \citenamefont {Nori}}]{qutip}%
  \BibitemOpen
  \bibfield  {author} {\bibinfo {author} {\bibfnamefont {J.~R.}\ \bibnamefont
  {Johansson}}, \bibinfo {author} {\bibfnamefont {P.~D.}\ \bibnamefont
  {Nation}},\ and\ \bibinfo {author} {\bibfnamefont {F.}~\bibnamefont {Nori}},\
  }\bibfield  {title} {\bibinfo {title} {{QuTiP: An open-source Python
  framework for the dynamics of open quantum systems}},\ }\href
  {https://doi.org/10.1016/j.cpc.2012.02.021} {\bibfield  {journal} {\bibinfo
  {journal} {Comput. Phys. Commun.}\ }\textbf {\bibinfo {volume} {183}},\
  \bibinfo {pages} {1760} (\bibinfo {year} {2012})}\BibitemShut {NoStop}%
\end{thebibliography}%

\appendix

\vspace{1cm}

\onecolumngrid

\section{Derivations of coherence and fidelity in the circuit} \label{apdx:derivations}
\subsection{Evaluating states in the circuit}

The quantum addition circuit  consists of two different unitaries which are Hadamard gate and controlled rotation.
In this section, we compute explicitly the state after  each step of the circuit, i.e., after the QFT and the SUM circuit in order to calculate the fidelity and the coherence when noise is inherently present in the circuit.

First, we consider the effect of phase damping channel (PDC) in the quantum addition circuit. We consider PDC after the application of each controlled rotation, and we evaluate the state after the QFT and the SUM circuit. Any general qudit state is of the form $\sum_{k,l=0}^{d-1}C_{kl}\ket{k}\bra{l}$ where \(C_{kl}\) follows all the properties to be a state. By passing through \(d\)-dimensional phase damping channel, the state evolves as 
\begin{equation*}   
\begin{split}
\sum_{k,l=0}^{d-1}C_{kl}&\ket{k}\bra{l}\xrightarrow{PDC} \sum_{E = 0}^{d}M_E\left(\sum_{k,l=0}^{d-1}C_{kl}\ket{k}\bra{l}\right)M_E^{\dagger}\\
& = (1-p)\left(\sum_{k,l=0}^{d-1}C_{kl}\ket{k}\bra{l} \right) + p C_{00}\ket{0}\bra{0}+...+ pC_{(d-1)(d-1)} \ket{d-1}\bra{d-1}  = \sum_{k,l=0}^{d-1}(1-p)^{1-\delta_{k,l}}C_{kl}\ket{k}\bra{l}, 
\end{split}
\end{equation*}
where \(\delta_{k,l}\) is the delta function and \(p\) is the  strength of the noise in the channel. Let the initial state be \(\ket{a_ta_{t-1}\ldots a_0}\).  Since the total state remains product before and after the application of PDC,   the state of the single subsystem after \(t^{th}\) controlled rotation is given by (see Fig. \ref{fig:schematics_qft_adder})

\begin{eqnarray}
&&\ket{a_t}  \xrightarrow{H_d} \frac{1}{{d}}\sum_{k,l = 0}^{d-1}e^{i 2 \pi (0.a_t)(k-l)}\ket{k}\bra{l}
\xrightarrow{R_d^{(2)}} \frac{1}{{d}}\sum_{k,l = 0}^{d-1}e^{i 2 \pi (0.a_ta_{t-1})(k-l)}\ket{k}\bra{l}\nonumber \\
&&\xrightarrow{PDC} \frac{1}{{d}}\sum_{k,l = 0}^{d-1}(1-p)^{(1-\delta_{k,l})}e^{i 2 \pi (0.a_ta_{t-1})(k-l)}\ket{k}\bra{l} \nonumber\\
&&\vdotswithin{=}\nonumber \\
&&\xrightarrow{R_d^{(t+1)}} \frac{1}{{d}}\sum_{k,l = 0}^{d-1}(1-p)^{(t-1)(1-\delta_{k,l})}e^{i 2 \pi (0.a_ta_{t-1}...a_0)(k-l)}\ket{k}\bra{l} \nonumber\\
&&\xrightarrow{PDC}  \frac{1}{{d}}\sum_{k,l = 0}^{d-1}(1-p)^{(t)(1-\delta_{k,l})}e^{i 2 \pi (0.a_ta_{t-1}...a_0)(k-l)}\ket{k}\bra{l}.\nonumber\\
    \label{eq:after_QFT}
\end{eqnarray}   


Hence the state  after the noisy QFT channel is given as
\begin{equation} 
\label{eq:qft_total_state}
    {\rho^{in}_{PDC}} = \bigotimes_{t=0}^{n-1}{\frac{1}{{d}}\sum_{k,l = 0}^{d-1}(1-p)^{t(1-\delta_{k,l})}e^{i 2 \pi (0.a_ta_{t-1}...a_0)(k-l)}\ket{k}\bra{l}}.
\end{equation}
In the SUM circuit, there is also noise after the application of each controlled rotation and in this case,  we need to consider the other integer, \(b\) too. The final state after the SUM circuit reads as
\begin{equation*}    
\label{eq:app_SUM}
\begin{split}
&\frac{1}{{d}}\sum_{k,l = 0}^{d-1}(1-p)^{t(1-\delta_{k,l})}e^{i 2 \pi (0.a_ta_{t-1}...a_0)(k-l)}\ket{k}\bra{l}\\
&\vdotswithin{=} \\
&\xrightarrow{R_d^{(m(q,t))}} \frac{1}{{d}}\sum_{k,l = 0}^{d-1}(1-p)^{(t+m(q,t)-1)(1-\delta_{k,l})}e^{i 2 \pi (0.a_ta_{t-1}...a_0+0.b_tb_{t-1}...b_{t-m(q,t)+1})(k-l)}\ket{k}\bra{l}\\
&\xrightarrow{PDC} \frac{1}{{d}}\sum_{k,l = 0}^{d-1}(1-p)^{(t+m(q,t))(1-\delta_{k,l})}e^{i 2 \pi (0.a_ta_{t-1}...a_0+0.b_tb_{t-1}...b_{t-m(q,t)+1})(k-l)}\ket{k}\bra{l},\\
\end{split}
\end{equation*}
where \(m(q,t)=\min (q,t+1)\) and \(q\) is the number of controlled rotation. Therefore, the total state after the SUM circuit is given as
\begin{equation}   
\rho_{PDC}^{out(q)} = \bigotimes_{t=0}^{n-1}\frac{1}{d}\sum_{k,l = 0}^{d-1}(1-p)^{(t+m(q,t))(1-\delta_{k,l})}e^{i 2 \pi (0.a_t...a_0+0.b_t...b_{t-m(q,t)+1})(k-l)}\ket{k}\bra{l}
\label{eq:gen_stateaftersum}
\end{equation} 
We are now ready to evaluate the fidelity and the coherence present in these states.

\subsection{Evaluation of fidelity at different points of circuit}

One of the figure of merit used in this work is the fidelity which quantifies the deviation of  the output state due to noise from the state in the ideal scenario. Let $\ket{\Psi^\alpha}$ be the desired state in absence of any noise and banding. The fidelity is then defined as
\begin{equation}    
f = \left\langle \Psi^\alpha | \rho_{er} | \Psi^\alpha \right\rangle,
\end{equation}
where $\alpha \in \{in,out\}$ and \(\rho_{er}=\bigotimes_{t=0}^{n-1}\rho_{er_{t}} \), and $\rho_{er_{t}}$ are the states of qudits in the noisy circuit. The fidelity can be re-expressed as
\begin{eqnarray}
    f &=&  \left(\bigotimes_{j = 0}^{n-1}\langle\Psi_j^{\alpha} |\right) \left(
 \bigotimes_{k=0}^{n-1}\rho_{er_{t}} \right) \left( \bigotimes_{l=0}^{n-1}|\Psi_l^{\alpha} \rangle \right) = \prod_{t=0}^{n-1}{\langle \Psi_t^{\alpha} | \rho_{er_t} | \Psi_t^{\alpha} \rangle}=\prod_{t=0}^{n-1} f_{er_t}.
\end{eqnarray}

This is possible because the state remains fully separable in every step of the circuit and hence finding the change in one of the qudits is enough  to obtain the state of the entire system.  


We derive the change of fidelity with the noise strength of phase damping channel. The fidelity of the subsystem after the application of Hadamard gate and the controlled rotation followed by a phase damping channel is given as

\begin{multline}
\begin{split}
f_{PDC_t} &= \frac{1}{d^2}  \sum_{k' = 0}^{d-1}e^{-i 2 \pi (0.a_ta_{t-1}...a_0)k'}\bra{k'} \sum_{k,l = 0}^{d-1}(1-p)^{t(1-\delta_{k,l})}e^{i 2 \pi (0.a_ta_{t-1}...a_0)(k-l)}\ket{k}\bra{l} \sum_{l' = 0}^{d-1}e^{i 2 \pi (0.a_ta_{t-1}...a_0)(l')}\ket{l'}\\
&=\frac{1}{d^2}\left(\sum_{k,l = 0}^{d-1}(1-p)^{t(1-\delta_{k,l})}\right)=\frac{1}{d} \left( 1 + (d-1)(1-p)^{t} \right).
\end{split}
\end{multline}
The subscript \(t\) in \(PDC\) denotes that the \(t^{th}\) qudit state is affected by noise and the fidelity is computed of the affected state with the desired \(t^{th}\) qudit. Therefore, the fidelity of the composite system after the QFT circuit is given by \(f_{PDC}^{in}=\prod_{t=0}^{n-1}{\frac{1}{d}\left( 1 + (d-1)(1-p)^{t} \right)}\). Similarly, we can calculate the fidelity at the end of the SUM circuit which reads as

\begin{equation}
\begin{split}
f_{PDC_t}^{out} &= \frac{1}{d^2}\left(\sum_{k,l = 0}^{d-1}(1-p)^{(t+m(q,t))(1-\delta_{kl})}e^{i 2 \pi (0.\overbrace{00...00}^{m(q,t)}b_{t-m(q,t)}...b_{0}(l-k))}\right)\\
&= \frac{1}{d^2} \left( d + \sum_{l>k ,~ l = 0}^{d-1}{2(1-p)^{(t+m(q,t))}\cos{(2\pi\times0.00...00b_{t-m(q,t)}...b_{0}(l-k))}} \right)\\
&= \frac{1}{d} \left( 1 + (1-p)^{(t+m(q,t))}\frac{1}{d}\sum_{k = 0}^{d-1}{2k\cos{(2\pi\times0.00...00b_{t-m(q,t)}...b_{0}(d-k))}} \right).\\
\end{split}
\end{equation}

Therefore, the  fidelity expression after considering the total output state takes the form as

\begin{equation}   
\label{eq:genfidelity}
\begin{split}
    &f_{PDC}^{out} =\prod_{t=0}^{n-1}{\frac{1}{d} \left( 1 + (1-p)^{(t+m(q,t))}\frac{1}{d}\sum_{k = 0}^{d-1}{2k\cos{(2\pi\times0.\overbrace{00...00}^{m(q,t)}b_{t-m(q,t)}...b_{0}(d-k))}}\right)}.
\end{split}
\end{equation}

\subsection{Qudit after different noisy channels}
\label{app:matrix_element}

In the quantum addition circuit, we deal only with the product state such that \(\rho=\bigotimes_{t=0}^{n-1}\rho_t\), where \(\rho_t\) be any general qudit state given as $\rho_t=\sum_{k,l=0}^{d-1}a_{kl}\ket{k}\bra{l}$. So for our calculation, it is enough to handle the effect of noise on individual qudit.

\subsubsection{Phase damping noise}
After passing through the phase damping channel, the individual state, \(\rho_t\) transforms as
\begin{eqnarray}  
&&\sum_{k,l=0}^{d-1}a_{kl}\ket{k}\bra{l}\xrightarrow{PDC}\sum_{E = 0}^{d}M_E\left(\sum_{k,l=0}^{d-1}a_{kl}\ket{k}\bra{l}\right)M_E^{\dagger}\nonumber \\
&&= (1-p)\left(\sum_{k,l=0}^{d-1}a_{kl}\ket{k}\bra{l} \right) + p a_{00}\ket{0}\bra{0} + ...+ pa_{(d-1)(d-1)} \ket{d-1}\bra{d-1} = \sum_{k,l=0}^{d-1}(1-p)^{1-\delta_{k,l}}a_{kl}\ket{k}\bra{l}.
\end{eqnarray}
Hence all the elements of the state scale by the factor \((1-p)^{1-\delta_{k,l}}\). A tensor product over all such qudit subsystem, the matrix elements of the entire state of the circuit scaled by a positive coefficient $p_{jk}$. Therefore, the noise satisfies Eq. (\ref{eq:noise_model}), following our assumption.

\subsubsection{Amplitude damping noise} \label{apdx:ampdamp_generalcriteria}

In Sec. \ref{sec:noiseandbanding}, we define the Kraus operators of the amplitude-damping channel in arbitrary dimension.  If \(\rho\) is the initial qudit state, after passing through ADC, the state becomes

\begin{align}
&\sum_{k,l=0}^{d-1}a_{kl}\ket{k}\bra{l}\xrightarrow{ADC} \sum_{E = 0}^{d}M_E\left(\sum_{k,l=0}^{d-1}a_{kl}\ket{k}\bra{l}\right)M_E^{\dagger}\nonumber\\
& = \sum_{k,l=0}^{d-1}\sqrt{(1-kp)(1-lp)}a_{kl}\ket{k}\bra{l}+ \sum_{k,l = 0}^{d-2}p a_{(k+1)(l+1)}\ket{k}\bra{l}+...+ pa_{(d-1)(d-1)} \ket{0}\bra{0}\nonumber\\
& =\sum_{k,l=0}^{d-1}\sqrt{(1-kp)(1-lp)}a_{kl}+ \sum_{m = 1}^{d-1-\max(k,l)}pa_{(k+m)(l+m)}\ket{k}\bra{l}.\nonumber
&. 
\end{align}

In the particular states that the Hadamard gate produces in the circuit, i.e., $\rho = \frac{1}{{d}}\sum_{k = 0}^{d-1}e^{i 2 \pi \phi(k-l)}\ket{k}\bra{l}$ with $a_{kl} = a_{(k+m)(l+m)}$, the above equation reduces to \begin{equation}\sum_{k,l=0}^{d-1}a_{kl}\ket{k}\bra{l}\xrightarrow{ADC} \sum_{k,l=0}^{d-1}\Biggl(\sqrt{(1-kp)(1-lp)} \nonumber + p(d-1-max(k,l))\Biggr)a_{kl}\ket{k}\bra{l}  \end{equation}

Hence, the net effect of  noise is to scale the elements of the density matrix by the factor $$ \sqrt{(1-kp)(1-lp)} + p(d-1-max(k,l)).$$ Again, by doing tensor product on all such subsystem of qudits, we construct the density matrix of the circuit where each individual element of the entire set get scaled up by positive coefficients $p_{kl}$ thus satisfying Eq. (\ref{eq:noise_model}).

\end{document}